# Second-Order Bi-Scalar-Vector-Tensor Field Equations Compatible with the Conservation of Charge in a Space of Four-Dimensions

by

Gregory W. Horndeski
Adjunct Associate Professor of Applied Mathematics
University of Waterloo
200 University Avenue
Waterloo, Ontario
Canada
N2L 3G1

email:
horndeskimath@gmail.com
ghorndeski@uwaterloo.ca

April 20, 2026

**ABSTRACT**

The purpose of this paper is to explore, in a space of four-dimensions, the possible forms that second-order, bi-scalar-vector-tensor field equations derivable from a variational principle can assume. In order to restrict this enormous class of field equations I shall first require that the equations governing the vector field (which will be identified with the vector potential of an electromagnetic field) be consistent with the notion of conservation of charge. Secondly I shall require that these vector equations reduce to Maxwell's equations in a flat space when the scalar fields are constant. Unfortunately even with these two powerful restrictions on the form of the field equations I have not been able to construct a Lagrangian which yields all possible field equations of this nature. This situation will lead to a discussion of other ways in which the field equations can be restricted to obtain viable bi-scalar-vector-tensor field equations. Lastly I shall make a few remarks on how the results obtained can be used to show that the Higgs field can generate electromagnetic fields in the early Universe, and how to couple bi-scalar fields to gauge-tensor fields to construct second-order, bi-scalar-Yang-Mills-tensor field theories compatible with the conservation of gauge charge.

## Section 1: Introduction

In the course of preparing my paper on second-order, bi-scalar-tensor field equations, Horndeski [1], I realized that it was possible to couple two scalar fields to the electromagnetic field tensor in such a way as to obtain Maxwell type equations for the vector field, with the scalar fields acting as the source of electromagnetism. To see how this can be done let $\varphi$ and $\xi$ denote the scalar fields, and $F_{ij} := A_{j,i} - A_{i,j}$ be the electromagnetic field tensor, where $A_i$ denotes the vector potential of the electromagnetic field and a comma denotes partial differentiation with respect to the local coordinates $x^i$ of a chart x. Let the Maxwell and bi-scalar Lagrangians $L_M$ and $L_{BS}$ be defined by

$$L_M := \tfrac{1}{4} g^{1/2} F_{ab} F^{ab} \tag{1}$$

and

$$L_{BS} := \kappa_{BS} g^{1/2} \varphi_a \xi_b F^{ab}, \tag{2}$$

where $\kappa_{BS}$ is a constant, $g:=|\det(g_{ab})|$, indices are lowered and raised by the metric tensor $g_{ab}$ and its inverse $g^{ab}$, with the summation convention being used throughout this paper. The vector and scalar equations generated by $L_M+L_{BS}$ are given by

$$E^i(L_M+L_{BS}) \equiv \frac{\delta(L_M+L_{BS})}{\delta A_i} = g^{1/2}\{F^{ij}{}_{|j} - \kappa_{BS}[\xi^i \Box\varphi - \varphi^i \Box\xi + \varphi^j \xi^i{}_j - \xi^j \varphi^i{}_j]\} = 0 \tag{3}$$

$$E_\varphi(L_M+L_{BS}) \equiv \frac{\delta(L_M+L_{BS})}{\delta\varphi} = \kappa_{BS} g^{1/2} \xi_b F^{bj}{}_{|j} = 0 \tag{4}$$

$$E_\xi(L_M+L_{BS}) \equiv \frac{\delta(L_M+L_{BS})}{\delta\xi} = -\kappa_{BS} g^{1/2} \varphi_b F^{bj}{}_{|j} = 0\,, \tag{5}$$

where $\Box\varphi := g^{ab}\varphi_{ab}$ and a vertical bar denotes covariant differentiation except on $\varphi$ and $\xi$, where it is omitted, so $\varphi_{ab} := \varphi_{|ab}$ . Combining (3), (4) and (5) we find that the scalar equations can be rewritten

$$\xi_a\xi^a\Box\varphi - \xi_a\varphi^a\Box\xi + \varphi^a\xi^b\xi_{ab} - \xi^a\xi^b\varphi_{ab} = 0 \qquad (6)$$

$$\xi_a\varphi^a\Box\varphi - \varphi_a\varphi^a\Box\xi + \varphi^a\varphi^b\xi_{ab} - \varphi^a\xi^b\varphi_{ab} = 0\,, \qquad (7)$$

Thus we see that the equations governing $\varphi$ and $\xi$ can be reformulated to be independent of $A_i$ and its derivatives (which I regard as very interesting), with the electromagnetic field equation expressible as

$$F^{ij}{}_{|j} = \kappa_{BS}J_{BS}{}^{i} \qquad (8)$$

where the bi-scalar current $J_{BS}{}^{i}$ is defined by

$$J_{BS}{}^{i} := \xi^i\Box\varphi - \varphi^i\Box\xi + \varphi^j\xi^i{}_j - \xi^j\varphi^i{}_j\,. \qquad (9)$$

So these equations would have the bi-scalar fields act as a source for electromagnetism in regions devoid of charge. At first sight this might seem ridiculous, since we all "know" that electromagnetic fields are only created by charged particles. But do we really? After all, this knowledge was acquired during the past several hundred years in an environment where gravity was fairly weak and virtually constant, with the Higgs field (a pair of bi-scalar fields) being constant. So why should we conclude from these observations that Maxwell's vacuum field equations, $F^{ij}{}_{|j}=0$, should apply in all situations throughout the history of our universe? We shouldn't. We should look to see what possible second-order bi-scalar-vector-tensor field equations are out there, and then look for reasons to cull out those that seem inappropriate. Two good restrictions on such theories are:

(i) the field equations should be compatible with the conservation of charge. This requires that $E^i(L)$ should be identically divergence-free since when charges are present the electromagnetic equation usually has the form $E^i(L) = (\text{constant})g^{1/2}J^i$, where $J^i$ is the charge-current vector which must be divergence-free; and

(ii) when the curvature tensor vanishes and the scalar fields are constant the vacuum field equations

for the electromagnetic field should reduce to Maxwell's equations $F^{ij}{}_{|j}=0$.

The Lagrangian $L_M+L_{BS}$ presented above satisfies these criteria.

In Horndeski [2], I show that in the absence of scalar fields the most general second-order vector-tensor field equations in a space of four-dimensions which satisfy (i) and (ii) above can be obtained from the generalized Einstein-Maxwell Lagrangian

$$L_{GEM} := g^{1/2}\{-\lambda R + \tfrac{1}{4}\gamma F_{ab}F^{ab} + 2\mu + \tfrac{1}{4}\tau_{VT}\ \delta^{abcd}_{pqrs} F_{ab}F^{pq}R_{cd}{}^{rs}\}\ , \tag{10}$$

with $\lambda$, $\gamma$, $\mu$ and $\tau_{VT}$ being constants and $\delta^{abcd}_{pqrs}$ being the 4x4 generalized Kronecker delta. (Here I define the components $R_a{}^b{}_{cd}$ of the curvature tensor so that when $V^b$ denotes the components of an arbitrary contravariant vector, $R_a{}^b{}_{cd}$ are defined by the Ricci identity, $V^b{}_{|cd}-V^b{}_{|dc} = V^aR_{abcd}$, with the Ricci tensor $R_{ac} := R_a{}^b{}_{cb}$ , curvature scalar $R:= g^{ac}R_{ac}$ , and the Einstein tensor $G_{ac} := R_{ac}-\tfrac{1}{2}g_{ac}R$.) When $\lambda=\gamma=1$, and $\tau_{VT}=0$, the Lagrangian (10) becomes the usual Einstein-Maxwell Lagrangian with cosmological term. When $\tau_{VT} \neq 0$, the vacuum electromagnetic equations turn out to be

$$F^{ij}{}_{|j} = -\tau_{VT}\ \delta^{iabc}_{pqrs} F^{pq}{}_{|a}R^{rs}{}_{bc}\ . \tag{11}$$

So we have curvature coupled to $F^{pq}{}_{|a}$ acting as the source of electromagnetism, and hence we have $F_{ab}$ interacting with itself via the curvature tensor. Upon examining the Lagrangians $L_M$ and $L_{BS}$ given in (1) and (2) we see that this self-interacting problem could be avoided if we replace the term with coefficient $\tau_{VT}$ in (11) with the bi-scalar-vector-tensor Lagrangian

$$L_{BSVT} := \tfrac{1}{2}\kappa_C g^{1/2}\ \delta^{abcd}_{pqrs}\ \varphi_a\xi_b F^{pq}R_{cd}{}^{rs}\ , \tag{12}$$

where $\kappa_C$ is a constant. This Lagrangian will generate second-order, bi-scalar-vector-tensor field equations satisfying (i) and (ii) above with

$$E^i(L_{BSVT}) = -\kappa_C g^{1/2}\, J_C{}^i \tag{13}$$

where

$$J_C{}^i := \delta^{iabc}_{pqrs}\, \xi^p{}_a \varphi^q\, R_{bc}{}^{rs} - \delta^{iabc}_{pqrs}\, \xi^q \varphi^p{}_a R_{bc}{}^{rs} =$$

$$= 2\{\xi^i{}_a\varphi^a R - \varphi^i{}_a\xi^a R - 2\,\xi^i{}_a\varphi_b R^{ab} + 2\,\varphi^i{}_a\xi_b R^{ab} - R\varphi^i\Box\xi + R\xi^i\Box\varphi + 2\varphi^a R^i{}_a\Box\xi - 2\xi^a R^i{}_a\Box\varphi +$$

$$+ 2\xi_{ab}R^{ab}\varphi^i - 2\varphi_{ab}R^{ab}\xi^i - 2\xi_{ab}\varphi^a R^{ib} + 2\varphi_{ab}\xi^a R^{ib} + 2\xi_{ab}\varphi_c R^{iabc} - 2\varphi_{ab}\xi_c R^{iabc}\}\,. \tag{14}$$

So the bi-scalar Einstein-Maxwell Lagrangian with cosmological term, $L_{BSEM}$, defined by

$$L_{BSEM} := g^{1/2}\,\{-R + 2\mu + L_M + L_{BS} + L_{BSVT}\} \tag{15}$$

will yield second-order field equations with the electromagnetic field equation $E^i(L_{BSEM})=0$ becoming

$$F^{ij}{}_{|j} = \kappa_{BS}J_{BS}{}^i + \kappa_C J_C{}^i \tag{16}$$

where $J_{BS}{}^i$ and $J_C{}^i$ are given by (9) and (14). Thus we see that when using the Lagrangian $L_{BSEM}$ the electromagnetic field has two blocks of terms as its source in vacuum. One block involving the first and second covariant derivatives of the bi-scalar fields and a second block involving the first and second covariant derivatives of the scalar field coupled to the curvature tensor. (16) shows us that when the bi-scalar fields are constant the electromagnetic equation reduces to Maxwell's equation in vacuum. Note that each of the new "currents" $J_{BS}{}^i$ and $J_C{}^i$ have identically vanishing divergences and hence are conserved currents in their own right.

The Lagrangians presented in (1), (2), (10) and (15) are of the form

$$L = L(g_{ab};\, g_{ab,c};\, g_{ab,cd};\, \varphi;\, \varphi_{,a};\, \varphi_{,ab};\, \xi;\, \xi_{,a};\, \xi_{,ab};\, A_{a,b})\,. \tag{17}$$

Before I can introduce formulas to express the variational derivatives of this Lagrangian in a manifestly tensorial form I have to introduce some notation, which will also be required in the next section.

If $T^{\cdots}$ denotes the components of a second-order bi-scalar-vector-tensor concomitant, let

$$T^{\cdots;ab} := \frac{\partial T^{\cdots}}{\partial g_{ab}}; \quad T^{\cdots;ab,c} := \frac{\partial T^{\cdots}}{\partial g_{ab,c}}; \quad T^{\cdots;ab,cd} := \frac{\partial T^{\cdots}}{\partial g_{ab,cd}}; \tag{18}$$

$$T_{\varphi}{}^{\cdots} := \frac{\partial T^{\cdots}}{\partial \varphi}; \quad T_{\varphi}{}^{\cdots;c} := \frac{\partial T^{\cdots}}{\partial \varphi_{,c}}; \quad T_{\varphi}{}^{\cdots;cd} := \frac{\partial T^{\cdots}}{\partial \varphi_{,cd}}; \quad T_{\xi}{}^{\cdots} := \frac{\partial T^{\cdots}}{\partial \xi}; \quad T_{\xi}{}^{\cdots;c} := \frac{\partial T^{\cdots}}{\partial \xi_{,c}}; \quad T_{\xi}{}^{\cdots;cd} := \frac{\partial T^{\cdots}}{\partial \xi_{,cd}}; \tag{19}$$

$$T^{\cdots;a} := \frac{\partial T^{\cdots}}{\partial A_a}; \quad T^{\cdots;a,b} := \frac{\partial T^{\cdots}}{\partial A_{a,b}}; \text{ and } \quad T^{\cdots;a,bc} := \frac{\partial T^{\cdots}}{\partial A_{a,bc}}. \tag{20}$$

One should note that $T^{\cdots;ab,cd}$ is symmetric in a,b and in c,d, while $T_{\varphi}{}^{\cdots;cd}$, $T_{\xi}{}^{\cdots;cd}$ and $T^{\cdots;a,cd}$ are symmetric in c,d. In general the partial derivatives given in (18)-(20) are non-tensorial. In fact the only tensorial derivatives amongst that collection of partial derivatives are

$$T^{\cdots;ab,cd};\ T_{\varphi}{}^{\cdots};\ T_{\varphi}{}^{\cdots;cd};\ T_{\xi}{}^{\cdots};\ T_{\xi}{}^{\cdots;cd} \text{ and } T^{\cdots;a,bc}.$$

However Rund [3], du Plessis [4] and Horndeski [5] have demonstrated that it is possible to construct tensor densities from them. When L has the form given in (17) these tensor densities are expressible as:

$$\Pi^{hi}(L) = \tfrac{1}{2}g^{hi}L - R_{r}{}^{i}{}_{ks}L^{;rs,hk} + \tfrac{1}{3}R_{r}{}^{h}{}_{ks}L^{;rs,ik} + \tfrac{1}{2}L^{;r,h}F_{r}{}^{i} - \tfrac{1}{2}\varphi^{i}\Phi^{h}(L) - \varphi^{i}{}_{r}L_{\varphi}{}^{;rh} - \tfrac{1}{2}\xi^{i}\Xi^{h}(L) - \xi^{i}{}_{r}L_{\xi}{}^{;rh} \tag{21}$$

$$\Pi^{hi,j}(L) = \tfrac{1}{2}(L_{\varphi}{}^{;hi}\varphi^{j} - L_{\varphi}{}^{;ji}\varphi^{h} - L_{\varphi}{}^{;jh}\varphi^{i} + L_{\xi}{}^{;hi}\xi^{j} - L_{\xi}{}^{;ji}\xi^{h} - L_{\xi}{}^{;jh}\xi^{i}), \tag{22}$$

$$\Phi^{i}(T^{\cdots}) := T^{\cdots}{}_{\varphi}{}^{;i} + \Gamma^{i}{}_{ab}T^{\cdots}{}_{\varphi}{}^{;ab} \text{ and } \Xi^{i}(T^{\cdots}) := T^{\cdots}{}_{\xi}{}^{;i} + \Gamma^{i}{}_{ab}T^{\cdots}{}_{\xi}{}^{;ab}, \tag{23}$$

where $\Gamma^{i}{}_{ab}$ are the components of the Christoffel symbols of the second kind.

It is possible to use the results of my thesis (see pages 46 and 116 of Horndeski [5]) to show that the Euler-Lagrange tensor densities derivable from (17) are given by

$$E^{ij}(L) = \Pi^{ij}(L) - \Pi^{ij,k}(L)_{|k} + L^{;ij,hk}{}_{|hk}; \tag{24}$$

$$E_\varphi(L) = L_\varphi - \Phi^i(L)_{|i} + L_\varphi{}^{;ij}{}_{|ij}\,; \tag{25}$$

$$E_\xi(L) = L_\xi - \Xi^i(L)_{|i} + L_\xi{}^{;ij}{}_{|ij} \tag{26}$$

and

$$E^i(L) = -\, L^{;i,j}{}_{|j}\,. \tag{27}$$

Using (24)-(27) it can be shown that for the Lagrangian $L_{BSVT}$ given in (12)

$$\begin{aligned}
&E^{hi}(L_{BSVT}) = \\
&= \kappa_C g^{1/2}\{F^{ha}{}_{|a}\Box\varphi\xi^i + F^{ia}{}_{|a}\Box\varphi\xi^h - F^{ha}{}_{|a}\Box\xi\varphi^i - F^{ia}{}_{|a}\Box\xi\varphi^h - F^{ba}{}_{|a}\varphi_b{}^h\xi^i - F^{ba}{}_{|a}\varphi_b{}^i\xi^h + F^{ba}{}_{|a}\xi_b{}^h\varphi^i + F^{ba}{}_{|a}\xi_b{}^i\varphi^h + \\
&- F^{ha}{}_{|a}\varphi^{ib}\xi_b - F^{ia}{}_{|a}\varphi^{hb}\xi_b + F^{ha}{}_{|a}\xi^{ib}\varphi_b + F^{ia}{}_{|a}\xi^{hb}\varphi_b + 2F^{ba}{}_{|a}\varphi^{hi}\xi_b - 2F^{ba}{}_{|a}\xi^{hi}\varphi_b + 2F^{ba}{}_{|a}\varphi_{bc}\xi^c g^{hi} - 2F^{ba}{}_{|a}\xi_{bc}\varphi^c g^{hi} + \\
&- 2F^{ba}{}_{|a}\xi_b\Box\varphi g^{hi} + 2F^{ba}{}_{|a}\varphi_b\Box\xi g^{hi} - F^{ha|b}\varphi_{ab}\xi^i - F^{ia|b}\varphi_{ab}\xi^h + F^{ha|b}\xi_{ab}\varphi^i + F^{ia|b}\xi_{ab}\varphi^h + F^{hb}{}_{|a}\varphi^{ai}\xi_b + F^{ib}{}_{|a}\varphi^{ah}\xi_b + \\
&- F^{hb}{}_{|a}\xi^{ai}\varphi_b - F^{ib}{}_{|a}\xi^{ah}\varphi_b + F^{ha|i}\varphi_{ab}\xi^b + F^{ia|h}\varphi_{ab}\xi^b - F^{ha|i}\xi_{ab}\varphi^b - F^{ia|h}\xi_{ab}\varphi^b - F^{ha|i}\Box\varphi\xi_a - F^{ia|h}\Box\varphi\xi_a + F^{ha|i}\Box\xi\varphi_a + \\
&+ F^{ia|h}\Box\xi\varphi_a + F^{ab|i}\varphi_a{}^h\xi_b + F^{ab|h}\varphi_a{}^i\xi_b - F^{ab|i}\xi_a{}^h\varphi_b - F^{ab|h}\xi_a{}^i\varphi_b - 2F^{ab|c}\varphi_{ac}\xi_b g^{hi} + 2F^{ab|c}\xi_{ac}\varphi_b g^{hi}\} + \\
&+ \kappa_C\{-\varphi_a\xi_b F^{bc}R_c{}^{hia} - \varphi_a\xi_b F^{bc}R_c{}^{iha} + \varphi_a\xi_b F^{ac}R_c{}^{hib} + \varphi_a\xi_b F^{ac}R_c{}^{ihb} - 2g^{hi}\varphi_a\xi_b F^{bc}R^a{}_c + 2g^{hi}\varphi_a\xi_b F^{ac}R^b{}_c - 2g^{hi}\varphi_a\xi_b F^{ab}R \\
&+ 4\varphi_a\xi_b F^{ab}R^{hi} - \xi^i\varphi_a F^{ab}R^h{}_b - \xi^h\varphi_a F^{ab}R^i{}_b + \varphi^i\xi_a F^{ab}R^h{}_b + \varphi^h\xi_a F^{ab}R^i{}_b + \varphi_a\xi_b F^{bh}R^{ia} + \varphi_a\xi_b F^{bi}R^{ha} - \varphi_a\xi_b F^{ah}R^{bi} + \\
&- \varphi_a\xi_b F^{ai}R^{bh} + \tfrac{1}{2}\xi^i\varphi_a F^{ah}R + \tfrac{1}{2}\xi^h\varphi_a F^{ai}R - \tfrac{1}{2}\varphi^i\xi_a F^{ah}R - \tfrac{1}{2}\varphi^h\xi_a F^{ai}R\}\,.
\end{aligned} \tag{28}$$

$$E_\varphi(L_{BSVT}) = -\tfrac{1}{2}\kappa_C\ \delta^{abcd}_{pqrs}\,\xi_b F^{pq}{}_{|a}R_{cd}{}^{rs} = -2\,\kappa_C\{\xi_b F^{ab}{}_{|a}R + 2\xi_b F^{ca}{}_{|a}R^b{}_c + 2\xi_b F^{bc}{}_{|a}R^a{}_c + F_{ab|c}\xi_d R^{abcd}\}, \tag{29}$$

$$E_\xi(L_{BSVT}) = \tfrac{1}{2}\kappa_C\ \delta^{abcd}_{pqrs}\,\varphi_b F^{pq}{}_{|a}R_{cd}{}^{rs} = 2\,\kappa_C\{\varphi_b F^{ab}{}_{|a}R + 2\varphi_b F^{ca|a}R^b{}_c + 2\varphi_b F^{bc}{}_{|a}R^a{}_c + F_{ab|c}\varphi_d R^{abcd}\}, \tag{30}$$

with $E^i(L_{BSVT})$ given in (13). In deriving the result presented in (28) I made use of the fact that in a space of four-dimensions

$$0 = \delta^{habcd}_{tpqrs}\, g^{ti}\varphi_a\xi_b F^{pq}R_{cd}{}^{rs} + \delta^{iabcd}_{tpqrs}\, g^{th}\varphi_a\xi_b F^{pq}R_{cd}{}^{rs}\,.$$

So the expression given in (28) is not valid in spaces of dimension n>4.

In the next section we shall study the possible forms which second-order bi-scalar-vector-tensor field equations can assume. That investigation will show that even with the demand that the equations be compatible with conservation of charge there exists a cornucopia of possible field equations, which will not be appreciably simplified by the requirement that the vector equation reduces to Maxwell's equation when the space is flat and the bi-scalar fields are constant. One reason for this is that if we have a Lagrangian that yields such field equations we can always add to it a Lagrangian of the form presented in (17) which is independent of $A_{a,b}$ which yields second-order bi-scalar-tensor field equations, and due to [1], there are a copious supply of such Lagrangians. This will lead us to consider Lagrangians L that are the sum of Lagrangians of the form given in (17), with each term in that sum having a non-zero derivative with respect to $A_{a,b}$. The goal of this investigation will be to determine what further restrictions need to be imposed on field theories generated by Lagrangians of this form in order that the $E^i(L)$ equation be independent of $A_a$ and its derivatives. The Lagrangians of such theories could then be added to $L_M$ to produce non-trivial second-order bi-scalar-vector-tensor field theories consistent with charge conservation and such that the electromagnetic field equation reduces to Maxwell's in a flat space when the scalar fields are constant, while in curved space the electromagnetic field equations have the form $F^{ij}{}_{|j} = J^i$ , where the conserved current vector $J^i$ has the functional form

$$J^i = J^i(g_{ab}; g_{ab,c}; g_{ab,cd}; \varphi; \varphi_{,a}; \varphi_{,ab}; \xi; \xi_{,a}; \xi_{,ab}). \qquad (31)$$

Hence for such theories we shall not have the electromagnetic field serving as a source for itself.

The Lagrangians $L_{BS}$ and $L_{BSVT}$ given in (2) and (12) are Lagrangians of the required type, but they are not alone. To see why note that for these two Lagrangians $\varphi_a$ and $\xi_b$, are always summed into something that is antisymmetric. So we can replace the $\varphi_a\xi_b$ combination in these Lagrangians by the exact two form $\omega := d\varphi \wedge d\xi = d(\varphi d\xi)$, with local components $\omega_{ab} = ½(\varphi_a\xi_b - \varphi_b\xi_a)$, without affecting the value of these Lagrangians. Consequently the bi-scalar fields enter these Lagrangians in a manner similar to which the electromagnetic field tensor $F_{ab} = A_{b,a} - A_{a,b}$ enters these Lagrangians. This suggests that we should look at which of the bi-scalar-tensor Lagrangians presented in [1] can be represented in terms of $\omega_{ab}$. Due to the **Theorem** and the discussion in **Section 3** of [1], three such Lagrangians are

$$L_0 := K_0 g^{1/2} \omega_{ab}\, \omega^{ab}, \tag{32}$$

$$L_{15} := g^{1/2}\,(K_1\, \delta^{abc}_{pqr}\ \omega_{ab}\omega^{pq}\ \varphi_c{}^r + K_2\ \delta^{abc}_{pqr}\ \omega_{ab}\omega^{pq}\xi_c{}^r)\ , \tag{33}$$

$$L_{*} := g^{1/2}\{K\, \delta^{abcd}_{pqrs}\ \omega_{ab}\omega^{pq}R_{cd}{}^{rs} - 4K_X\, \delta^{abcd}_{pqrs}\ \omega_{ab}\omega^{pq}\varphi_c{}^r\varphi_d{}^s - 4K_Y\, \delta^{abcd}_{pqrs}\ \omega_{ab}\omega^{pq}\xi_c{}^r\xi_d{}^s +$$

$$-\ 4K_Z\, \delta^{abcd}_{pqrs}\ \omega_{ab}\omega^{pq}\ \varphi_c{}^r\xi_d{}^s\ \} \tag{34}$$

where $K_0$, $K_1$, $K_2$ and K are real valued scalar functions of φ, ξ, X, Y and Z with

$$X := g^{ab}\varphi_a\varphi_b\ ,\ Y := g^{ab}\xi_a\xi_b\ \text{and}\ \ Z := g^{ab}\varphi_a\xi_b\ . \tag{35}$$

$K_0$ is an arbitrary function , while K is a bi-scalar-tensor (=:BST) function which requires $K_{xy}=\frac{1}{4}K_{zz}$ and the pair $(K_1,K_2)$ is a bi-scalar-tensor conjugate pair (=:BSTCP) of functions which requires $K_{1Z}=2K_{2X}$ and $K_{2Z}=2K_{1Y}$. The Lagrangians $L_{15}$ and $L_{*}$ appeared in Ohashi, *et al.,* [6], Kobayashi, *et al.,* [7], and Akama & Kobayashi [8].

There are other Lagrangians of the $L_0$ form that we could consider such as $K_0 g^{1/2}\omega_{ab}\omega^{bc}\omega_{cd}\,\omega^{da} = \frac{1}{2}K_0 g^{1/2}(\omega_{ab}\omega^{ab})^2$, or, more generally, $K_0 g^{1/2} S(g_{ab},\omega_{ab})$, where S is a scalar concomitant of the indicated variables. However, we shall just stick with the simplest such Lagrangian presented in (32).

Due to my remarks above we are led to consider the following seven Lagrangians

$$L_{01} := K_{01} g^{1/2}\omega_{ab}\, F^{ab},\ L_{02} := K_{02} g^{1/2} F_{ab}\, F^{ab},\ L_{03} := K_{03} g^{1/2}\omega_{ab} {*F^{ab}},\ L_{04} := K_{04} g^{1/2} F_{ab} {*F^{ab}}, \tag{36}$$

$$L_{15}' := g^{1/2}\,(K_1'\, \delta^{abc}_{pqr}\ \omega_{ab}F^{pq}\ \varphi_c{}^r + K_2'\ \delta^{abc}_{pqr}\ \omega_{ab}F^{pq}\xi_c{}^r) \tag{37}$$

$$L_{15}'' := g^{1/2}\,(K_1''\,\delta^{abc}_{pqr}\;F_{ab}F^{pq}\,\varphi_c{}^r + K_2''\,\delta^{abc}_{pqr}\;F_{ab}F^{pq}\xi_c{}^r) \tag{38}$$

$$L_{*}' := g^{1/2}\{K'\,\delta^{abcd}_{pqrs}\,\omega_{ab}F^{pq}R_{cd}{}^{rs} - 4K_X'\,\delta^{abcd}_{pqrs}\,\omega_{ab}F^{pq}\varphi_c{}^r\varphi_d{}^s - 4K_Y'\;\delta^{abcd}_{pqrs}\,\omega_{ab}F^{pq}\xi_c{}^r\xi_d{}^s +$$

$$-\;4K_Z'\,\delta^{abcd}_{pqrs}\,\omega_{ab}F^{pq}\;\varphi_c{}^r\xi_d{}^s\;\} \tag{39}$$

$$L_{*}'' := g^{1/2}\{K''\,\delta^{abcd}_{pqrs}\,F_{ab}F^{pq}R_{cd}{}^{rs} - 4K_X''\,\delta^{abcd}_{pqrs}\,F_{ab}F^{pq}\varphi_c{}^r\varphi_d{}^s - 4K_Y''\;\delta^{abcd}_{pqrs}\,F_{ab}F^{pq}\xi_c{}^r\xi_d{}^s +$$

$$-\;4K_Z''\,\delta^{abcd}_{pqrs}\,F_{ab}F^{pq}\;\varphi_c{}^r\xi_d{}^s\;\}, \tag{40}$$

where $K_{01}$,..., $K_{04}$ are arbitrary functions of φ, ξ, X, Y and Z; $(K_1',K_2')$ and $(K_1'',K_2'')$ are BSTCPs; K' and K'' are BST functions. $L_0'$ is a generalization of the Lagrangian $L_{BS}$ given in (2) and $L_{*}'$ is a generalization of the Lagrangian $L_{BSVT}$ given in (12). It is obvious that $L_0'$, $L_0''$ and $L_0'''$ yield second-order bi-scalar-vector-tensor field equations. Showing that this is also true for the other Lagrangians appearing in (37)-(40) is a bit more work. Nevertheless, it is possible to use (21)-(23) and (24)-(27) to establish that all of the Lagrangians appearing in (37)-(40) yield second-order field equations. One of the key things required in confirming this requires noting that since $\omega_{[ab|c]} = F_{[ab|c]} = 0$, quantities such as $\delta^{abc}_{pqr}\,\omega_{ab}{}^{|p}{}_{|c}$ are of second-order.

The following theorem recapitulates some of the work done in this introduction:

**Theorem 1:** If any one (or more) of the Lagrangians appearing in (36)-(40) is added to the Maxwell Lagrangian $L_M$ given in (1) then the resulting bi-scalar-vector-tensor field equations generated by that Lagrangian will be of second-order, consistent with the conservation of charge and such that the electromagnetic field equation reduces to Maxwell's equations in a flat space when the scalar fields are constant.■

One should note that when the Lagrangians $L_{02}$, $L_{04}$, $L_{15}''$ and $L_{*}''$ are added to $L_M$ the resultant vector equations would have $F_{ab}$ appearing in the conserved current that acts as the source for $F^{ij}{}_{|j}$.

I would now like to conclude this introduction with two remarks.

First, unlike $L_M$ given in (1) the Lagrangian given in (2) is not invariant (modulo sign) under the duality transformation that replaces $F^{ab}$ by $*F^{ab}$. However, we can define a bi-scalar duality transformation $*_{BS}$ by:

$$*_{BS}\colon\ \omega^{ab}\rightarrow *\omega^{ab} := \tfrac{1}{2}g^{-\frac{1}{2}}\varepsilon^{abcd}\omega_{cd}\ ;\ F^{ab}\rightarrow *F^{ab} := \tfrac{1}{2}g^{-\frac{1}{2}}\varepsilon^{abcd}F_{cd}\ ;\ \text{and } R^{abcd}\rightarrow *R*^{abcd} := \tfrac{1}{4}g^{-1}\varepsilon^{abpq}\varepsilon^{cdrs}R_{pqrs}. \quad (41)$$

The bi-scalar duality transformation will leave the Lagrangians given in (36)-(40) invariant modulo sign. In demonstrating this note that*; e.g.,* $*F_{cd} := g_{ca}g_{db}*F^{ab} = -\tfrac{1}{2}g^{\frac{1}{2}}\varepsilon_{cdlm}F^{lm}$ when the metric is Lorentzian.

I admit that the $*_{BS}$ transformation is a bit strange since as defined it does not act on $\varphi$, $\xi$, X, Y or Z in the coefficient functions, nor on the second derivatives of $\varphi$ and $\xi$ where they appear in the Lagrangians. So it only seems to make sense when the coefficient functions in the Lagrangians in (36)-(40) are constants, which is a severe restriction.

The second thing that I would like to point out is that for the case where there is only one scalar field, there exist scalar-vector-tensor Lagrangians compatible with the conservation of charge and such that they yield Maxwell's equations in a flat space when the scalar field is constant. For let's consider the Lagrangian

$$L_s := \tfrac{1}{2}g^{\frac{1}{2}}\kappa_S\varphi_a\varphi_b F^{ac}F^{b}{}_{c} \qquad (42)$$

where $\kappa_S$ is a constant. This Lagrangian yields second-order scalar-vector-tensor field equations with

$$E^i(L_S) = g^{\frac{1}{2}}\kappa_S\{\Box\varphi\varphi_a F^{ai} + \varphi_a\varphi^{ab}F_b{}^i + \varphi_a\varphi_b F^{ai|b} + \varphi^{ia}\varphi^b F_{ab} - \varphi^i\varphi_a F^{ab}{}_{|b}\}\ . \qquad (43)$$

In my opinion this Lagrangian has the same problem that the Lagrangians appearing in (36)-(40) that are quadratic in $F_{ab}$ have; *viz.,* when you add one to $L_M$ the field equations for the electromagnetic field have the form $F^{ij}{}_{|j}$ equals terms involving $F_{ab}$ and its covariant derivative and not just the scalar and tensor fields. Consequently $F_{ab}$ will be acting as its own source for these Lagrangians.

**Section 2: The Structure of the Field Tensor Densities**

In order to determine the most general second order bi-scalar-vector-tensor field equations in a space of four-dimensions let us consider a quartet of tensor density concomitants $(A^{ij},B,C,D^i)$ which satisfy the following conditions**:**

(i) functionally $A^{ij}$, B, C and $D^i$ have the form

$$A^{ij}= A^{ij}(g_{ab};g_{ab,c};g_{ab,cd};\varphi;\varphi_{,a};\varphi_{,ab};\xi;\xi_{a};\xi_{,ab};A_a;A_{a,b};A_{a,bc}) \tag{44}$$

$$B = B(g_{ab};g_{ab,c};g_{ab,cd};\varphi;\varphi_{,a};\varphi_{,ab};\xi;\xi_{a};\xi_{,ab};A_a;A_{a,b};A_{a,bc}) \tag{45}$$

$$C = C(g_{ab};g_{ab,c};g_{ab,cd};\varphi;\varphi_{,a};\varphi_{,ab};\xi;\xi_{a};\xi_{,ab};A_a;A_{a,b};A_{a,bc}) \tag{46}$$

$$D^i = D^i(g_{ab};g_{ab,c};g_{ab,cd};\varphi;\varphi_{,a};\varphi_{,ab};\xi;\xi_{a};\xi_{,ab};A_a;A_{a,b};A_{a,bc})\ ; \tag{47}$$

(ii) there exists a class $C^\infty$ Lagrange scalar density L of the form

$$L = L(g_{ab};g_{ab,c};\ldots;\varphi;\varphi_{,a};\ldots:\xi;\xi_{a};\ldots;A_a;A_{a,b};\ldots)$$

which is of finite differential order in $g_{ab}$, $\varphi$, $\xi$, and $A_a$, and such that

$$A^{ij} = E^{ij}(L) \equiv \frac{\delta L}{\delta g_{ij}} := \frac{\partial L}{\partial g_{ij}} - \frac{d}{dx^k}\frac{\partial L}{\partial g_{ij,k}} + \ldots \tag{48}$$

$$B = E_\varphi(L) \equiv \frac{\delta L}{\delta \varphi} := \frac{\partial L}{\partial \varphi} - \frac{d}{dx^k}\frac{\partial L}{\partial \varphi_{,k}} + \ldots \tag{49}$$

$$C = E_\xi(L) \equiv \frac{\delta L}{\delta \xi} := \frac{\partial L}{\partial \xi} - \frac{d}{dx^k}\frac{\partial L}{\partial \xi_{,k}} + \ldots \tag{50}$$

and

$$D^i = E^i(L) \equiv \frac{\delta L}{\delta A_i} := \frac{\partial L}{\partial A_i} - \frac{d}{dx^k}\frac{\partial L}{\partial A_{i,k}} + \ldots \tag{51}$$

(iii) $D^i$ is divergence-free; *i.e.,*

$$D^i{}_{|i} = 0 \,, \tag{52}$$

and

(iv) when evaluated for a flat metric tensor and constant scalar fields

$$D^i = \gamma g^{\frac{1}{2}} F^{ij}{}_{|j} \tag{53}$$

where $\gamma$ is a constant.

If I say that a bi-scalar-vector-tensor field theory satisfies conditions (i)-(iv), then I shall mean that the quartet of Euler-Lagrange tensor densities of that theory satisfies these conditions.

The approach I shall take in studying the quartet of concomitants $(A^{ij},B,C,D^i)$ will be a generalization of the approach used to treat similar problems considered by Lovelock [9], Horndeski [1,2,10] and Ohashi, *et al.,* [6].

To begin since the concomitants of $(A^{ij},B,C,D^i)$ are derivable from a variational principle they must be related by the identity

$$A^{ij}{}_{|j} = \tfrac{1}{2}\varphi^i B + \tfrac{1}{2}\xi^i C + \tfrac{1}{2}F^i{}_j D^j - \tfrac{1}{2}A^i D^j{}_{|j} \,. \tag{54}$$

I derive this result in [5] for the scalar-vector-tensor case (see page 63), but it can also be derived by examining the effects of a coordinate transformation on the scalar density L, as is done in Ohashi, *et al.,* [6] for the bi-scalar-tensor case.

(54) is valid regardless of the differential order of $A^{ij}$, B, C and $D^i$. However in our case these concomitants are second order with $D^i{}_{|i} = 0$, and hence $A^{ij}{}_{|j}$ must be of second order and not third order. As usual this places severe restrictions on the form of $A^{ij}$ which I present in

**Lemma 1:** If the quartet of concomitants $(A^{ij},B,C,D^i)$ satisfies (i), (ii) and (iii) above then

$$A^{i(u;|rs|,tv)} = 0 \,, \quad A_{\varphi}{}^{i(t;uv)} = 0, \quad A_{\xi}{}^{i(t;uv)} = 0 \,, \text{ and } \quad A^{i(u;|s|,tv)} = 0 \quad , \tag{55}$$

where parentheses about a string of indices denotes symmetrization over the all the enclosed indices except for those surround by vertical bars. ■

**Proof:** After expanding the expression for $A^{ij}{}_{|j}$ in (54), differentiate that equation with respect to $g_{rs,tuv}$, $\varphi_{,tuv}$, $\xi_{,tuv}$ and $A_{i,tuv}$, to obtain (55).■

At this point in the construction of concomitants like our quartet ($A^{ij}$,B,C,$D^i$) it is customary to appeal to the invariance identities satisfied by these concomitants in order to roughly determine their form. These identities were introduced by Rund [3,11], and further developed by du Plessis [4], Horndeski & Lovelock [12], and Horndeski [5]. To obtain these identities for our situation suppose that x and x' are charts at an arbitrary point P of a four-dimensional manifold. So on neighborhood of P we can functionally write $x^r = x^r(x'^h)$ and $x'^r = x'^r(x^h)$. Then at P the x and x' components of the metric tensor and vector field are are related by

$$g'_{hk} = g_{rs}J^r{}_hJ^s{}_k \quad \text{and} \quad A'_h = A_rJ^r{}_h \tag{56}$$

where I let

$$J^r{}_h := \frac{\partial x^r}{\partial x'^h}\,,\ J^r{}_{hk} := \frac{\partial^2 x^r}{\partial x'^h \partial x'^k} \quad \text{and} \quad J^r{}_{hkl} := \frac{\partial^3 x^r}{\partial x'^h \partial x'^k \partial x'^l}\ . \tag{57}$$

Consequently at P we can write

$$g'_{hk,pq} = g_{rs}(J^r{}_{hpq}J^s{}_k + J^r{}_hJ^{sk}{}_{pq}) + \text{lower order terms}$$

$$A'_{h,pq} = A_rJ^r{}_{hpq} + \text{lower order terms}$$

where the "lower order terms" are those terms not involving $J^{ab}{}_{cd}$. Since $A^{ij}$ is a tensor density valued concomitant its components with respect to the charts x and x' at P are related by

$$A^{ij}(g'_{hk};\ g'_{hk,p};\ g'_{hk,pq};\ \varphi';\ \varphi'_{,p};\ \varphi'_{,pq};\ \xi';\ \xi'_{,p};\ \xi'_{,pq};\ A'_h;\ A'_{h,p};\ A'_{h,pq}) =$$

$$= \det(J^l{}_m)J'^i{}_eJ'^j{}_fA^{ef}(g_{rs};\ g_{rs,t};\ g_{rs,tu};\ \varphi;\ \varphi_{,t};\ \varphi_{,tu};\ \xi;\ \xi_{,t};\ \xi_{,tu};\ A_r;\ A_{r,t};\ A_{r,tu})\ , \tag{58}$$

where $J^{\prime\prime i}{}_{e}$ is the derivative of $x^{\prime\prime i}$ with respect to $x^e$, and is the matrix inverse of $J^i{}_e$. If we differentiate (58), along with the similar transformation equations for the concomitants B, C and $D^i$, with respect to $J^a{}_{bcd}$ and then evaluate the result for the identity coordinate transformation at P we obtain

**Lemma 2:** The tensorial concomitants $A^{ij}$, B, C and $D^i$ satisfy the following invariance identities:

$$2A^{ij;h(b,cd)}\,g_{ha} + A^{ij;(b,cd)}A_a = 0\ , \tag{59}$$

$$2B^{;h(b,cd)}\,g_{ha} + B^{;(b,cd)}A_a = 0\ , \tag{60}$$

$$2C^{;h(b,cd)}\,g_{ha} + C^{;(b,cd)}A_a = 0\ , \tag{61}$$

and

$$2D^{i;h(b,cd)}\,g_{ha} + D^{i;(b,cd)}A_a = 0\ . \ \blacksquare \tag{62}$$

From (59) we see that in the present case $A^{ij;a(b,cd)}$ might be different from 0. This is unfortunate since if it were zero, then we could use **Lemma 1** and an argument similar to the one employed to prove **Lemma 1** in [1] to prove that $A^{ij}$ must be a polynomial $g_{ab,cd}$, $\varphi_{,ab}$ and $\xi_{,ab}$. But what if we could prove that the charge conservation condition, $D^i{}_{|i} = 0$, implies that each concomitant of the quartet $(A^{ij},B,C,D^i)$ is independent of explicit dependence on $A_a$. Then by differentiating (59) with respect to $A_a$ we could show that $A^{ij;a(b,cd)} = 0$, among other things. So proving the quartets independence of $A_a$ will now be our quest, and its accomplishment won't be that easy, although we shall begin with something very simple.

**Lemma 3:** If the quartet of concomitants $(A^{ij},B,C,D^i)$ satisfies charge conservation then

$$D^{(c;|ab|,de)} = 0,\ D_{\varphi}{}^{(a;bc)} = 0,\ D_{\xi}{}^{(a;bc)} = 0 \text{ and } D^{(b;|a|,cd)} = 0\ . \tag{63}$$

**Proof:** Simply differentiate the equation $D^i{}_{|i} = 0$, with respect to $g_{ab,cde}$, $\varphi_{,abc}$, $\xi_{,abc}$ and $A_{a,bcd}$ and (63) will result. $\blacksquare$

To proceed further in our analysis of the concomitant quartet $(A^{ij},B,C,D^i)$ we shall have to make more use of condition (ii) as embodied by (48)-(51). When dealing with the bi-scalar-tensor case I only needed the analog of condition (ii) to give me (54) with $D^i = 0$. But here we shall require what I refer to after (60) in [1] as the "consistency conditions" generated by (48)-(51). These are modeled on the theory I developed in Horndeski [13] and presented in

**Lemma 4:** If the concomitant quartet $(A^{ij},B,C,D^i)$ satisfies (48)-(51) then its members must satisfy the following sixteen equations:

$$E^{ij}(A^{hk}) = A^{ij;hk};\ E_\varphi(A^{hk}) = B^{;hk};\ E_\xi(A^{hk}) = C^{;hk};\ E^i(A^{hk}) = D^{i;hk} \tag{64}$$

$$E^{ij}(B) = A_\varphi{}^{ij;};\ E_\varphi(B) = B_\varphi\,;\ E_\xi(B) = C_\varphi\,;\ E^i(B) = D_\varphi{}^i \tag{65}$$

$$E^{ij}(C) = A_\xi{}^{ij;};\ E_\varphi(C) = B_\xi\,;\ E_\xi(C) = C_\xi\,;\ E^i(C) = D_\xi{}^i \tag{66}$$

$$E^{ij}(D^h) = A^{ij;h};\ E_\varphi(D^h) = B^{;h};\ E_\xi(D^h) = C^{;h};\ E^i(D^h) = D^{i;h}. \tag{67}$$

**Proof:** The proof is actually very simple and relies heavily on the fact that when an Euler-Lagrange operator acts on anything expressible as a divergence, it annihilates that thing (for a proof of this fact, see, *e.g.,* Lovelock [14]).

To prove the first equation in (64) let L be the Lagrangian which generates $A^{ij}$. Then

$$E^{hk}(A^{ij}) = E^{hk}\left(\frac{\partial L}{\partial g_{ij}} + \text{a divergence}\right).$$

Since $E^{hk}$ is a linear operator that commutes with $\dfrac{\partial}{\partial g_{ij}}$ and annihilates divergences this equation tells us that

$$E^{hk}(A^{ij}) = A^{hk;ij}$$

as required.

The proof of all the other equations in the statement of the **Lemma** is similar.■

**Lemma 4** is valid no matter what the differential order of the concomitants $A^{ij}$, B, C and $D^i$ happens to be. However, in the present case since each of these concomitants is of second order, the right-hand-side of each of the sixteen equations presented in **Lemma 4** must also be of second order while the left-hand-sides are, in general, of fourth order. Hence these equations can be differentiated with respect to fourth and third order derivatives of the field variables to generate a cornucopia of conditions on the partial derivatives of the concomitants in our quartet. When working on the bi-scalar-tensor problem in [1] I found that the conditions generated in that way were not that valuable. However, that is not the case for the problem which we are currently considering, and later in this section we shall explore some of the implications of such an analysis. For the present we shall explore another way of garnering information from the equations given in (64)-(67), and that is by examining their transformation properties. Since B and C are scalar densities, each of the terms in (65) and (66) are tensorial, and hence provide no new information when we examine their transformation properties. On the other hand, each of the terms in (64) and (67) is non-tensorial, and hence their behavior under coordinate transformations is quite informative. Determining that information is the purpose of

**Lemma 5:** If the concomitant quartet $(A^{ij},B,C,D^{i})$ satisfies conditions (i) and (ii) given in (44)-(47) and (48)-(51) then

$$A^{ab;cd,ef} = A^{cd;ab,ef}, \tag{68}$$

$$2\frac{d}{dx^j}A^{ab;cd,ej} - A^{ab;cd,e} - A^{cd;ab,e} = 0 \tag{69}$$

$$A^{ab;c,de} = D^{c;ab,de}, \tag{70}$$

$$2\frac{d}{dx^j}A^{ab;c,dj} - A^{ab;c,d} - D^{c;ab,d} = 0 \tag{71}$$

$$D^{(a;|b|,cd)} = D^{b;(a,cd)} \tag{72}$$

$$2\frac{d}{dx^j}D^{a;b,cj} + 2\frac{d}{dx^j}D^{c;b,aj} - D^{a;b,c} - D^{c;b,a} - D^{b;a,c} - D^{b;c,a} = 0, \tag{73}$$

$$A_{\varphi}^{a(b;cd)} = B^{;a(b,cd)} \;,\; A_{\xi}^{a(b;cd)} = C^{;a(b,cd)} \tag{74}$$

$$2\frac{d}{dx^{j}} A_{\varphi}^{a(b;c)j} - A_{\varphi}^{a(b;c)} - B^{;a(b,c)} = 0 \tag{75}$$

$$2\frac{d}{dx^{j}} A_{\xi}^{a(b;c)j} - A_{\xi}^{a(b;c)} - C^{;a(b,c)} = 0 \tag{76}$$

$$A_{\varphi}^{a(b;c)d} + A_{\varphi}^{d(b;c)a} = B^{;a(b,c)d} + B^{;d(b,c)a}, \tag{77}$$

$$A_{\xi}^{a(b;c)d} + A_{\xi}^{d(b;c)a} = C^{;a(b,c)d} + C^{;d(b,c)a}, \tag{78}$$

$$D_{\varphi}^{(a;bc)} = 0 \;,\; D_{\xi}^{(a;bc)} = 0 \;, \tag{79}$$

$$2\frac{d}{dx^{j}} D_{\varphi}^{(a;b)j} - D_{\varphi}^{(a;b)} - B^{;(a,b)} = 0, \tag{80}$$

$$2\frac{d}{dx^{j}} D_{\xi}^{(a;b)j} - D_{\xi}^{(a;b)} - C^{;(a,b)} = 0 \;. \tag{81}$$

**Proof:** The results presented in (68)-(73) follow from an analysis of the transformation properties of the first and fourth equations in (64). This analysis does not depend upon the presence of the bi-scalar fields in the concomitants $A^{ij}$ and $B^{i}$ appearing in [2], but only on the vector-tensor part. Consequently we can draw upon the results given in (2.8)-(2.13) of [2] to establish the validity of (68)-(73) above.

To obtain the remaining results given in **Lemma 5** we need to examine the transformation properties of the terms appearing in the second and third equations appearing in (64) and (67). I shall briefly explain how this is done for the second equation in (64).

If x and x' are charts at a point P then in terms of x', the second equation in (64) is

$$A'^{hk}_{\varphi} - \frac{d}{dx'^{l}} A'^{hk;l}_{\varphi} + \frac{d^{2}}{dx'^{l} dx'^{m}} A'^{hk;lm} = B'^{;hk} \tag{82}$$

where

$$\varphi = \varphi' \;;\; \varphi_{,r} = \varphi'_{,i} J'^{i}_{\;r} \;;\; \varphi_{,rs} = \varphi'_{,ij} J'^{i}_{\;r} J'^{j}_{\;s} + \varphi'_{,i} J'_{rs} \;. \tag{83}$$

Using (83) along with the fact that $A'^{hk} = JA^{pq}J'^{h}{}_{p}J'^{k}{}_{q}$ ,where $J = \det(J^{a}{}_{b})$, we find that

$$A'_{\varphi}{}^{hk} = JA_{\varphi}{}^{pq}J'^{h}{}_{p}J'^{k}{}_{q} \qquad (84)$$

$$A'_{\varphi}{}^{hk;l} = JJ'^{h}{}_{p}J'^{k}{}_{q}\,\frac{\partial A^{pq}}{\partial \varphi_{,r}}\,\frac{\partial \varphi_{,r}}{\partial \varphi'_{,l}} + JJ'^{h}{}_{p}J'^{k}{}_{q}\,\frac{\partial A^{pq}}{\partial \varphi_{,rs}}\,\frac{\partial \varphi_{,rs}}{\partial \varphi'_{,l}}$$

$$= JA_{\varphi}{}^{pq;r}J'^{h}{}_{p}J'^{k}{}_{q}J'^{l}{}_{r} + JA_{\varphi}{}^{pq;rs}J'^{h}{}_{p}J'^{k}{}_{q}J'^{l}{}_{rs} \qquad (85)$$

$$A'_{\varphi}{}^{hk;lm} = JA_{\varphi}{}^{pq;rs}J'^{h}{}_{p}J'^{k}{}_{q}J'^{l}{}_{r}J'^{ms}\ , \qquad (86)$$

recalling that $J'^{h}{}_{p} = \dfrac{\partial x'^{h}}{\partial x^{p}}$. Similarly we find that

$$B'^{hk} = JB^{;pq}J'^{h}{}_{p}J'^{k}{}_{q} + JB^{;pq,r}J'^{h}{}_{pr}J'^{k}{}_{q} + JB^{;pq,r}J'^{k}{}_{pr}J'^{h}{}_{q} + JB^{;pq,rs}J'^{h}{}_{prs}J'^{k}{}_{q} + JB^{;pq,rs}J'^{k}{}_{prs}J'^{h}{}_{q} +$$

$$+ JB^{;pq,rs}J'^{h}{}_{pr}J'^{k}{}_{qs} + JB^{;pq,rs}J'^{k}{}_{pr}J'^{h}{}_{qs}\,. \qquad (87)$$

You should note that the right-hand sides of (84)-(87) are in terms of the chart x. We now insert the results from (84)-(87) into (82) converting all of the partial derivatives with respect to the chart x' into partials with respect to x. The end result is a massive equation involving $J'^{h}{}_{p}$, $J'^{h}{}_{pq}$ and $J'^{h}{}_{pqr}$ which simplifies a bit since we can use the second equation in (64) to eliminate a few terms. When this equation is differentiated four times, once with respect to $J'^{a}{}_{bcd}$, twice with respect to $J'^{h}{}_{ab}$, and $J'^{k}{}_{cd}$, and once with respect to $J'^{h}{}_{ab}$ we obtain three equations which provide us with the results I presented in (74) and (75). Note that the equations involving C there, follow from the $\varphi \leftrightarrow \xi$ symmetry in our problem.

A similar analysis produces the remainder of the results in **Lemma 5**.■

The above **Lemmas** will be instrumental in the construction of the basic form of the concomitants in the quartet $(A^{ij},B,C,D^{i})$. To that end I now need to introduce the notion of quantities with property S, which was first introduced by Lovelock in [9].

We shall say a quantity $Q = Q[i_1 i_2; \ldots ; i_{2p-1} i_{2p}]$ has Property S if it satisfies the following

three conditions:

(i) It is symmetric in the indicies $i_{2h-1}i_{2h}$ for h=1,...,p;

(ii) It is symmetric under the interchange of the pair $(i_1 i_2)$ with the pair $(i_{2h-1}i_{2h})$, for h=2,...,p; and

(iii) It satisfies the cyclic identity involving any three of the four indicies $(i_1 i_2)$, $(i_{2h-1}i_{2h})$, h=2,...,p;*e.g.,*

$$Q[i_1(i_2;|\ldots i_{2h-2}|;i_{2h-1}i_{2h});\ldots;i_{2p-1}i_{2p}] = 0.$$

A quantity $Q[i_1 i_2]$ is said to have Property S if it is symmetric in $i_1$, $i_2$.

One important fact about concomitants with Property S established by Lovelock [9] is that in a four-dimensional space any 10 index quantity with Property S must vanish. We shall be able to use this to demonstrate that each of the concomitants in the quartet $(A^{ij},B,C,D^i)$ must be polynomials of finite degree in the second derivatives of $g_{ij}$, $\varphi$, $\xi$ and $A_i$. To that end we shall begin by demonstrating how the various partial derivatives of the members of our concomitant quartet $(A^{ij},B,C,D^i)$ with respect to $g_{hi,jk}$, $\varphi_{,hi}$, $\xi_{,hi}$ and $A_{h,ij}$ generate tensorial concomitants with Property S.

From the first equation in (55) and (68) we see that $A^{ab;cd,ef}$ and $A^{ab;cd,ef;hi,jk}$ must have Property S, where one should note that $A^{a(b;|cd|,ef)} = 0$ implies that $A^{ab;cd.ef} = A^{ef;cd,ab}$. From the second and third equations in (55) we find that repeated partial derivatives, and mixed partial derivatives, of $A^{ab}$ with respect to $\varphi_{,cd}$ or $\xi_{,cd}$ generate tensorial concomitants with Property S. Consequently all the partial derivatives of $A^{ab}$ with respect to $g_{hi,jk}$, $\varphi_{,hi}$ and $\xi_{,hi}$ (regardless of how these partial derivatives are mixed) will generate tensorial concomitants with Property S.

The fourth equation in (55) tells us that $A^{ab;c,de}$ has Property S in the indices ab,de. So if we now differentiate $A^{ab;c,de}$ with respect to $g_{hi,jk}$, $\varphi_{,hi}$ or $\xi_{,hi}$ we shall be able to generate more tensorial concomitants with Property S. Since $A^{ab;cd,ef}$ has Property S we can use (59) to deduce that $A^{ab;(c,de)}=0$ and thus $A^{ab;c,de;h,ij}$ has Property S in the pairs ab,de,ij,ch. Consequently any partial derivative of $A^{ab;c,de;h,ij}$ with respect to $g_{pq,rs}$, $\varphi_{,pq}$, $\xi_{,pq}$ or $A_{p,qr}$ will vanish. These observations tell us that $A^{ij}$ can be expressed as follows:

$$A^{ij} = a_1{}^{ijabcd}g_{ab,cd} + a_2{}^{ijabcdef}\varphi_{,ab}\varphi_{,cd}\varphi_{,ef} + a_3{}^{ijabcdef}\varphi_{,ab}\varphi_{,cd}\xi_{,ef} + a_4{}^{ijabcdef}\varphi_{,ab}\xi_{,cd}\xi_{,ef} + a_5{}^{ijabcdef}\xi_{,ab}\xi_{,cd}\xi_{,ef} +$$

$$+a_6{}^{ijabcdef}A_{e,ab}A_{f,cd} + a_7{}^{ijabcde}A_{e,ab}\varphi_{,cd} + a_8{}^{ijabcde}A_{e,ab}\xi_{,cd} + a_9{}^{ijabcd}\varphi_{,ab}\varphi_{,cd} + a_{10}{}^{ijabcd}\varphi_{,ab}\xi_{,cd} + a_{11}{}^{ijabcd}\xi_{,ab}\xi_{,cd} +$$

$$+ a_{12}{}^{ijabc}A_{c,ab} + a_{13}{}^{ijab}\varphi_{,ab} + a_{14}{}^{ijab}\xi_{,ab} + a_{15}{}^{ij}, \qquad (88)$$

where each coefficient function $a_\alpha$ ($\alpha$=1,...,15) has the form

$$a_\alpha = a_\alpha(g_{ab};\ g_{ab,c};\ \varphi;\ \varphi_{,a};\ \xi;\ \xi_{,a};\ A_a;\ A_{a,b})$$

and they all have Property S with the exception of $a_7{}^{ijabcde}$, $a_8{}^{ijabcde}$ and $a_{12}{}^{ijabc}$ which only have Property S in the pairs ij,ab,cd and ij,ab along with the cyclic symmetry $a_7{}^{ijab(cde)}=0$, $a_8{}^{ijab(cde)}=0$ $a_{12}{}^{ij(abc)}=0$. At this juncture one should note that since

$$A_{c,ab} = A_{(a,bc)} + ⅓F_{ac,b} + ⅓F_{bc,a}$$

we can replace terms like $a_{12}{}^{ijabc}A_{c,ab}$ by $⅔a_{12}{}^{ijabc}\ F_{ac,b}$ and then absorb the ⅔ into $a_{12}{}^{ijabc}$. Similar remarks apply to all of the other terms involving the second derivatives of $A_a$ in (88). Hence these second derivatives of $A_a$ only appear in the form of $F_{ab,c}$. I shall have more to say about the structural form of the coefficient functions $a_\alpha$ later in this section.

We shall now turn our attention to B. Since $A_\varphi{}^{a(b;cd)} = 0$, we can use the first equation in (74) and (60) to deduce that

$$B^{;a(b,cd)} = 0, \text{ and } B^{;(a,bc)} = 0. \qquad (89)$$

Using (89) and the fact that $A_\varphi{}^{a(b;cd)} = 0$ in (77) shows us that

$$A_\varphi{}^{ab;cd} = B^{;ab,cd}. \qquad (90)$$

From this we are able to conclude that the partial derivatives of $B^{;ab,cd}$ with respect to $g_{hi,jk}$, $\varphi_{,hi}$ and $\xi_{,hi}$ will all have Property S. In addition $B^{;ab,cd;h,ij}$ will have Property S in ab,cd,ij, and so if we differentiate $B^{;ab,cd;h,ij}$ with respect to $g_{pq,rs}$, $\varphi_{,pq}$, $\xi_{,rs}$ or $A_{p,qr}$ we obtain 0. This information can give us a pretty good idea of what B must look like, but there are some very important things we do not know about B; *e.g.,* does $B_{\varphi\varphi}{}^{;ab;cd}$, $B_{\varphi\xi}{}^{;ab;cd}$, $B_{\xi\xi}{}^{;ab;cd}$ , $B_{\varphi}{}^{;ab;e,cd}$ or $B_{\xi}{}^{;ab;e,cd}$ have Property S in ab,cd? The answer to this question is provided by our next

**Lemma 6:** If the quartet of tensorial concomitants $(A^{ij},B,C,D^{i})$ satisfy conditions (i), (ii) and (iii) then $B_{\varphi\varphi}{}^{;ab;cd}$, $B_{\xi\xi}{}^{;ab;cd}$, $C_{\varphi\varphi}{}^{;ab;cd}$, $C_{\xi\xi}{}^{;ab;cd}$, $D_{\varphi\varphi}{}^{i;ab;cd}$, and $D_{\xi\xi}{}^{i;ab;cd}$ have Property S in ab,cd, while

$$B_{\varphi\xi}{}^{;a(b;cd)} + B_{\xi\varphi}{}^{;a(b;cd)} = 0,\ C_{\varphi\xi}{}^{;a(b;cd)} + C_{\xi\varphi}{}^{;a(b;cd)} = 0,\ \ D_{\varphi\xi}{}^{i;a(b;cd)} + D_{\xi\varphi}{}^{i;a(b;cd)} = 0. \tag{91}$$

$$B_{\varphi}{}^{;h,i(j;kl)} + B_{\varphi}{}^{;h,(jk;l)i} = 0,\ \ B_{\xi}{}^{;h,i(j;kl)} + B_{\xi}{}^{;h,(jk;l)i} = 0,\ C_{\varphi}{}^{;h,i(j;kl)} + C_{\varphi}{}^{;h,(jk;l)i} = 0,\ C_{\xi}{}^{;h,i(j;kl)} + C_{\xi}{}^{;h,(jk;l)i} = 0, \tag{92}$$

$$D_{\varphi}{}^{h;i(k;|j|,lm)} + D_{\varphi}{}^{h;(kl;|j|,m)i} = 0,\ D_{\xi}{}^{h;i(k;|j|,lm)} + D_{\xi}{}^{h;(kl;|j|,m)i} = 0, \quad D^{h;i,j(l;|k|,mn)} + D^{h;i,(lm;|k|,n)j} = 0, \tag{93}$$

$$B^{;h,i(k;|j|,lm)} + B^{;h,(kl;|j|,m)i} = 0 \text{ and } \ C^{;h,i(k;|j|,lm)} + C^{;h,(kl;|j|,m)i} = 0. \tag{2.30d}$$

**Proof:** The proofs for all of these results are very similar and follow from the second, third and fourth equations in each of (65), (66) and (67) upon differentiating these equations with respect to $\varphi_{,hijk}$, $\xi_{,hijk}$ and $A_{h,ijkl}$. To show you how it goes I shall only prove that $D_{\varphi\varphi}{}^{i;ab;cd}$ has Property S in ab,cd. From the second equation in (67) we know that $E_{\varphi}(D^{i}) = B^{;i}$, and hence $E_{\varphi}(D^{i})$ must be of second order, and not of fourth order. So if we differentiate $E_{\varphi}(D^{i})$ with respect to $\varphi_{,abcd}$ we obtain $D_{\varphi\varphi}{}^{i;(ab;cd)}$ $= 0$, which when written out gives us

$$D_{\varphi\varphi}{}^{i;ab;cd} + D_{\varphi\varphi}{}^{i;ac;db} + D_{\varphi\varphi}{}^{i;ad;bc} + D_{\varphi\varphi}{}^{i;bc;da} + D_{\varphi\varphi}{}^{i;bd;ac} + D_{\varphi\varphi}{}^{i;ba;cd} + D_{\varphi\varphi}{}^{i;cd;ab} + D_{\varphi\varphi}{}^{i;ca;bd} + D_{\varphi\varphi}{}^{i;cb;da} +$$
$$+ D_{\varphi\varphi}{}^{i;da;bc} + D_{\varphi\varphi}{}^{i;db;ca} + D_{\varphi\varphi}{}^{i;dc;ab} = 0.$$

This equation simplifies to $D_{\varphi\varphi}{}^{i;a(b;cd)} = 0$, and hence $D_{\varphi\varphi}{}^{i;ab;cd}$ has Property S in ab,cd as desired. The proof that $B_{\varphi\varphi}{}^{;ab;cd}$ and $C_{\varphi\varphi}{}^{;ab;cd}$ have Property S is similar, and the other Property S conditions for the $\xi\xi$ versions of B, C and $D^{i}$ follows from the $\varphi \leftrightarrow \xi$ transformation.■

I have to admit that I find the conditions presented in (91)-(94) to be bizarre. They are like weakened versions of Property S for various four index groupings, and nothing like this occurs in the study of second-order bi-scalar-tensor field theories. One might think that differentiating equations such as $E_\varphi(D^i) = B^{;i}$ with respect to (say) $\varphi_{,abd}$ and $\xi_{,pqr}$ could yield some new results, but that is not the case. And differentiating $E_\varphi(D^i) = B^{;i}$ with respect to $g_{pq,rstu}$ just yields constraints that we found previously by other means. So we seem to be stuck with (91)-(94) as the best we can do.

We could now attempt to construct the basic form of B in a manner analogous to our construction of $A^{ij}$ but it is not clear what the algebraic degree of B in $\varphi_{,ab}$, $\xi_{,ab}$ and $A_{a,bc}$ would be. Sure due to **Lemma 6** we could not have terms of the form $\varphi_{,ab}\ \varphi_{,cd}\ \varphi_{,ef}\ \varphi_{,hi}\ \varphi_{,jk}$, since $B_{\varphi\varphi}{}^{;ab;cd}$ has Property S. But it is not clear if there could be terms like $\varphi_{,ab}\xi_{,cd}A_{e,fh}A_{i,jk}A_{l,mn}$ in B. We shall return to the structural form of B and C in a while.

I shall now try to obtain the basic form of $D^i$. (70) tells us that $A^{ab;c,de} = D^{c;ab,de}$. Due to the above work with $A^{ab}$ we know that $A^{ab;c,de}$ has Property S in ab,de; and hence so does $D^{c;ab,de}$. In addition since $A^{ab;c,de;pq,rs}$ has Property S in ab,de,pq,rs; $D^{c;ab,de;pq,rs}$ must have Property S in ab,de,pq,rs. Since we know from (63) that $D^{(c;|ab|,de)} = 0$, we can conclude that

$$D^{c;ab,de;pq,rs} = 0\ . \tag{95}$$

Using (70) we find that

$$D_\varphi{}^{c;ab,de;pq} = A_\varphi{}^{ab;c,de;pq} \text{ and } D_\xi{}^{c;ab,de;pq} = A_\xi{}^{ab;c,de;pq}\ ,$$

and hence the second and third equations in (55) can be used to deduce that $D_\varphi{}^{c;ab,de;pq}$ and $D_\xi{}^{c;ab,de;pq}$ have Property S in ab,de,pq. Thus we can see that

$$D_{\varphi\varphi}{}^{c;ab,de;pq;rs} = D_{\varphi\xi}{}^{c;ab,de;pq;rs} = D_{\xi\xi}{}^{c;ab,de;pq;rs} = 0\ . \tag{96}$$

Using $A^{ab;c,de} = D^{c;ab,de}$ we find that $A^{ab;c,de;p,qr} = D^{c;ab,de;p,qr}$ and hence $D^{c;ab,de;p,qr}$ has Property S

in the indices ab,de,qr,cp. Consequently we have shown that

$$D^{c;ab,de;p,qr;h,ij}=0,\ D_{\varphi}^{\ c;ab,de;p,qr;ij} = D_{\xi}^{\ c;ab,de;p,qr;ij} = 0\ . \tag{97}$$

(95), (96) and (97) provide us with a wealth of information on $D^{a;bc,de}$. **Lemma 6** provides us with some information on the partial derivatives of $D^{a;b,cd}$, $D_{\varphi}^{\ a;bc}$ and $D_{\xi}^{\ a;bc}$ with respect to $\varphi_{,hi}$, $\xi_{,hi}$ and $A_{h,ij}$, but not enough to construct the general form of $D^a$. In a little while I shall show you how the general form of B, C and $D^i$ can be obtained. But before doing that we need to garner more information on the partial derivatives of the elements of our quartet ($A^{ij}$,B,C,$D^i$) with respect to $A_a$, $A_{a,b}$ and $A_{a,bc}$.

**Lemmas 3** and **4** in [2] allow us to deduce that

$$A^{ab;c,d} = -A^{ab;d,c}\ ,\ A^{ab;c} = 0,\ D^{a;b,c} = -\ D^{a;c,b} \text{ and } D^{a;b} = 0\ , \tag{98}$$

since the equations in [2] used to prove the vector-tensor version of (98) have precise analogues in what we have established so far. I shall now demonstrate that B and C satisfy similar conditions.

We start by writing out the equation $D^h_{\ ,h} = 0$ in full, noting (63), to obtain

$$D^h_{\ ,h} = D^{h;rs}g_{rs,h} + D^{h;rs,t}g_{rs,th} + D_{\varphi}^{\ h}\varphi_{,h} + D_{\varphi}^{\ h;r}\varphi_{,rh} + D_{\xi}^{\ h}\xi_{,h} + D_{\xi}^{\ h;r}\xi_{,rh} + D^{h;r}A_{r,h} + D^{h;r,s}A_{h,rs} = 0. \tag{99}$$

Upon differentiating this equation with respect to $\varphi_{,ab}$ we find that the resulting equation can be rewritten as

$$2\frac{d}{dx^h}D_{\varphi}^{\ h;ab} + D_{\varphi}^{\ a;b} + D_{\varphi}^{\ a;b} = 0\ . \tag{100}$$

But the first equation in (79) and (80) combine to tell us that

$$2\frac{d}{dx^h}D_{\varphi}^{\ h;ab} + D_{\varphi}^{\ a;b} + D_{\varphi}^{\ a;b} = -B^{;a,b}\ \ -B^{;b,a}\ \ . \tag{101}$$

(100) and (101), along with the $\varphi\leftrightarrow\xi$ symmetry of our calculations, permit us to conclude that

$$B^{;a,b} = -B^{;b,a} \text{ and } C^{;a,b} = -C^{;b,a} \quad . \tag{102}$$

Next we want to show that $B^{;a}=0$. To that end, since we want to isolate $B^{;a}$, let's consider the second equation in (67), which when written out gives us

$$\frac{d^2}{dx^i dx^j} D_\varphi^{\;a;ij} - \frac{d}{dx^i} D_\varphi^{\;a;i} + D_\varphi^{\;a} = B^{;a} \quad . \tag{103}$$

Next, since we now know from (102) that $B^{;(a,b)} = 0$, we can use (80) to write

$$2\frac{d}{dx^j} D_\varphi^{\;a;ij} + 2\frac{d}{dx^j} D_\varphi^{\;i;aj} = D_\varphi^{\;a;i} + D_\varphi^{\;i;a} \; .$$

When this is differentiated with respect to $x^i$, noting that $D_\varphi^{\;(i;aj)} = 0$, we obtain

$$\frac{d^2}{dx^i dx^j} D_\varphi^{\;a;ij} = \frac{d}{dx^i} D_\varphi^{\;a;i} + \frac{d}{dx^i} D_\varphi^{\;i;a} \; . \tag{104}$$

Combining (103) and (104) shows us that

$$\frac{d}{dx^i} D_\varphi^{\;i;a} + D_\varphi^{\;a} = B^{;a} \quad . \tag{105}$$

If we differentiate (99) with respect to $\varphi_{,a}$ we discover that the resulting expression can be written as

$$\frac{d}{dx^i} D_\varphi^{\;i;a} + D_\varphi^{\;a} = 0 \; ,$$

and hence (105) shows us that $B^{;a} = 0$. Due to the $\varphi \leftrightarrow \xi$ of our concomitants we can also deduce that $C^{;a} = 0$.

To recapitulate the above work we have

**Lemma 7:** If the quartet of tensorial concomitants $(A^{ij},B,C,D^i)$ satisfies conditions (i), (ii) and (iii) then they must be independent of explicit dependence on $A_a$ with

$$A^{ij;(a,b)} = 0,\ A^{ij;(a,bc)} = 0,\ B^{;(a,b)} = 0,\ B^{;(a,bc)} = 0,\ C^{;(a,b)} = 0,\ C^{;(a,bc)} = 0,\ D^{i;(a,b)} = 0 \text{ and } D^{i;(a,bc)} = 0. \ \blacksquare \tag{106}$$

An important implication of **Lemma 7** is that it implies that the functional dependence of the members of our quartet ($A^{ij}$,B,C,$D^{i}$) on $A_{a,b}$ and $A_{a,bc}$ must be expressible through $F_{ab}$ and $F_{ab,c}$. This follows from

**Lemma 8:** It $T^{\cdots}$ denotes the components of a tensorial concomitant which has the functional form

$$T^{\cdots} = T^{\cdots}(g_{ab};g_{ab,c};g_{ab,cd};\varphi;\varphi_{,a};\varphi_{,ab};\xi;\xi_{a};\xi_{,ab};A_{a};A_{a,b};A_{a,bc})$$

and is such that $T^{\ldots;a}=0$, $T^{\ldots;(a,b)} = 0$ and $T^{\ldots;(a,bc)} = 0$, then we can assume that the functional form of $T^{\cdots}$ is actually given by

$$T^{\cdots} = T^{\cdots}(g_{ab};g_{ab,c};g_{ab,cd};\varphi;\varphi_{,a};\varphi_{,ab};\xi;\xi_{a};\xi_{,ab};F_{ab};F_{ab,c}) .$$

**Proof:** It is clear that $T^{\cdots}$ is independent of explicit dependence on $A_{a}$. To determine its dependence on F we begin by noting that if we define

$$S_{ab}:= A_{(a,b)} \text{ and } S_{abc} := A_{(a,bc)} \tag{107}$$

then we may write

$$A_{a,b} = S_{ab} + \tfrac{1}{2}F_{ba} \text{ and } A_{a,bc} = S_{abc}+ \tfrac{1}{3}F_{ba,c} +\tfrac{1}{3}F_{ca,b} \ . \tag{108}$$

I now define a new tensorial concomitant $\tau^{\cdots}$ with local functional form

$$\tau^{\cdots} = \tau^{\cdots}(g_{ab};g_{ab,c};g_{ab,cd};\varphi;\varphi_{,a};\varphi_{,ab};\xi;\xi_{a};\xi_{,ab};S_{ab};F_{ab};S_{abc};F_{ab,c}) \tag{109}$$

by

$$\tau^{\cdots} = T^{\cdots}(g_{ab};g_{ab,c};g_{ab,cd};\varphi;\varphi_{,a};\varphi_{,ab};\xi;\xi_{a};\xi_{,ab};0;S_{ab}+\tfrac{1}{2}F_{ba} ;S_{abc}+\tfrac{1}{3}F_{ac,b}+\tfrac{1}{3}F_{bc,a}) . \tag{110}$$

Evidently $\tau^{\cdots}$ and $T^{\cdots}$ have the same numerical values when $S_{ab}$ and $S_{abc}$ are given by (107). If we take the partial derivative of $\tau^{\cdots}$ with respect to $S_{ab}$ we find using (108) and (110) that

$$\frac{\partial\tau^{\cdots}}{\partial S_{ab}} = \frac{\partial T^{\cdots}}{\partial A_{r,s}}\frac{\partial A_{r,s}}{\partial S_{ab}} = T^{\ldots;(a,b)} .$$

Hence due to our assumptions about $T^{\cdots}$ , $\tau^{\cdots}$ must be independent of $S_{ab}$. In a similar way we can prove that the partial of $\tau^{\cdots}$ with respect to $S_{abc}$ vanishes and so $\tau^{\cdots}$ is independent of $S_{abc}$ . These observations establish the **Lemma.**■

**Lemma 8** tells us that whenever a concomitant satisfies the conditions of the **Lemma** then it is gauge invariant: *i.e.,* its value is unchanged whenever the vector potential $A_a$ is replaced by $A_a+\chi_{,a}$, where $\chi$ is an arbitrary scalar field. I discuss the relationship between gauge invariance and conservation of charge in Horndeski [15].

The next result we require is a "replacement theorem." These theorems were developed by Thomas's [16] using a differentiaton process he called "extensions," which were similar to covariant differentiation.

**Lemma 9 (Thomas's Replacement Theorem for Bi-Scalar-Vector-Tensor Field Theories).** If $T^{\cdots} = T^{\cdots}(g_{ab};g_{ab,c};g_{ab,cd};\varphi;\varphi_{,a};\varphi_{,ab};\xi;\xi_{a};\xi_{,ab};F_{ab};F_{ab,c})$ denotes the components of a tensorial concomitant of the indicated functional form then its value is unaffected if we replace $g_{ab,c}$ by 0; $g_{ab,cd}$ by $⅓(R_{acdb} + R_{adcb})$; $\varphi_{,ab}$ by $\varphi_{ab}$; $\xi_{,ab}$ by $\xi_{ab}$ and $F_{ab,c}$ by $F_{ab|c}$.

**Proof:** The proof is similar to that given for second-order, scalar-tensor concomitants given in **Lemma 4** of Horndeski [17]. In **Appendix B** of that paper I also explain Thomas's notion of "tensor extensions," and demonstrate how they can be used. ■

Due to **Lemmas 7**, **8** and **9** we can immediately rewrite the expression for $A^{ij}$ given in (88) in a manifestly tensorial form. That formulation is presented in

**Lemma 10:** If the quartet of tensorial concomitants $(A^{ij},B,C,D^{i})$ satisfy conditions (i), (ii) and (iii) then

$$A^{ij}= a_1{}^{ijabcd}R_{cabd} + a_2{}^{ijabcdef}R_{cabd}\varphi_{ef} + a_3{}^{ijabcdef}R_{cabd}\xi_{ef} + a_4{}^{ijabcdef}\varphi_{ab}\varphi_{cd}\varphi_{ef} + a_5{}^{ijabcdef}\varphi_{ab}\varphi_{cd}\xi_{ef} + a_6{}^{ijabcdef}\varphi_{ab}\xi_{cd}\xi_{ef}$$
$$+ a_7{}^{ijabcdef}\xi_{ab}\xi_{cd}\xi_{ef} + a_8{}^{ijabcdef}F_{ae|b}F_{cf|d} + a_9{}^{ijabcde}F_{ae|b}\varphi_{cd} + a_{10}{}^{ijabcde}F_{ae|b}\xi_{cd} + a_{11}{}^{ijabcd}\varphi_{ab}\varphi_{cd} + a_{12}{}^{ijabcd}\varphi_{ab}\xi_{cd} +$$
$$+a_{13}{}^{ijabcd}\xi_{ab}\xi_{cd} + a_{14}{}^{ijabc}F_{ac|b} + a_{15}{}^{ijab}\varphi_{ab} + a_{16}{}^{ijab}\xi_{ab} + a_{17}{}^{ij}, \qquad (111)$$

where each coefficient function $a_\alpha^{\cdots}$ ($\alpha$=1,...,15) is a tensorial concomitant of the form

$$a_\alpha^{\cdots} = a_\alpha^{\cdots}(g_{ab};\varphi;\ \varphi_{,a};\ \xi;\ \xi_{,a};\ F_{ab})$$

and they all have Property S, with the exception of $a_9^{ijabcde}$, $a_{10}^{ijabcde}$ and $a_{14}^{ijabc}$ which only have Property S in the pairs ij,ab,cd and ij,ab along with the cyclic symmetry $a_9^{ij(ab|cd|e)}=0$, $a_{10}^{ij(ab|cd|e)}=0$ $a_{14}^{ij(abc)}=0$. ■

We previously saw that we were having difficulties trying to construct expressions for B, C and $D^i$ similar to what we have for $A^{ij}$. I shall now show how we can obtain such expressions.

From (54) we see that our concomitant quartet ($A^{ij}$,B,C,$D^i$) satisfies

$$A^{ij}{}_{|j} = \tfrac{1}{2}\varphi^i B + \tfrac{1}{2}\xi^i C + \tfrac{1}{2}F^i{}_j D^j \ . \qquad (112)$$

In the generic situation the set of three vectors $\{\varphi^i B,\ \xi^i C,\ F^i{}_j D^j\}$ is a linearly independent set. Thus by constructing $A^{ij}{}_{|j}$ ,using (111) we can determine the basic forms for B, C and $D^i$. In order to accomplish this task we need to know what things like $a^{ijabcd}R_{cabd|j}$, $a^{ijab}\varphi_{abj}$ and $a^{ijabc}F_{ac|bj}$ are equal to, when $a^{ijabcd}$, $a^{ijab}$ and $a^{ijabc}$ have Property S in ij,ab,cd ; ij,ab and ij,ab respectively, and in addition $a^{ijabc}$ vanishes upon cycling on i,j,c or a,b,c. We know that each of these tensors are at most of second-order, but we need to know their basic form to determine the constituents of $A^{ij}{}_{|j}$. So let's examine each of these tensors in turn.

To that end I would like to begin by pointing out that due to the symmetries enjoyed by $a^{ijabcd}$ and $a^{ijabc}$ each of these quantites vanishes when we symmetrize on i,a,c which is not obvious at the outset. For example symmetrizing on i,a,c in $a^{ijabc}$ gives us

$$6\, a^{(i|j|a|b|c)} = a^{ijabc} + a^{ijcba} + a^{ajcbi} + a^{ajibc} + a^{cjiba} + a^{cjabi} = -(a^{ijacb} + a^{ajcib} + a^{cjaib}) = 0$$

where Property S has been used several times to regroup terms. This observation also allows us to deduce that if $Q = Q[i_1 i_2; \ldots ; i_{2p-1} i_{2p}]$ ($p \geq 2$) has Property S then it vanishes when symmetrized on any three indices chosen from $i_1 i_2; \ldots ; i_{2p-1} i_{2p}$.

Let's consider $a^{ijabcd}R_{cabd|j}$, which due to the definition of covariant differentiation is given by

$$a^{ijabcd}R_{cabd|j} = a^{ijabcd}R_{cabd,j} + \text{terms, each of which involves } g_{pq,r}\,.$$

Since $R_{cabd} = \tfrac{1}{2}(g_{ab,cd}+g_{cd,ab}-g_{cb,ad}-g_{ad,bc})$ + terms quadratic in $g_{pq,r}$ we have

$$a^{ijabcd}R_{cabd|j} = \tfrac{1}{2}a^{ijabcd}(g_{ab,cdj}+g_{cd,abj}-g_{cb,adj}-g_{ad,bcj}) + \text{terms involving } g_{pq,r}\;.$$

Due to Property S and Thomas's replacement Theorem, **Lemma 9**, we see that

$$a^{ijabcd}R_{cabd|j} = 0. \tag{113}$$

In a similar way we can use the symmetries of $a^{ijab}$ and $a^{ijabc}$ to show that

$$a^{ijab}\varphi_{abj} = \tfrac{2}{3}a^{ijab}\varphi_m R_a{}^{mj}{}_b \tag{114}$$

and, with somewhat more effort, that

$$a^{ijabc}F_{ac|bj} = -\tfrac{2}{3}a^{ijabc}F_{pc}R_a{}^p{}_{bj} - \tfrac{2}{3}a^{ijabc}F_{ap}R_c{}^p{}_{bj} + \tfrac{1}{3}a^{ijabc}F_{ap}R_j{}^p{}_{bc}\,. \tag{115}$$

We can now employ (111), (113) and (114) to deduce that $A^{ij}{}_{|j}$ can be expressed in terms of the elements of the following set

$$\begin{aligned}
&\{R_{abcd}R_{efhj};\ R_{abcd}\varphi_{ef}\varphi_{hj};\ R_{abcd}\xi_{ef}\xi_{hj};\ R_{abcd}\varphi_{ef}\xi_{hj};\ R_{abcd}\varphi_{ef}F_{hj|k};\ R_{abcd}\xi_{ef}F_{hj|k};\ R_{abcd}\varphi_{ef};\ R_{abcd}\xi_{ef};\ R_{abcd}F_{ef|h};\ R_{abcd};\\
&\varphi_{ab}\varphi_{cd}\varphi_{ef}\varphi_{hj};\varphi_{ab}\varphi_{cd}\varphi_{ef}\xi_{hj};\varphi_{ab}\varphi_{cd}\xi_{ef}\xi_{hj};\varphi_{ab}\xi_{cd}\xi_{ef}\xi_{hi};\xi_{ab}\xi_{cd}\xi_{ef}\xi_{hj};\varphi_{ab}\varphi_{cd}\varphi_{ef}F_{hj|k};\varphi_{ab}\varphi_{cd}\xi_{ef}F_{hj|k};\varphi_{ab}\xi_{cd}\xi_{ef}F_{hj|k};\\
&\xi_{ab}\xi_{cd}\xi_{ef}F_{hj|k};\ \varphi_{ab}\varphi_{cd}\varphi_{ef};\ \varphi_{ab}\varphi_{cd}\xi_{ef};\ \varphi_{ab}\xi_{cd}\xi_{ef};\ \xi_{ab}\xi_{cd}\xi_{ef};\ \varphi_{ab}\varphi_{cd}F_{ef|h};\ \varphi_{ab}\xi_{cd}F_{ef|h};\ \xi_{ab}\xi_{cd}F_{ef|h};\ \varphi_{ab}F_{cd|e}F_{fh|j};\xi_{ab}F_{cd|e}F_{fh|j};\\
&F_{ab|c}F_{ef|h}F_{jk|m};\ \varphi_{ab}\varphi_{cd};\ \varphi_{ab}\xi_{cd};\ \xi_{ab}\xi_{cd};\ \varphi_{ab}F_{cd|e};\ \xi_{ab}F_{cd|e};\ F_{ab|c}F_{de|f};\ \varphi_{ab};\ \xi_{ab};\ F_{ab|c}\}
\end{aligned} \tag{116}$$

with coefficients which are functions of $g_{ab}$, $\varphi$, $\varphi_{,a}$, $\xi$; $\xi_{,a}$ and $F_{ab}$. Not all elements of the set (116) will appear in the general expression for B, C and $D^i$. *E.g.,* $R_{abcd}R_{efhi}$ will not appear in $D^i$ due to (95), while it will appear in B and C. Similarly, $R_{abcd}\varphi_{ef}F_{hi|j}$ and $R_{abcd}\xi_{ef}F_{hi|j}$ can not appear in B, C or $D^i$ due to (97) and the remarks I made after (90), while $\varphi_{ab}\varphi_{cd}\varphi_{ef}\varphi_{hi}$ and $\xi_{ab}\xi_{cd}\xi_{ef}\xi_{hi}$ can not appear in $D^i$ due to **Lemma 6**.

Recall that when we were previously studying the properties of B, C and $D^i$ we were unable to determine their algebraic degree in $\varphi_{ab}$, $\xi_{ab}$ and $A_{a,bc}$ (or $F_{ab|c}$). The set presented in (116) tells us what those degrees are. In view of (88) and (116) we see that in general B, C and $D^i$ will be functionally more complex than $A^{ij}$. Since we shall be more interested in $D^i$ than in B or C, I shall present its form in

**Lemma 11:** If the quartet of tensorial concomitants $(A^{ij},B,C,D^i)$ satisfy conditions (i), (ii) and (iii) then

$$
\begin{aligned}
D^i = {} & d_1{}^{iabcdef}R_{acdb}\varphi_{ef}+d_2{}^{iabcdef}R_{acdb}\xi_{ef}+d_3{}^{ihabcdef}R_{abcd}F_{eh|f}+d_4{}^{iabcd}R_{acdb}+d_5{}^{iabcdefhk}\varphi_{ab}\varphi_{cd}\varphi_{ef}\xi_{hk}+d_6{}^{iabcdefhk}\varphi_{ab}\varphi_{cd}\xi_{ef}\xi_{hk} \\
& +d_7{}^{iabcdefhk}\varphi_{ab}\xi_{cd}\xi_{ef}\xi+d_8{}^{iabcdefhjk}\varphi_{ab}\varphi_{cd}\varphi_{ef}F_{hk|j}+d_9{}^{iabcdefhjk}\varphi_{ab}\varphi_{cd}\xi_{ef}F_{hk|j}+d_{10}{}^{iabcdefhjk}\varphi_{ab}\xi_{cd}\xi_{ef}F_{hk|j}+ \\
& +d_{11}{}^{iabcdefhjk}\xi_{ab}\xi_{cd}\xi_{ef}F_{hk|j}+d_{12}{}^{iabcdef}\varphi_{ab}\varphi_{cd}\varphi_{ef}+d_{13}{}^{iabcdef}\varphi_{ab}\varphi_{cd}\xi_{ef}+d_{14}{}^{iabcdef}\varphi_{ab}\xi_{cd}\xi_{ef}+d_{15}{}^{iabcdef}\xi_{ab}\xi_{cd}\xi_{ef}+ \\
& +d_{16}{}^{iabcdefh}\varphi_{ab}\varphi_{cd}F_{eh|f}+d_{17}{}^{iabcdefh}\varphi_{ab}\xi_{cd}F_{ef|h}+d_{18}{}^{iabcdefh}\xi_{ab}\xi_{cd}F_{ef|h}+d_{19}{}^{iabcdfhek}\varphi_{ab}F_{ce|d}F_{fk|h}+d_{20}{}^{iabcdfhek}\xi_{ab}F_{ce|d}F_{fk|h}+ \\
& +d_{21}{}^{iabcdefhjk}F_{ac|b}F_{df|e}F_{hk|j}+d_{22}{}^{iabcd}\varphi_{ab}\varphi_{cd}+d_{23}{}^{iabcd}\varphi_{ab}\xi_{cd}+d_{24}{}^{iabcd}\xi_{ab}\xi_{cd}+d_{25}{}^{iabcde}\varphi_{ab}F_{ce|d}+d_{26}{}^{iabcde}\xi_{ab}F_{ce|d}+ \\
& +d_{27}{}^{iabcdef}F_{ac|b}F_{df|e}+d_{28}{}^{iab}\varphi_{ab}+d_{29}{}^{iab}\xi_{ab}+d_{30}{}^{iabc}F_{ac|b}+d_3{}^i.
\end{aligned}
\tag{117}
$$

The tensorial concomitants $d_\alpha{}^{\cdots} = d_\alpha{}^{\cdots}(g_{ab};\varphi;\varphi_{,a};\xi;\xi_{,a};F_{ab})$, for $\alpha=1,\ldots,31$, will enjoy various symmetries which can be ascertained from the symmetries of the derivatives of $D^i$ with respect to $g_{ab,cd}$, $\varphi_{,ab}$ and $\xi_{,ab}$. ■

At this stage in the construction of the concomitant quartet satisfying conditions (i)-(iii) stated at the outset of this section it is customary (*see,* Horndeski [1]) to construct the coefficient functions appearing in (111) to obtain a more explicit form of $A^{ij}$, which due to (112), acts as the generator of B, C and $D^i$. Constructing these coefficient functions is an incredibly difficult task, even if we assume that all of them can be expressed as polynomials in their tensorial indices, with coefficients being scalar density functions of $\varphi$, $\xi$, $\varphi_{,i}$, $\xi_{,i}$ and $F_{ij}$. I shall now give you some idea of how these various coefficient concomitants can be constructed.

From the work done in [1] we can say that if $\alpha_{abcdefhi}$ is an 8 index tensorial concomitant of $\varphi$, $\xi$, $\varphi_{,i}$, $\xi_{,i}$ and $F_{ij}$ then

$$\alpha_{abcdefhi} = \alpha\Theta_{Labcdefhi} \tag{118}$$

where $\Theta_{Labcdefhi}$ is the Lovelock tensor defined by

$$\Theta_{Labcdefhi} := g\Sigma_{ab}\Sigma_{cd}\Sigma_{ef}\Sigma_{hi}\varepsilon_{aceh}\varepsilon_{bdfi} \tag{119}$$

and if $T_{...a...b...}$ are the components of a tensor then $\Sigma_{ab}\, T_{...a...b...} := ½(T_{...a...b...} + T_{...b...a...})$. In (118) $\alpha$ is an arbitrary scalar concomitant with functional form $\alpha=\alpha(\varphi, \xi, \varphi_{,i}, \xi_{,i}, F_{ij})$.

In **Appendix B** of [1] I demonstrate that there exists a one-to-one correspondence between the 2 and 6 index tensorial concomitants with Property S and that if $\alpha_{hi}$ is he most general 2-index tensorial concomitant of $\varphi$, $\xi$, $\varphi_{,i}$, $\xi_{,i}$ and $F_{ij}$ with Property S, then the most general 6-index tensorial concomitant of $\varphi$, $\xi$, $\varphi_{,i}$, $\xi_{,i}$ and $F_{ij}$ with Property S is given by

$$\alpha_{abcdef} = \Theta_{Labcdefhi}\alpha^{hi} \quad . \tag{120}$$

Unfortunately there are a lot of tensorial concomitants like $\alpha^{hi}$. If we just consider concomitants which algebrically are no worse than second degree in $F_{ab}$ then $\alpha^{hi}$ could be built from elements of the set

$\{g^{hi}; \varphi^h\varphi^i; \xi^h\xi^i; (\varphi^h\xi^i+\varphi^i\xi^h); F^{ia}F^{h}{}_{a}; F^{ia}F^{hb}\varphi_{,a}\varphi_{,b}; F^{ia}F^{hb}\xi_{,a}\xi_{,b}; F^{ia}F^{hb}(\varphi_{,a}\xi_{,b}+\varphi_{,b}\xi_{,,a}); {}^*F^{ia}{}^*F^{hb}\varphi_{,a}\varphi_{,b}; {}^*F^{ia}{}^*F^{hb}\xi_{,a}\xi_{,b};$
${}^*F^{ia}{}^*F^{hb}(\varphi_{,a}\xi_{,b}+\varphi_{,b}\xi_{,,a}); {}^*F^{ia}F^{hb}\varphi_{,a}\varphi_{,b}+{}^*F^{ha}F^{ib}\varphi_{,a}\varphi_{,b}; {}^*F^{ia}F^{hb}\xi_{,a}\xi_{,b}+{}^*F^{ha}F^{ib}\xi_{,a}\xi_{,b}; ({}^*F^{ia}F^{hb}+{}^*F^{ha}F^{ib})(\varphi_{,a}\xi_{,b}+\varphi_{,b}\xi_{,,a})\}$

with coefficient functions that are scalar concomitants of $\varphi$, $\xi$, $\varphi_{,i}$, $\xi_{,i}$ and $F_{ij}$. This could then be combined with (119) and (120) to build all of the 6-index tensorial concomitants of $\varphi$, $\xi$, $\varphi_{,i}$, $\xi_{,i}$ and $F_{ij}$ with Property S that are tensorially at most of second degree in $F^{ab}$. Truly a monstrous tensorial concomitant. And of course there is no reason to stop at second degree in $F^{ab}$. However, when you start going to higher degrees you have to be aware that there are various dimensionally dependent identities relating the $F_{ab}$'s, $\varphi_{,a}$'s and $\xi_{,a}$'s such as

$$0 = \delta^{abcde}_{pqrst} F_b{}^q F_c{}^r F_d{}^s F_e{}^t \,, 0 = \delta^{abcde}_{pqrst} F_b{}^q F_c{}^r F_d{}^s \varphi_e \varphi^t \,, 0 = \delta^{abcde}_{pqrst} F_b{}^q F_c{}^r \varphi_d \varphi^s \xi_e \xi^t \,, \text{ etc..}$$

Building the four index tensorial concomitant of $\varphi$, $\xi$, $\varphi_{,i}$, $\xi_{,i}$ and $F_{ij}$ with Property S is even worse, since we can assume that such a concomitants is built from all possible products of the set $\{g_{ab};\varphi_{,a};\xi_{,a};F_{ab};\varepsilon_{abcd}\}$ with the appropriate symmetries. This was hard enough to do in [1] and [6] when $F_{ab}$ was absent, and is much harder when $F_{ab}$ is thrown into the mix.

To complete the formal construction of $A^{ij}$ we require the form of $a_9{}^{ijabcde}$, $a_{10}{}^{ijabcde}$ and $a_{14}{}^{ijabc}$ where each of these concomitants has Property S on all of their indices except the last one with $a_9{}^{ijab(cde)} = a_{10}{}^{ijab(cde)} = 0$ and $a_{14}{}^{ij(abc)} = 0$. To help construct these concomitants. We shall say a quantity $Q = Q[i_1 i_2; \ldots ; i_{2p-1} i_{2p}; i]$ has Property T if it has Property S in the indices $i_1 i_2; \ldots ; i_{2p-1} i_{2p}$; and it satisfies the cyclic identity involving the three indices $i_{2h-1} i_{2h}$, and i for h=2,...,p;*e.g.,*

$$Q[i_1 i_2; \ldots i_{2h-2};(i_{2h-1} i_{2h};|. \, . \, . ;i_{2p-1} i_{2p}|;i)] = 0.$$

The concomitants $a_9{}^{ijabcde}$, $a_{10}{}^{ijabcde}$ and $a_{14}{}^{ijabc}$ have Property T. Now it is fairly clear that if $\alpha_f$ is a co-vector concomitant of $\varphi$, $\xi$, $\varphi_{,i}$, $\xi_{,i}$ and $F_{ij}$ then $\alpha_f \Theta_L{}^{ijabcdef}$ will be a tensorial concomitant of $\varphi$, $\xi$, $\varphi_{,i}$, $\xi_{,i}$ and $F_{ij}$ with Property T. Conversely if $\alpha_{ijabcde}$ is a 7 index tensorial concomitant of $\varphi$, $\xi$, $\varphi_{,i}$, $\xi_{,i}$ and $F_{ij}$ with Property T then $\alpha_{ijabcde} \Theta_L{}^{ijabcdef}$ will be a vector concomitant of $\varphi$, $\xi$, $\varphi_{,i}$, $\xi_{,i}$ and $F_{ij}$. It can be shown that this map between one index concomitants with Property T and seven index concomitants with Property T is an isomorphism and hence we can easily build concomitants such as $a_9{}^{ijabcde}$ and $a_{10}{}^{ijabcde}$. Similarly we can build the five index concomitants of $\varphi$, $\xi$, $\varphi_{,i}$, $\xi_{,i}$ and $F_{ij}$ from the three index concomitants of these variables using the Lovelock tensor $\Theta_L{}^{ijabcdef}$. Unfortunately building the three index concomitants with Property T is not that trivial. In any case we can, with some effort, build all seventeen of the coefficients functions required in (111).

Once the $A^{ij}$ of (111) is expressed using the general forms for the coefficient functions $a_\alpha$ ($\alpha$=1,...,17), we obtain an immense equation involving numerous scalar concomitants of $\varphi$, $\xi$, $\varphi_{,i}$, $\xi_{,i}$ and $F_{ij}$, which need to be constrained. These scalar concomitants are constrained by computing $A^{ij}{}_{|j}$ and requiring the resultant equation to be expressible as in (112) for some concomitants B, C and $D^i$. Once this Herculean task is complete the Lagrangian $L_A := g_{ij}A^{ij}$ will be our candidate for the Lagrangian that yields the most general quartet of concomitants ($A^{ij}$,B,C,$D^i$) that satisfies conditions (i)-(iii). The reason why $L_A$ is our choice for such a Lagrangian follows from:

**Lemma 12:** If the tensorial concomitant quartet ($A^{ij}$,B,C,$D^i$) satisfies conditions (i)-(iii) then all of the variational derivatives of $L_A := g_{ij}A^{ij}$ are of second order.

**Proof:** For the bi-scalar-tensor case I prove this result in **Lemma 7** of [1]. I shall now show that the variational derivative of $L_A$ with respect to $A_i$ is of second order. To that end let L be the Lagrangian which is such that $E^{ij}(L) = A^{ij}$, where L need not be of second order. Then we have

$$E^i(L_A) = E^i(g_{ab}A^{ab}) = \frac{\partial}{\partial A_i}(g_{ab}A^{ab}) - \frac{d}{dx^j}\frac{\partial}{\partial A_{i,j}}(g_{ab}A^{ab}) + \frac{d^2}{dx^j dx^k}\frac{\partial}{\partial A_{i,jk}}(g_{ab}A^{ab})$$

which can be rewritten as follows:

$$E^i(L_A) = g_{ab}\{A^{ab;i} - \frac{d}{dx^j}A^{ab;i,j} + \frac{d^2}{dx^j dx^k}A^{ab;i,jk}\} - g_{ab,j}A^{ab;i,j} + g_{ab,jk}A^{ab;i,jk} + 2g_{ab,j}\frac{d}{dx^k}A^{ab;i,jk}. \qquad (121)$$

The term within curly brackets in (121) is just $E^i(A^{ab})$, which due to the fourth equation in (64) is $D^{i;ab}$. Hence the term within curly brackets in (121) is

$$g_{ab}E^i(A^{ab}) = g_{ab}D^{i;ab},$$

which clearly is of second order. We could now apply Thomas's replacement Theorem to (121) to deduce that $E^i(L_A)$ is indeed of second order. This result could also be obtained by using the Property S conditions satisfied by the various partial derivatives of $A^{ij}$ to demonstrate that the last term on the right-hand side of (121) is of second order.■

In a similar way we could prove that each of the Lagrangians $\varphi_{,a}\varphi_{,b}A^{ab}$, $\varphi_{,a}\xi_{,b}A^{ab}$, $\xi_{,a}\xi_{,b}A^{ab}$, $\varphi_{,a}A_bA^{ab}$, $\xi_{,a}A_bA^{ab}$, $A_{,a}A_bA^{ab}$, B and C generate second-order bi-scalar-vector field equations when the concomitant quartet $(A^{ij},B,C,D^i)$ satisfies conditions (i)-(iii). From my experience working on problems involving the construction of second-order field equations the Lagrangians listed in the previous sentence are much more complex then the Lagrangian obtained from $g_{ab}A^{ab}$. There are several reasons for that. One is that the Lagrangians involving B and C in general are built from the 38 elements of the set presented in (116), unlike the 16 terms that arise in the $A^{ij}$ of (111). Secondly almost all of the coefficient functions that arise in the construction of the quartet members involve mxm generalized Kronecker deltas. These Kronecker deltas usually arise in such a way that their $g_{ab}$ contraction turns them into (m−1)x(m−1) generalized Kronecker deltas, which does not happen for the other contractions, and, of course, the smaller generalized Kronecker deltas are easier to work with in practice. You are probably wondering what about the $D^a$ contractions? The answer to this question is provided by

**Lemma 13:** Suppose that the tensorial concomitant quartet $(A^{ij},B,C,D^i)$ satisfies conditions (i)-(iii), and let $L_{\varphi D} := \varphi_{,a}D^a$, $L_{\xi D} := \xi_{,a}D^a$, and $L_{AD} := A_aD^a$. The Lagrangians $L_{\varphi D}$ and $L_{\xi D}$ yield field theories that are at most of second-order and are compatible with conservation of charge. However the Lagrangian $L_{AD}$ yields field theories that are compatible with conservation of charge with $E^{ij}(L_{AD})$ being at most of second-order, while the other Euler-Lagrange tensors are at most of third-order and first degree in $\varphi_{,rst}$, $\xi_{,rst}$ and $A_{a,rst}$.

**Proof:** To begin let us look at $E^i(L_{AD})$ which is given by

$$E^i(L_{AD}) = \frac{\partial}{\partial A_i}(A_a D^a) - \frac{d}{dx^j}(A_a D^{a;i,j}) + \frac{d^2}{dx^j dx^k}(A_a D^{a;i,jk}) .$$

When we expand this equation as we did for $E^{ij}(L_A)$ in the proof of **Lemma 12**, we find that

$$E^i(L_{AD}) = D^i + A_a E^i(D^a) - A_{a,j} D^{a;i,j} + A_{a,jk} D^{a;i,jk} + 2A_{a,k} \frac{d}{dx^j} D^{a;i,jk} . \tag{122}$$

Due to the fourth equation in (67) we know that $E^i(D^a) = D^{i;a}$ which vanishes, and hence (122) tells us that $E^i(L_{AD})^{;a}=0$. If we expand the last term appearing on the right-hand side of (122) we see that the term involving $g_{pq,rsj}$ must vanish, but there is no reason why the terms involving $\varphi_{,pqj}$, $\xi_{,pqj}$ or $A_{p,qrj}$ must vanish (even when (93) is taken into account). As a result of this observation we see that in general $E^i(L_{AD})$ is at most of third-order. In a similar way we can show that the same conclusion also applies to $E^{ij}(L_{AD})$, $E_\varphi(L_{AD})$ and $E_\xi(L_{AD})$, and that each of these field tensor densities is independent of explicit dependence on $A_a$.■

**Section 3: The Search for Simpler Formulations of the Electromagnetic Field Equations**

(117) has shown us that in general $D^i$ is redoubtable. From my discussion in **Section 1** it seems like it would be interesting to know the form of the most general $D^i$ that is independent of $A_a$ and its derivatives. From (117) we can say that the basic form of such a reduced concomitant, $\text{Đ}^i$, will be

$$\text{Đ}^i =$$

$$d_1{}^{iabcdef} R_{acdb} \varphi_{ef} + d_2{}^{iabcdef} R_{acdb} \xi_{ef} + d_3{}^{iabcd} R_{acdb} + d_4{}^{iabcdefhk} \varphi_{ab} \varphi_{cd} \varphi_{ef} \xi_{hk} + d_5{}^{iabcdefhk} \varphi_{ab} \varphi_{cd} \xi_{ef} \xi_{hk} +$$

$$+ d_6{}^{iabcdefhk} \varphi_{ab} \xi_{cd} \xi_{ef} \xi_{hk} + d_7{}^{iabcdef} \varphi_{ab} \varphi_{cd} \varphi_{ef} + d_8{}^{iabcdef} \varphi_{ab} \varphi_{cd} \xi_{ef} + d_9{}^{iabcdef} \varphi_{ab} \xi_{cd} \xi_{ef} + d_{10}{}^{iabcdef} \xi_{ab} \xi_{cd} \xi_{ef} +$$

$$+d_{11}{}^{iabcd}\varphi_{ab}\varphi_{cd} + d_{12}{}^{iabcd}\varphi_{ab}\xi_{cd} + d_{13}{}^{iabcd}\xi_{ab}\xi_{cd} + d_{14}{}^{iab}\varphi_{ab} + d_{15}{}^{iab}\xi_{ab} + d_{16}{}^{i}, \tag{123}$$

where $d_\alpha{}^{\cdots} = d_\alpha{}^{\cdots}(g_{ij};\varphi;\varphi_{,i};\xi;\xi_{,i})$ for every $\alpha$=1,..,16. Due to the absence of $F_{ij}$, these coefficient functions are easier to build than the ones appearing in (117), but are no means trivial. In fact the only really simple ones are $d_1{}^{iabcdef}$, $d_2{}^{iabcdef}$, $d_3{}^{iabcd}$, $d_{16}{}^{iab}$, $d_{17}{}^{iab}$ and $d_{18}{}^{i}$. The others are much more difficult since they do not have enough symmetries. So rather than try to build the general $Đ^i$, let's come at the problem from the other direction and try to build the simplest possible $Đ^i$'s and see how far we can get.

In view of (123) we see that the simplest $Đ^i$ would be the one devoid of all second order terms in the scalar field and metric tensor and given by

$$Đ_1{}^{i} = d_{18}{}^{i}. \tag{124}$$

In **Appendix B** of [1] (*see*, **Lemma B.1**) I show that the most general coefficient function $d_{18}{}^{i}$ is given as a linear combination of $g^{½}\varphi^i$ and $g^{½}\xi^i$ and so (124) becomes

$$Đ_1{}^{i} = g^{½}(\theta_1\, \varphi^i + \theta_2\, \xi^i\,) \tag{125}$$

where $\theta_\alpha = \theta_\alpha(\varphi,\xi,X,Y,Z)$ (for $\alpha$=1,2) with X, Y, Z defined in (35). To guarantee that charge is conserved we require $Đ_1{}^{i}{}_{|i} = 0$. Using (125) we see that this condition implies that

$$\begin{aligned}&\theta_{1\varphi}X + \theta_{1\xi}Z + 2\theta_{1X}\varphi^i\varphi^j\varphi_{ij} + 2\theta_{1Y}\varphi^i\xi^j\xi_{ij} + \theta_{1Z}\varphi^i\varphi^j\xi_{ij} + \theta_{1Z}\varphi^i\xi^j\varphi_{ij} + \\ &+ \theta_{2\varphi}Z + \theta_{2\xi}Y + 2\theta_{2X}\xi^i\varphi^j\varphi_{ij} + 2\theta_{2Y}\xi^i\xi^j\xi_{ij} + \theta_{2Z}\xi^i\varphi^j\xi_{ij} + \theta_{2Z}\xi^i\xi^j\varphi_{ij} + \theta_1\Box\varphi + \theta_2\Box\xi = 0\end{aligned} \tag{126}$$

where I have denoted partial derivatives with respect to $\varphi$, $\xi$, X, Y and Z by subscripts. This equation must hold for all $\varphi$ and $\xi$. So if we differentiate (126) with respect to $\varphi_{rs}$ we obtain

$$2\theta_{1X}\varphi^r\varphi^s + ½\theta_{1Z}(\varphi^r\xi^s + \varphi^s\xi^r) + \theta_{2X}(\varphi^r\xi^s + \varphi^s\xi^r) + \theta_{2Z}\xi^r\xi^s + \theta_1 g^{rs} = 0\,. \tag{127}$$

To see that (127) implies that $\theta_1 = 0$, let P be an arbitrary point in the four-dimensional space in which we are considering a bi-scalar-vector-tensor field theory. At P we can find a unit vector $V^r$ perpendicular to the (in general) 2-plane spanned by $\{\varphi^r,\xi^r\}$. If we contract (127) with $V_rV_s$ we see that $\theta_1 = 0$. In a similar way we find that $\theta_2 = 0$. Thus our first candidate for a simplest $\text{Đ}^i$ does not work. Now for our next candidate.

Let $\text{Đ}_{Si}$ denote the vector density we obtain from (123) when we consider only those terms which are algebraically at most of first degree in the second derivatives of $g_{ab}$, $\varphi$ and $\xi$. Thus we have

$$\text{Đ}_S{}^i = d_{S1}{}^{iabcd}R_{acdb} + d_{S2}{}^{iab}\varphi_{ab} + d_{S3}{}^{iab}\xi_{ab} + d_{S4}{}^i, \tag{128}$$

where $d_{S\alpha}{}^{\cdots} = d_{S\alpha}{}^{\cdots}(g_{ij};\varphi;\varphi_{,i};\xi;\xi_{,i})$ and each have Property T. From our work in **Section 1**, in particular our work with the Lagrangian $L_{BS}$ given in (2), we know that $\text{Đ}_{Si} \neq 0$. We shall now determine exactly what it does equal.

**Lemma B.4** in [1] tells us that the most general 3 index scalar density concomitant of the required variables with Property T is given by

$$\theta_{abc} = g^{½}\{\alpha_1(g_{ac}\varphi_b + g_{ab}\varphi_c - 2g_{bc}\varphi_a) + \alpha_2(g_{ac}\xi_b + g_{ab}\xi_c - 2g_{bc}\xi_a) + \alpha_3(\varphi_a\xi_b\varphi_c + \varphi_a\varphi_b\xi_c - 2\xi_a\varphi_b\varphi_c) +$$
$$+ \alpha_4(\xi_a\varphi_b\xi_c + \xi_a\xi_b\varphi_c - 2\varphi_a\xi_b\xi_c) + \alpha_5({}^*\omega_{ab}\varphi_c + {}^*\omega_{ac}\varphi_b) + \alpha_6({}^*\omega_{ab}\xi_c + {}^*\omega_{ac}\xi_b)\}, \tag{129}$$

where recall that $\omega_{ab} := ½(\varphi_a\xi_b - \varphi_b\xi_a)$ and ${}^*\omega_{ab} := ½g^{½}\varepsilon_{abcd}\,\omega^{cd}$. (For the case when the metric is Lorentzian ${}^*\omega^{ab} := g^{ap}g^{bq}\,{}^*\omega_{pq} = -½g^{-½}\varepsilon^{abrs}\,\omega_{rs}$.) Note $\theta_{abc}$ is symmetric in b and c, cycles to zero on a,b,c. I previously pointed out that the most general 5 index tensor density concomitant $\theta^{defhi}$ of $g_{ij}$, $\varphi$, $\varphi_{,i}$, $\xi$ and $\xi_{,i}$ can be obtained by contracting the Lovelock tensor, $\Theta_L{}^{abcdefhi}$ given in (119) with $\theta_{cab}$. This $\theta^{defhi}$ will be symmetric in e,f and h,i, and cycle to zero on d,e,f and d,h,i. Thus we can use (129) and (118) to obtain

$$\theta^{defhi} =$$

$$g^{½}\{\beta_1[\varphi_a\,\delta^{aeh}_{pqr}\,g^{pd}g^{qf}g^{ri} + \varphi_a\,\delta^{afi}_{pqr}\,g^{pd}g^{qe}g^{rh} + \varphi_a\,\delta^{afh}_{pqr}\,g^{pd}g^{qe}g^{ri} + \varphi_a\,\delta^{aei}_{pqr}\,g^{pd}g^{qf}g^{rh}] +$$

$$+ \beta_2[\xi_a\,\delta^{aeh}_{pqr}\,g^{pd}g^{qf}g^{ri} + \xi_a\,\delta^{afi}_{pqr}\,g^{pd}g^{qe}g^{rh} + \xi_a\,\delta^{afh}_{pqr}\,g^{pd}g^{qe}g^{ri} + \xi_a\,\delta^{aei}_{pqr}\,g^{pd}g^{qf}g^{rh}] +$$

$$+\ \beta_3[\varphi_a\xi_b\varphi^c\,\delta^{abeh}_{cpqr}\,g^{pd}g^{qf}g^{ri} + \varphi_a\xi_b\varphi^c\,\delta^{abfi}_{cpqr}\,g^{pd}g^{qe}g^{rh} + \varphi_a\xi_b\varphi^c\,\delta^{abfh}_{cpqr}\,g^{pd}g^{qe}g^{ri} + \varphi_a\xi_b\varphi^c\,\delta^{abei}_{cpqr}\,g^{pd}g^{qf}g^{rh}] +$$

$$+\ \beta_4[\xi_a\varphi_b\xi^c\,\delta^{abeh}_{cpqr}\,g^{pd}g^{qf}g^{ri} + \xi_a\varphi_b\xi^c\,\delta^{abfi}_{cpqr}\,g^{pd}g^{qe}g^{rh} + \xi_a\varphi_b\xi^c\,\delta^{abfh}_{cpqr}\,g^{pd}g^{qe}g^{ri} + \xi_a\varphi_b\xi^c\,\delta^{abei}_{cpqr}\,g^{pd}g^{qf}g^{rh}]\ \} +$$

$$+\beta_5[\omega^{eh}\varphi_b\varepsilon^{bdfi} + \omega^{fi}\varphi_b\varepsilon^{bdeh} + \omega^{fh}\varphi_b\varepsilon^{bdei} + \omega^{ei}\varphi_b\varepsilon^{bdfh}] + \beta_6[\omega^{eh}\xi_b\varepsilon^{bdfi} + \omega^{fi}\xi_b\varepsilon^{bdeh} + \omega^{fh}\xi_b\varepsilon^{bdei} + \omega^{ei}\xi_b\varepsilon^{bdfh}], \quad (130)$$

where for $\gamma=1,...,6$, $\beta_\gamma = \beta_\gamma(\varphi,\xi,X,Y,Z)$. We can use (125), (129) and (130) to determine the general form of the coefficient functions $d_{S1}{}^{iabcd}$, $d_{S2}{}^{iab}$, $d_{S3}{}^{iab}$ and $d_{S4}{}^{i}$ appearing in (128). In this way we find that after some simplification that our preliminary version for $\text{Đ}_{Si}$ is given by

$$\text{Đ}_S{}^i =$$

$$g^{1/2}\{\alpha_1\varphi_a\,\delta^{aeh}_{pqr}\,g^{pi}R_{eh}{}^{qr} + \alpha_2\xi_a\,\delta^{aeh}_{pqr}\,g^{pi}R_{eh}{}^{qr} + \alpha_3\varphi_a\xi_b\varphi^c\,\delta^{abeh}_{cpqr}\,g^{pi}R_{eh}{}^{qr} + \alpha_4\xi_a\varphi_b\xi^c\,\delta^{abeh}_{cpqr}\,g^{pi}R_{eh}{}^{qr} +$$

$$+\,\alpha_5\omega^{eh}\varphi_b\varepsilon^{bifj}R_{ehfj} + \alpha_6\omega^{eh}\xi_b\varepsilon^{bifj}R_{ehfj}\} + g^{1/2}\{\beta_1(\varphi^i{}_a\varphi^a - \varphi^i\Box\varphi) + \beta_2(\varphi^i{}_a\xi^a - \xi^i\Box\varphi) + \beta_3(\varphi^i\varphi^a\xi^b\varphi_{ab} - \xi^i\varphi^a\varphi^b\varphi_{ab}) +$$

$$+\,\beta_4(\xi^i\xi^a\varphi^b\varphi_{ab} - \varphi^i\xi^a\xi^b\varphi_{ab}) + \beta_5{*\omega}^{ia}\varphi^b\varphi_{ab} + \beta_6{*\omega}^{ia}\xi^b\varphi_{ab}\} + g^{1/2}\{\gamma_1(\xi^i{}_a\varphi^a - \varphi^i\Box\xi) + \gamma_2(\xi^i{}_a\xi^a - \xi^i\Box\xi) +$$

$$+\gamma_3(\varphi^i\varphi^a\xi^b\xi_{ab} - \xi^i\varphi^a\varphi^b\xi_{ab}) + \gamma_4(\xi^i\xi^a\varphi^b\xi_{ab} - \varphi^i\xi^a\xi^b\xi_{ab}) + \gamma_5{*\omega}^{ia}\varphi^b\xi_{ab} + \gamma_6{*\omega}^{ia}\xi^b\xi_{ab}\} + g^{1/2}(\delta_1\varphi^i + \delta_2\xi^i), \quad (131)$$

where the coefficient functions $\alpha$, $\beta$, $\gamma$ and $\delta$ are concomitants of $\varphi$, $\xi$, X, Y and Z, with X, Y and Z as in (35). We require $\text{Đ}_S{}^i$ to have identically vanishing divergence. You might think that due to our work with $\text{Đ}_1{}^i$ (*see,* (125)) we can conclude that $\delta_1=\delta_2=0$ in (131). However that conclusion is premature since some of the terms found in the $\text{Đ}_1{}^i{}_{|i}$ part of $D_S{}^i{}_{|i}$ occur in other terms in $D_S{}^i{}_{|i}$ such as the divergence of the term $g^{1/2}\beta_1\varphi^i\Box\varphi$. An in depth analysis of $D_S{}^i{}_{|i}$ is given in the **Appendix** where it is shown that its identical vanishing requires that

$$\text{Đ}_S{}^i = \quad E^i(L_1) + E^i(L_2)\,, \qquad (132)$$

where

$$L_1 := g^{1/2}\sigma_1\omega^{ab}F_{ab} \qquad (133)$$

$$L_2 := g^{1/2}\sigma_2\omega^{ab}{*F}_{ab} \qquad (134)$$

with $\sigma_1$ and $\sigma_2$ being arbitrary functions of $\varphi$, $\xi$, X, Y and Z. (Note that $L_1$ and $L_2$ are just the first and third Lagrangians we encountered in (36).) Thus we have established the following

**Theorem 2:** If $(A^{ij}, B, C, \text{Đ}^i)$ denotes a quartet of tensorial concomitants which satisfy conditions (i)-(iii) presented in (2.1)-(2.3), then $\text{Đ}^i$ must be independent of explicit dependence on $A_a$. If we also assume that $\text{Đ}^i$ is independent of the derivatives of $A_a$ then it must be a polynomial in $R_{abcd}$, $\varphi_{ab}$ and $\xi_{ab}$ whose basic form is given by (123). If we let $\text{Đ}_S{}^i$ denote the form of that polynomial when it is at most of first degree in $R_{abcd}$, $\varphi_{ab}$ and $\xi_{ab}$ then it is given by (132), which can be expressed as

$$\begin{aligned}\text{Đ}_S{}^i = g^{1/2}\{&\sigma_1[\varphi^i{}_a\xi^a - \xi^i\Box\varphi + \varphi^i\Box\xi - \xi^i{}_a\varphi^a] + 2\sigma_{1X}[\varphi^i\varphi^a\xi^b\varphi_{ab} - \xi^i\varphi^a\varphi^b\varphi_{ab}] + 2\sigma_{1Y}[\varphi^i\xi^a\xi^b\xi_{ab} - \xi^i\xi^a\varphi^b\xi_{ab}] + \\ &+\sigma_{1Z}[\varphi^i\xi^a\xi^b\varphi_{ab} - \xi^i\xi^a\varphi^b\varphi_{ab} + \varphi^i\varphi^a\xi^b\xi_{ab} - \xi^i\varphi^a\varphi^b\xi_{ab}] + \varphi^i[\sigma_{1\varphi}Z + \sigma_{1\xi}Y] - \xi^i[\sigma_{1\varphi}X + \sigma_{1\xi}Z] + \\ &+ 4\sigma_{2X}{}^*\omega^{ij}\varphi^c\varphi_{cj} + 4\sigma_{2Y}{}^*\omega^{ij}\xi^c\xi_{cj} + 2\sigma_{2Z}[{}^*\omega^{ij}\varphi^c\xi_{cj} + {}^*\omega^{ij}\xi^c\varphi_{cj}]\}\,, \end{aligned} \tag{135}$$

and hence independent of $R_{abcd}$. $\text{Đ}_S{}^i$ vanishes when the scalar fields are constant.■

If L is any bi-scalar-tensor Lagrangian that yields second-order field equations then let $\mathscr{L}$ denote the Larangian defined by

$$\mathscr{L} := L + \gamma L_M - L_1 - L_2\,, \tag{136}$$

where $L_M$, $L_1$ and $L_2$ are defined by (1), (133) and (134). Using $\mathscr{L}$ we obtain a second-order, bi-scalar-vector-tensor field theory compatible with conditions (i)-(iv) given in (44)-(53) and such that the field equation $E^i(\mathscr{L}) = 0$, has the form

$$\gamma F^{ij}{}_{|j} = E^i(L_1) + E^i(L_2)\,. \tag{137}$$

Due to **Theorem 2**, we know that the field equation presented in (137) is the most general such second-order, bi-scalar-vector-tensor field equation expressible in the form $F^{ij}{}_{|j}$ equal to a term independent of the vector potential and its derivatives that is also at most of first degree in the second derivatives of the metric tensor and scalar fields. Since $E^i(L_1) + E^i(L_2)$ is given by (135) we see that the equation given in (137) is not quasi-linear in the second derivatives of the scalar fields since there are always terms involving either $\varphi_a$ or $\xi_a$ multiplying $\varphi_{,ab}$ or $\xi_{,ab}$. These multiplying factors are either of at least (algebraically) first degree or third degree in $\varphi_a$ or $\xi_a$. To get them to be of first degree we would require that $\sigma_1$ and $\sigma_2$ be functions only of $\varphi$ and $\xi$, in which case $L_2$ would be a divergence and not participate in any of the field equations.

I would like to conclude this section with the following

**Conjecture:** If $(A^{ij},B,C,Đ^i)$ denotes a quartet of tensorial concomitants which satisfy conditions (i)-(iii) presented in (44)-(52), then $Đ^i$ must be independent of explicit dependence on $A_a$. If we also assume that $Đ^i$ is independent of the derivatives of $A_a$ then it is expressible as the variational derivative with respect to $A_i$ of a Lagrangian which is the sum of the four Lagrangians $L_{01}$, $L_{03}$, $L'_{15}$ and $L_*'$ given in (36), (37) and (39).■

To show that this conjecture is indeed a Theorem will be quite difficult. I would begin by first considering the case when $Đ^i$ is at most of second degree in $R_{abcd}$, $\varphi_{,ab}$ and $\xi_{,ab}$. Your analysis should just follow the steps that were used in this section to study $Đ_S{}^i$ when it was of first degree. Once you have finished that go back and consider the case where $Đ^i$ is of third degree in $R_{abcd}$, $\varphi_{,ab}$ and $\xi_{,ab}$, using the same approach again. At some point you will have to demonstrate that the fourth-degree terms in $\varphi_{,ab}$ and $\xi_{,ab}$ that appear in $Đ^i$ vanish. This will be the case if you can show that things like $Đ_{\varphi\varphi\varphi\xi}{}^{i;ab;cd;ef;jk}$ have Property T.

**Section 4: Ramifications of the Theory Developed**

The primary purpose of this paper has been to show how we can use bi-scalar fields to generate second-order bi-scalar-vector-tensor field theories for which the electromagnetic field equations have the form $F^{ij}{}_{|j}$ equals terms not involving the vector potential or its derivatives. For these theories we have the bi-scalar-tensor fields acting as the source of electromagnetism. The general form of these equations would be $F^{ij}{}_{|j} = \not{D}^i$ where $\not{D}^i$ is given by (3.1). I have also been able to show that when we assume that $\not{D}^i$ is at most of first degree in the second derivatives of the metric and bi-scalar fields then $\not{D}^i$ is given by (3.13); *viz.,*

$$\not{D}_S{}^i = g^{1/2}\{\sigma_1[\varphi^i{}_a\xi^a - \xi^i\Box\varphi + \varphi^i\Box\xi - \xi^i{}_a\varphi^a] + 2\sigma_{1X}[\varphi^i\varphi^a\xi^b\varphi_{ab} - \xi^i\varphi^a\varphi^b\varphi_{ab}] + 2\sigma_{1Y}[\varphi^i\xi^a\xi^b\xi_{ab} - \xi^i\xi^a\varphi^b\xi_{ab}] +$$

$$+\sigma_{1Z}[\varphi^i\xi^a\xi^b\varphi_{ab} - \xi^i\xi^a\varphi^b\varphi_{ab} + \varphi^i\varphi^a\xi^b\xi_{ab} - \xi^i\varphi^a\varphi^b\xi_{ab}] + \varphi^i[\sigma_{1\varphi}Z + \sigma_{1\xi}Y] - \xi^i[\sigma_{1\varphi}X + \sigma_{1\xi}Z] +$$

$$+ 4\sigma_{2X}{}^*\omega^{ij}\varphi^c\varphi_{cj} + 4\sigma_{2Y}{}^*\omega^{ij}\xi^c\xi_{cj} + 2\sigma_{2Z}[{}^*\omega^{ij}\varphi^c\xi_{cj} + {}^*\omega^{ij}\xi^c\varphi_{cj}]\} \ . \qquad (138)$$

The natural question to ask is: if we consider a *scalar*-vector-tensor field theory of second-order could we obtain an electromagnetic field equation of the form $F^{ij}{}_{|j} = \not{D}^i$ where $\not{D}^i$ is independent of the vector potential and its derivatives, and is built solely from the scalar and tensor fields? The answer is no. To see why I say this first note that the mathematical machinery necessary to handle the scalar-vector-tensor field theory case can be obtained from what we did in **Sections 2** and **3** by just setting $\xi = 0$. Thus we can use (3.1) to deduce that the general form of the reduced $\not{D}^i$ for the scalar-vector-tensor theory will be given by

$$\not{D}^i =$$

$$d_1{}^{iabcdef}R_{acdb}\varphi_{ef} + d_3{}^{iabcd}R_{acdb} + d_7{}^{iabcdef}\varphi_{ab}\varphi_{cd}\varphi_{ef} + d_{11}{}^{iabcd}\varphi_{ab}\varphi_{cd} + d_{14}{}^{iab}\varphi_{ab} + d_{16}{}^i, \qquad (139)$$

where the $d_{\cdot}{}^{i\ldots}$ satisfy the same symmetries as they did in (123), and are concomitants of $g_{ab}$, $\varphi$ and $\varphi_a$. However, we can appeal to **Lemmas 2**, **3**, **5** and **6** to deduce that all of the $d_\alpha{}^{i\ldots}$ appearing in (139) have Property T. Now due to my remarks in **Section 3**, we know that if we can build $\theta^i$, the most general one index concomitant of $g_{ij}$, $\varphi$ and $\varphi_{,i}$, then the most general seven index concomitant of the same variables with Property T will be given by $\theta^i\Theta_{Lijabcdef}$. Similarly if we can build the most general three index concomitant $\theta^{jkl}$, of $g_{ij}$, $\varphi$ and $\varphi_{,i}$ that has Property T then the most general five index concomitant of these arguments would be given by $\theta^{jkl}\Theta_{Lijklabcd}$. From my work in **Appendix B** [1] we know that

$$\theta^i = \alpha\varphi^i \text{ and } \theta^{ijk} = \beta(g^{ij}\varphi^k + g^{ik}\varphi^j - 2g^{jk}\varphi^i) \tag{140}$$

where $\alpha$ and $\beta$ are arbitrary functions of $\varphi$ and X. Using (119), (139) and (140) it is a straight forward matter to show that our preliminary version of $Đ^i$ is given by

$$Đ^i = g^{½}\{\alpha_1\,\delta^{iabc}_{defh}\,\varphi^d\varphi_a{}^e R_{bc}{}^{fh} + \alpha_2\,\delta^{iab}_{cde}\,\varphi^c R_{ab}{}^{de} + \alpha_3\,\delta^{iabc}_{defh}\,\varphi^d\varphi_a{}^e\varphi_b{}^f\varphi_c{}^h + \alpha_4\,\delta^{iab}_{cde}\,\varphi^c\varphi_a{}^d\varphi_b{}^e +$$

$$+ \alpha_5\,\delta^{ia}_{bc}\,\varphi^b\varphi_a{}^c + \alpha_6\varphi^i\} \tag{141}$$

where $\alpha_1$,..., $\alpha_6$ are concomitants of $\varphi$ and X. We require that $Đ^i{}_{|i} = 0$. Using calculations similar to those employed in the **Appendix**, it is not too difficult to prove that the demand that $Đ^i$ be divergence-free implies that all of the $\alpha$ coefficients in (141) must vanish. Thus we have established the following

**Theorem 3:** If in a space of four-dimensions the triple $(A^{ij},B,D^i)$ of second-order scalar-vector-tensor concomitants arise from a variational principle which yields a field theory consistent with the conservation of charge and such that the vector equation has the form $g^{½}F^{ij}{}_{|j} = Đ^i$ where $Đ^i$ is independent of the vector potential and its derivatives, then $Đ^i = 0$.■

So we have shown that if we want to adjoin to a vector-tensor theory some scalar fields for which the electromagnetic field equation has the form $F^{ij}{}_{|j}$ equals a current built from the scalar fields and the metric tensor, then we have to have at least two scalar fields.

I shall now explain how we can use the theory developed so far to introduce the Higgs field into a vector-tensor theory. In the electroweak theory the Higgs field, which I shall unconventionally denote by $\sigma = \begin{pmatrix}\sigma_1 \\ \sigma_2\end{pmatrix}$ where $\sigma_1$ and $\sigma_2$ are complex valued scalar fields which can be expressed as $\sigma_1 := \varphi_1 + i\xi_1$ and $\sigma_2 := \varphi_2 + i\xi_2$, with $\varphi_1$, $\xi_1$, $\varphi_2$ and $\xi_2$ being real scalar fields as we have considered throughout this paper. $\sigma_1$ is called the electric part of $\sigma$ and interacts with the $W^+$ and $W^-$ fields, while $\sigma_2$ is called the neutral part of $\sigma$, which interacts with the Z field and produces the experimentally observed Higgs particle. Although I have not actually discussed the case of theories involving four scalar fields, a vector field and a tensor field in this paper we still can use the theory that was developed to treat the Higgs field. This is so since it conveniently breaks up into two bi-scalar fields. To that end I define $\omega_1 := d\varphi_1 \wedge d\xi_1$ and $\omega_2 := d\varphi_2 \wedge d\xi_2$, so that locally

$$\omega_{1ab} := \tfrac{1}{2}(\varphi_{1a}\xi_{1b} - \varphi_{1b}\xi_{1a}) \text{ and } \omega_{2ab} := \tfrac{1}{2}(\varphi_{2a}\xi_{2b} - \varphi_{2b}\xi_{2a})\ . \tag{142}$$

The work done in **Section 1**, in particular the first equation in (36) and (39), suggest that we now consider the two Lagrangians

$$L_{H1} := \kappa_1 g^{1/2} \omega_1{}^{ab} F_{ab} \tag{143}$$

and

$$L_{H2} := \kappa_2 g^{1/2}\, \delta^{abcd}_{pqrs}\, \omega_2{}^{pq} F_{ab} R_{cd}{}^{rs}\ , \tag{144}$$

where $\kappa_1$ and $\kappa_2$ are constants. The combined Maxwell-Higgs Lagrangian

$$\mathfrak{L}_{MH} := L_M + L_{H1} + L_{H2} \tag{145}$$

will provide us with a way for the electric part of the Higgs field to produce a conserved current source for the electromagnetic field through $L_{H1}$. While the Lagrangian $L_{H2}$ will provide a way for the neutral part of the Higgs field to allow the (neutral) curvature tensor to also produce a conserved current source for the electromagnetic field. Of course, there exists other numerous alternate ways to use (36)-(40) to couple the Higgs field to the electromagnetic field. The Lagrangian $\mathfrak{L}_{MH}$ could now be added to the traditional Einstein Lagrangian with cosmological term or to one of the scalar-tensor, or bi-scalar-tensor Lagrangians of [1] or [10], to obtain a complete second-order multi-scalar-vector- tensor field theory. An important thing to note is that when applying any of these combined Lagrangians to cosmological considerations the Lagrangians $L_{H1}$ and $L_{H2}$ will vanish once the Universe gets older than about $10^{-12}$ seconds, since the Higgs field will have assumed its constant vacuum expectation value after that time and $\omega_1$ and $\omega_2$ will be zero. However before that time the Higgs field will be able to create electromagnetic fields, vestiges of which might still be observable. It should be noted that unlike the Lagrangian $L_M$, the Lagrangians $L_{H1}$ and $L_{H2}$ are not invariant (modulo sign) under the transformation that replaces $F_{ab}$ by $*F_{ab}$. In fact $L_{H1}$ becomes a divergence when transformed in this way. Consequently one might suspect that the combined Lagrangian could produce more magnetic fields than electric fields (or vice versa) during the period before the Universe was $10^{-12}$ seconds old.

I would now like to conclude with a few brief remarks about bi-scalar-gauge-tensor field theories. To that end let $P(M,G,\pi)$ be a principal fibre bundle over our spacetime M, with n dimensional structure group G, projection $\pi$ and total space P. A connection $\eta$ in P will be called a gauge field or Yang-Mills field [18]. (Please see Kobayashi & Nomizu [19] for a discussion of principal fibre bundles and connections in them.) If $U \subset M$ is an open subset, then by a choice of gauge in M, or gauge in M, with domain U, I shall mean a pair $(U,\gamma)$, where $\gamma: U \subset M \to P$ is a section of $\pi$. If $(U,\gamma)$ is a gauge in M then $\eta_\gamma := \gamma^*\eta$ is an LG-valued 1 form on U, where LG is the Lie algerbra of the Lie group G. Suppose that (V,x) is a chart of M and that $U \cap V$ is non-empty, and let $\{e_\beta\}$ be a basis for LG. On $U \cap V$ we can locally express $\eta_\gamma = A^\beta{}_i e_\beta \otimes dx^i$ where the repeated small Greek indices are summed from 1 to n =dim G. The functions $A^\beta{}_i$ are the gauge potentials of $\eta$ with respect to the gauge $(U,\gamma)$, chart (V,x) and basis $\{e_\beta\}$, or just the gauge potentials for short. In terms of these gauge potentials the local components of the LG valued curvature two form F associated with the connection $\eta$ are defined by

$$F^{\beta}{}_{ab} := A^{\beta}{}_{b,a} - A^{\beta}{}_{a,b} + C^{\beta}{}_{\mu\nu} A^{\mu}{}_{a} A^{\nu}{}_{b}, \tag{146}$$

where $C^{\beta}{}_{\mu\nu}$ denotes the components of the structure constants of LG with respect to the basis $\{e_{\beta}\}$ and are defined by $C^{\beta}{}_{\mu\nu}\, e_{\beta} := [e_{\mu}, e_{\nu}]$. Locally the Yang-Mills Lagrangian is given by

$$L_{YM} := \tfrac{1}{4}\kappa g^{\frac{1}{2}} K_{\mu\nu} F^{\mu}{}_{ab} F^{\nu ab}, \tag{147}$$

where $\kappa$ is a constant and $K_{\mu\nu}$ denotes the components of any symmetric bilinear form on LG which is Ad invariant, and usually taken to be the Cartan-Killing bilinear form with components $K_{\mu\nu} := C^{\lambda}{}_{\mu\tau} C^{\tau}{}_{\nu\lambda}$. By $K_{\mu\nu}$ being Ad invariant I mean that it is invariant under the adjoint representation of G in the group of automorphisms of LG, which requires that $\forall\, g \in G$, $K_{\mu\nu} = Ad(g)^{\lambda}{}_{\mu}\, Ad(g)^{\tau}{}_{\nu} K_{\lambda\tau}$, where $Ad(g)e_{\mu} := Ad(g)^{\lambda}{}_{\mu}\, e_{\lambda}$.

Now suppose that $(U,\gamma)$ and $(U',\gamma')$ are overlapping gauges. Since $\gamma$ and $\gamma'$ map into P there exists a function $\psi_{UU'}: U \cap U' \subset M \rightarrow G$ which is such that $\forall\, m \in U \cap U'$, $\gamma(m) \bullet \psi_{UU'}(m) = \gamma'(m)$, where $\bullet$ denotes the right action of G on P. $\psi_{UU'}$ is called a gauge transformation. Under this gauge transformation the components of the gauge curvature field given in (146) transforms as

$$F'^{\beta}{}_{ab} = Ad^{\beta}{}_{\lambda}(\psi_{UU'}{}^{-1}) F^{\lambda}{}_{ab} \tag{148}$$

Thus we see that the Yang-Mills Lagrangian in (147) is invariant under a gauge transformation, which guarantees conservation of gauge charge.

Let $\varphi$ and $\xi$ be two scalar fields on M. What we would like to do is modify the Yang-Mills Lagrangian as we did the Maxwell Lagrangian given in (1). This would entail replacing $K_{\mu\nu} F^{\mu}{}_{ab}$ by something like $V_{\nu}\omega_{ab}$ where $V_{\nu}$ would denote the components of a vector in LG*, the dual space of LG. We would also require $V_{\nu}$ to be Ad invariant so that the new Lagrangian would be invariant under a gauge transformation. Fortunately such vector exists; *viz.,* $V_\nu := \kappa_G C^\lambda{}_{\nu\lambda}$, where $\kappa_G$ is a constant and $C^\lambda{}_{\nu\lambda}$ denotes the components of the the Cartan-Killing vector. $V_\nu$ is Ad invariant since the structure constants are Ad invariant, satisfying

$$Ad(g)^\alpha{}_\mu \, Ad(g)^\beta{}_\mu C^\gamma{}_{\alpha\beta} = Ad(g)^\gamma{}_\delta C^\delta{}_{\mu\nu}$$

for every $g \in G$. $V_\nu$ also has the property that $V_\nu C^\nu{}_{\alpha\beta} = 0$, and so from (146) we see that $V_\nu F^\nu{}_{ab}$ will be devoid of terms involving the undifferentiated gauge potentials. Unfortunately $C^\lambda{}_{\nu\lambda}$ is zero for all semi-simple Lie groups which are the groups usually used in Yang-Mills theory. The Lie groups for which $C^\lambda{}_{\nu\lambda}$ is non-zero are the non-unimodular, non-nilpotent Lie groups, such as the invertible upper trianglar $n \times n$ matrices. So for such groups we could go to the Lagrangians $L_{01}$, $L_{03}$, $L_{15}'$ and $L_*'$ presented in (36), (37) and (39) and replace $F_{ij}$ with $V_\nu F^\nu{}_{ij}$ to obtain bi-scalar-gauge-tensor Lagrangians which were Ad invariant. These Lagrangians would be capable of yielding second-order, bi-scalar-gauge-tensor field equations compatible with the conservation of gauge charge. So when any one of these Lagrangians is added to the Yang-Mills Lagrangian given in (147) the field equations obtained by varying the gauge fields $A^\alpha{}_i$ will have the form $F_\alpha{}^{ij}{}_{||j}$ equals a current independent of the gauge potentials or their derivatives, where $_{||j}$ denotes the gauge covariant derivative. Thus in this case the Yang-Mills fields will have as their source an Ad invariant current built from the bi-scalar fields and metric tensor, along with their first and second derivatives. The only problem with this is: of what physical significance are gauge fields associated with non-unimodular, non-nilpotent Lie Groups? If we could find any physical reason for studying such fields then we could couple them to the Higgs field in a manner analogous to how we coupled the electromagnetic field to Higgs.

In [20] and [21] I demonstrate that the Einstein-Yang-Mills equations (with cosmological term) are not the most general second-order gauge-tensor field equations derivable from a variational principle which is consistent with conservation of gauge charge and such that the gauge field equation reduces to the Yang-Mills equations when the curvature tensor vanishes. There exists another Lagrangian which can be added to the Einstein-Yang-Mills Lagrangian to yield a theory of the aforementioned type, and that Lagrangian is

$$L := g^{1/2}\tau\,\delta^{abcd}_{pqrs}\,K_{\alpha\beta}\,F^{\alpha}{}_{ab}F^{\beta pq}R_{cd}{}^{rs} \tag{149}$$

where $\tau$ is a constant and $K_{\alpha\beta}$ denotes the components of an Ad invariant symmetric bilinear form on LG. Note that the Lagrangian given in (149) is the obvious generalization to gauge field theory of the new term appearing in the Lagrangian given in (10).

In conclusion I would like to point out that if you do not mind $F^{\alpha}{}_{ab}$ appearing in the gauge-current that acts as the source for the gauge fields in the field equation obtained by varying $A^{\alpha}{}_{i}$ then you could add any one (or more) of the following Lagrangians to the Yang-Mills Lagrangian given in (147) and obtain such a theory

$L_{YM\varphi} := \frac{1}{4}\kappa g^{1/2}K_{\mu\nu}\varphi_a\varphi_b F^{\mu a}{}_{c}\,F^{\nu bc}$; $L_{YM\xi} := \frac{1}{4}\kappa g^{1/2}K_{\mu\nu}\xi_a\xi_b F^{\mu a}{}_{c}\,F^{\nu bc}$; $L_{YM\varphi\xi} := \frac{1}{4}\kappa g^{1/2}K_{\mu\nu}\varphi_a\xi_b F^{\mu a}{}_{c}\,F^{\nu bc}$;

$L_{\varphi} := g^{1/2}\tau\,\delta^{abcd}_{pqrs}\,K_{\alpha\beta}\varphi_a\varphi^p\,F^{\alpha}{}_{be}F^{\beta qe}R_{cd}{}^{rs}$; $L_{\xi} := g^{1/2}\tau\,\delta^{abcd}_{pqrs}\,K_{\alpha\beta}\xi_a\xi^p\,F^{\alpha}{}_{be}F^{\beta qe}R_{cd}{}^{rs}$; and

$L_{\varphi\xi} := g^{1/2}\tau\,\delta^{abcd}_{pqrs}\,K_{\alpha\beta}\varphi_a\xi^p\,F^{\alpha}{}_{be}F^{\beta qe}R_{cd}{}^{rs}$.

Each of the above Lagrangians will generate second-order bi-scalar-gauge-tensor field tensors that are gauge invariant. Evidently the Lagrangians $L_{YM\varphi}$, $L_{YM\xi}$ $L_{\varphi}$ and $L_{\xi}$ could be used in a scalar-gauge-tensor field theory.

**Acknowledgments**

I would like to thank Drs. Stephen J. Aldersley and James S. Barber, along with University of Waterloo Professor Emeritus J.Wainwright for discussions on the Higgs field and other topics addressed in this paper.

**Appendix: Determining the Coefficient Functions of $\text{Đ}_S{}^{i}$**

The purpose of this appendix is to determine the conditions which the coefficient functions appearing in (131) must satisfy in order for $\text{Đ}_S{}^{i}{}_{|i}=0$. Using (131) a straightforward calculation shows

$$\text{Đ}_S{}^{i}{}_{|i} =$$

$$g^{1/2}\{\alpha_{1\varphi}\varphi_a\varphi^p \delta^{aeh}_{pqr} R_{eh}{}^{qr} + \alpha_{1\xi}\varphi_a\xi^p \delta^{aeh}_{pqr} R_{eh}{}^{qr} + 2\alpha_{1X}\varphi_a\varphi_b\varphi^{bp} \delta^{aeh}_{pqr} R_{eh}{}^{qr} + 2\alpha_{1Y}\varphi_a\xi_b\xi^{bp} \delta^{aeh}_{pqr} R_{eh}{}^{qr} +$$

$$+ \alpha_{1Z}\varphi_a\varphi_b\xi^{bp} \delta^{aeh}_{pqr} R_{eh}{}^{qr} + \alpha_{1Z}\varphi_a\xi_b\varphi^{bp} \delta^{aeh}_{pqr} R_{eh}{}^{qr} + \alpha_1\varphi^p{}_a \delta^{aeh}_{pqr} R_{eh}{}^{qr} + \alpha_{2\varphi}\xi_a\varphi^p \delta^{aeh}_{pqr} R_{eh}{}^{qr} + \alpha_{2\xi}\xi_a\xi^p \delta^{aeh}_{pqr} R_{eh}{}^{qr}$$

$$+ 2\alpha_{2X}\xi_a\varphi_b\varphi^{bp} \delta^{aeh}_{pqr} R_{eh}{}^{qr} + 2\alpha_{2Y}\xi_a\xi_b\xi^{bp} \delta^{aeh}_{pqr} R_{eh}{}^{qr} + \alpha_{2Z}\xi_a\varphi_b\xi^{bp} \delta^{aeh}_{pqr} R_{eh}{}^{qr} + \alpha_{2Z}\xi_a\xi_b\varphi^{bp} \delta^{aeh}_{pqr} R_{eh}{}^{qr} +$$

$$+ \alpha_2\xi^p{}_a \delta^{aeh}_{pqr} R_{eh}{}^{qr} + \alpha_{3\xi}\varphi_a\xi_b\varphi^c\xi^p \delta^{abeh}_{cpqr} R_{eh}{}^{qr} + 2\alpha_{3X}\varphi_a\xi_b\varphi^c\varphi_d\varphi^{dp} \delta^{abeh}_{cpqr} R_{eh}{}^{qr} + 2\alpha_{3Y}\varphi_a\xi_b\varphi^c\xi_d\xi^{dp} \delta^{abeh}_{cpqr} R_{eh}{}^{qr}$$

$$+\alpha_{3Z}\varphi_a\xi_b\varphi^c\varphi_d\xi^{dp} \delta^{abeh}_{cpqr} R_{eh}{}^{qr} + \alpha_{3Z}\varphi_a\xi_b\varphi^c\xi_d\varphi^{dp} \delta^{abeh}_{cpqr} R_{eh}{}^{qr} + \alpha_3\varphi_a{}^p\xi_b\varphi^c \delta^{abeh}_{cpqr} R_{eh}{}^{qr} + \alpha_3\varphi_a\xi_b{}^p\varphi^c \delta^{abeh}_{cpqr} R_{eh}{}^{qr} +$$

$$+\alpha_{4\varphi}\xi_a\varphi_b\xi^c\varphi^p \delta^{abeh}_{cpqr} R_{eh}{}^{qr} + 2\alpha_{4X}\xi_a\varphi_b\xi^c\varphi_d\varphi^{dp} \delta^{abeh}_{cpqr} R_{eh}{}^{qr} + 2\alpha_{4Y}\xi_a\varphi_b\xi^c\xi_d\xi^{dp} \delta^{abeh}_{cpqr} R_{eh}{}^{qr} +$$

$$+\alpha_{4Z}\xi_a\varphi_b\xi^c\varphi_d\xi^{dp} \delta^{abeh}_{cpqr} R_{eh}{}^{qr} + \alpha_{4Z}\xi_a\varphi_b\xi^c\xi_d\varphi^{dp} \delta^{abeh}_{cpqr} R_{eh}{}^{qr} + \alpha_4\xi_a{}^p\varphi_b\xi^c \delta^{abeh}_{cpqr} R_{eh}{}^{qr} + \alpha_4\xi_a\varphi_b{}^p\xi^c \delta^{abeh}_{cpqr} R_{eh}{}^{qr} +$$

$$+\alpha_{5\xi}\omega^{eh}\varphi_b\xi_i\varepsilon^{bifj}R_{ehfj} + 2\alpha_{5X}\omega^{eh}\varphi_b\varphi^a\varphi_{ai}\varepsilon^{bifj}R_{ehfj} + 2\alpha_{5Y}\omega^{eh}\varphi_b\xi^a\xi_{ai}\varepsilon^{bifj}R_{ehfj} + \alpha_{5Z}\omega^{eh}\varphi_b\varphi^a\xi_{ai}\varepsilon^{bifj}R_{ehfj} +$$

$$+\alpha_{5Z}\omega^{eh}\varphi_b\xi^a\varphi_{ai}\varepsilon^{bifj}R_{ehfj} + \alpha_5\varphi^e{}_i\xi^h\varphi_b\varepsilon^{bifj}R_{ehfj} + \alpha_5\varphi^e\xi^h{}_i\varphi_b\varepsilon^{bifj}R_{ehfj} + \alpha_{6\varphi}\omega^{eh}\xi_b\varphi_i\varepsilon^{bifj}R_{ehfj} + 2\alpha_{6X}\omega^{eh}\xi_b\varphi^a\varphi_{ai}\varepsilon^{bifj}R_{ehfj}$$

$$+2\alpha_{6Y}\omega^{eh}\xi_b\xi^a\xi_{ai}\varepsilon^{bifj}R_{ehfj}+\alpha_{6Z}\omega^{eh}\xi_b\varphi^a\xi_{ai}\varepsilon^{bifj}R_{ehfj}+\alpha_{6Z}\omega^{eh}\xi_b\xi^a\varphi_{ai}\varepsilon^{bifj}R_{ehfj}+\alpha_6\varphi^{ei}\xi^h\xi_b\varepsilon^{bifj}R_{ehfj}+\alpha_6\varphi^e\xi^h{}_i\xi_b\varepsilon^{bifj}R_{ehfj}\}$$

$$+g^{1/2}\{\beta_{1\varphi}\varphi^a\varphi^i\varphi_{ia}+\beta_{1\xi}\varphi^a\xi^i\varphi_{ia}+2\beta_{1X}\varphi^a\varphi^b\varphi_{bi}\varphi^i{}_a+2\beta_{1Y}\varphi^a\xi^b\xi_{bi}\varphi^i{}_a+\beta_{1Z}\varphi^a\varphi^b\xi_{bi}\varphi^i{}_a+\beta_{1Z}\varphi^a\xi^b\varphi_{bi}\varphi^i{}_a+\beta_1\varphi^{ia}\varphi_{ia}+$$

$$-\beta_{1\varphi}X\square\varphi-\beta_{1\xi}Z\square\varphi-2\beta_{1X}\varphi^b\varphi^i\varphi_{bi}\square\varphi-2\beta_{1Y}\xi^b\varphi^i\xi_{bi}\square\varphi-\beta_{1Z}\varphi^b\varphi^i\xi_{bi}\square\varphi-\beta_{1Z}\xi^b\varphi^i\varphi_{bi}\square\varphi-\beta_1(\square\varphi)^2+$$

$$+\beta_1\varphi^a\varphi^bR_{ab}+\beta_{2\varphi}\xi^a\varphi^i\varphi_{ia}+\beta_{2\xi}\xi^a\xi^i\varphi_{ia}+2\beta_{2X}\xi^a\varphi^b\varphi_{bi}\varphi^i{}_a+2\beta_{2Y}\xi^a\xi^b\xi_{bi}\varphi^i{}_a+\beta_{2Z}\xi^a\varphi^b\xi_{bi}\varphi^i{}_a+\beta_{2Z}\xi^a\xi^b\varphi_{bi}\varphi^i{}_a+\beta_2\xi^{ia}\varphi_{ia}$$

$$-\beta_{2\varphi}Z\square\varphi-\beta_{2\xi}Y\square\varphi-2\beta_{2X}\xi^b\varphi^i\varphi_{bi}\square\varphi-2\beta_{2Y}\xi^b\xi^i\xi_{bi}\square\varphi-\beta_{2Z}\xi^b\varphi^i\xi_{bi}\square\varphi-\beta_{2Z}\xi^b\xi^i\varphi_{bi}\square\varphi-\beta_2\square\xi\square\varphi+$$

$$+\beta_2\varphi^a\xi^bR_{ab}+\beta_{3\varphi}X\varphi^a\xi^b\varphi_{ab}+\beta_{3\xi}Z\varphi^a\xi^b\varphi_{ab}+2\beta_{3X}\varphi^i\varphi^c\varphi_{ic}\varphi^a\xi^b\varphi_{ab}+2\beta_{3Y}\varphi^i\xi^c\xi_{ic}\varphi^a\xi^b\varphi_{ab}+\beta_{3Z}\varphi^i\varphi^c\xi_{ic}\varphi^a\xi^b\varphi_{ab}+$$

$$+\beta_{3Z}\varphi^i\xi^c\varphi_{ic}\varphi^a\xi^b\varphi_{ab}+\beta_3\varphi^a\xi^b\varphi_{ab}\square\varphi+\beta_3\varphi^i\varphi^a{}_i\xi^b\varphi_{ab}+\beta_3\varphi^i\varphi^a\xi^b{}_i\varphi_{ab}-\beta_{3\varphi}Z\varphi^a\varphi^b\varphi_{ab}-\beta_{3\xi}Y\varphi^a\varphi^b\varphi_{ab}-2\beta_{3X}\xi^i\varphi^c\varphi_{ci}\varphi^a\varphi^b\varphi_{ab}$$

$$-2\beta_{3Y}\xi^i\xi^c\xi_{ci}\varphi^a\varphi^b\varphi_{ab}-\beta_{3Z}\xi^i\varphi^c\xi_{ci}\varphi^a\varphi^b\varphi_{ab}-\beta_{3Z}\xi^i\xi^c\varphi_{ci}\varphi^a\varphi^b\varphi_{ab}-\beta_3\varphi^a\varphi^b\varphi_{ab}\square\xi-2\beta_3\xi^i\varphi^a{}_i\varphi^b\varphi_{ab}+\beta_{4\varphi}Z\xi^a\varphi^b\varphi_{ab}$$

$$+\beta_{4\xi}Y\xi^a\varphi^b\varphi_{ab}+2\beta_{4X}\xi^i\varphi^c\varphi_{ic}\xi^a\varphi^b\varphi_{ab}+2\beta_{4Y}\xi^i\xi^c\xi_{ic}\xi^a\varphi^b\varphi_{ab}+\beta_{4Z}\xi^i\varphi^c\xi_{ic}\xi^a\varphi^b\varphi_{ab}+\beta_{4Z}\xi^i\xi^c\varphi_{ic}\xi^a\varphi^b\varphi_{ab}+\beta_4\xi^a\varphi^b\varphi_{ab}\square\xi+$$

$$+\beta_4\xi^i\xi^a{}_i\varphi^b\varphi_{ab}+\beta_4\xi^i\xi^a\varphi^b{}_i\varphi_{ab}-\beta_{4\varphi}X\xi^a\xi^b\varphi_{ab}-\beta_{4\xi}Z\xi^a\xi^b\varphi_{ab}-2\beta_{4X}\varphi^i\varphi^c\varphi_{ic}\xi^a\xi^b\varphi_{ab}+$$

$$-2\beta_{4Y}\varphi^i\xi^c\xi_{ic}\xi^a\xi^b\varphi_{ab}-\beta_{4Z}\varphi^i\varphi^c\xi_{ic}\xi^a\xi^b\varphi_{ab}-\beta_{4Z}\varphi^i\xi^c\varphi_{ic}\xi^a\xi^b\varphi_{ab}-\beta_4\xi^a\xi^b\varphi_{ab}\square\varphi-2\beta_4\varphi^i\xi^a\xi^b{}_i\varphi_{ab}+\beta_4\varphi^a\xi^b\varphi^c\xi^dR_{abcd}\}+$$

$$-\tfrac{1}{2}\{2\beta_{5Y}\varepsilon^{iapq}\xi^c\xi_{ci}\varphi_p\xi_q\varphi^b\varphi_{ab}+\beta_{5Z}\varepsilon^{iapq}\varphi^c\xi_{ci}\varphi_p\xi_q\varphi^b\varphi_{ab}+\beta_{5Z}\varepsilon^{iapq}\xi^c\varphi_{ci}\varphi_p\xi_q\varphi^b\varphi_{ab}\}-\tfrac{1}{2}\{2\beta_{6X}\varepsilon^{iapq}\varphi^c\varphi_{ci}\varphi_p\xi_q\xi^b\varphi_{ab}+$$

$$+2\beta_{6Y}\varepsilon^{iapq}\xi^c\xi_{ci}\varphi_p\xi_q\xi^b\varphi_{ab}+\beta_{6Z}\varepsilon^{iapq}\varphi^c\xi_{ci}\varphi_p\xi_q\xi^b\varphi_{ab}+\beta_6\varepsilon^{iapq}\varphi_p\xi_q\xi^b{}_i\varphi_{ab}-\tfrac{1}{2}\beta_6\varepsilon^{iapq}\varphi_p\xi_q\varphi^r\xi^bR_{rbia}\}+$$

$$+g^{1/2}\{\gamma_{1\varphi}\varphi^a\varphi^i\xi_{ia}+\gamma_{1\xi}\varphi^a\xi^i\xi_{ia}+2\gamma_{1X}\varphi^a\varphi^b\varphi_{bi}\xi^i{}_a+2\gamma_{1Y}\varphi^a\xi^b\xi_{bi}\xi^i{}_a+\gamma_{1Z}\varphi^a\varphi^b\xi_{bi}\xi^i{}_a+\gamma_{1Z}\varphi^a\xi^b\varphi_{bi}\xi^i{}_a+\gamma_1\varphi^{ia}\xi_{ia}-\gamma_{1\varphi}X\square\xi+$$

$$-\gamma_{1\xi}Z\square\xi-2\gamma_{1X}\varphi^b\varphi^i\varphi_{bi}\square\xi-2\gamma_{1Y}\xi^b\varphi^i\xi_{bi}\square\xi-\gamma_{1Z}\varphi^b\varphi^i\xi_{bi}\square\xi-\gamma_{1Z}\xi^b\varphi^i\varphi_{bi}\square\xi-\gamma_1\square\varphi\square\xi+\gamma_1\varphi^a\xi^bR_{ab}+\gamma_{2\varphi}\xi^a\varphi^i\xi_{ia}$$

$$+\gamma_{2\xi}\xi^a\xi^i\xi_{ia}+2\gamma_{2X}\xi^a\varphi^b\varphi_{bi}\xi^i{}_a+2\gamma_{2Y}\xi^a\xi^b\xi_{bi}\xi^i{}_a+\gamma_{2Z}\xi^a\varphi^b\xi_{bi}\xi^i{}_a+\gamma_{2Z}\xi^a\xi^b\varphi_{bi}\xi^i{}_a+\gamma_2\xi^{ia}\xi_{ia}-\gamma_{2\varphi}Z\square\xi-\gamma_{2\xi}Y\square\xi+$$

$$-2\gamma_{2X}\xi^b\varphi^i\varphi_{bi}\square\xi-2\gamma_{2Y}\xi^b\xi^i\xi_{bi}\square\xi-\gamma_{2Z}\xi^b\varphi^i\xi_{bi}\square\xi-\gamma_Z\xi^b\xi^i\varphi_{bi}\square\xi-\gamma_2(\square\xi)^2+\gamma_2\xi^a\xi^bR_{ab}+\gamma_{3\varphi}X\varphi^a\xi^b\xi_{ab}+$$

$$+\gamma_{3\xi}Z\varphi^a\xi^b\xi_{ab}+2\gamma_{3X}\varphi^i\varphi^c\varphi_{ic}\varphi^a\xi^b\xi_{ab}+2\gamma_{3Y}\varphi^i\xi^c\xi_{ic}\varphi^a\xi^b\xi_{ab}+\gamma_{3Z}\varphi^i\varphi^c\xi_{ic}\varphi^a\xi^b\xi_{ab}+\gamma_{3Z}\varphi^i\xi^c\varphi_{ic}\varphi^a\xi^b\xi_{ab}+\gamma_3\varphi^a\xi^b\xi_{ab}\square\varphi$$

$$+\gamma_3\varphi^i\varphi^a{}_i\xi^b\xi_{ab}+\gamma_3\varphi^i\varphi^a\xi^b{}_i\xi_{ab}-\gamma_{3\varphi}Z\varphi^a\varphi^b\xi_{ab}-\gamma_{3\xi}Y\varphi^a\varphi^b\xi_{ab}-2\gamma_{3X}\xi^i\varphi^c\varphi_{ci}\varphi^a\varphi^b\xi_{ab}-2\gamma_{3Y}\xi^i\xi^c\xi_{ci}\varphi^a\varphi^b\xi_{ab}+$$

$$-\gamma_{3Z}\xi^i\varphi^c\xi_{ci}\varphi^a\varphi^b\xi_{ab}-\gamma_{3Z}\xi^i\xi^c\varphi_{ci}\varphi^a\varphi^b\xi_{ab}-\gamma_3\varphi^a\varphi^b\xi_{ab}\square\xi-2\gamma_3\xi^i\varphi^a{}_i\varphi^b\xi_{ab}+\gamma_3\varphi^a\xi^b\varphi^c\xi^dR_{abcd}+\gamma_{4\varphi}Z\xi^a\varphi^b\xi_{ab}+$$

$$+\gamma_{4\xi}Y\xi^a\varphi^b\xi_{ab}+2\gamma_{4X}\xi^i\varphi^c\varphi_{ic}\xi^a\varphi^b\xi_{ab}+2\gamma_{4Y}\xi^i\xi^c\xi_{ic}\xi^a\varphi^b\xi_{ab}+\gamma_{4Z}\xi^i\varphi^c\xi_{ic}\xi^a\varphi^b\xi_{ab}+\gamma_{4Z}\xi^i\xi^c\varphi_{ic}\xi^a\varphi^b\xi_{ab}+$$

$$+\gamma_4\xi^a\varphi^b\xi_{ab}\square\xi+\gamma_4\xi^i\xi^a{}_i\varphi^b\xi_{ab}+\gamma_4\xi^i\xi^a\varphi^b{}_i\xi_{ab}-\gamma_{4\varphi}X\xi^a\xi^b\xi_{ab}-\gamma_{4\xi}Z\xi^a\xi^b\xi_{ab}-2\gamma_{4X}\varphi^i\varphi^c\varphi_{ic}\xi^a\xi^b\xi_{ab}+$$

$$- 2\gamma_{4Y}\varphi^{i}\xi^{c}\xi_{ic}\xi^{a}\xi^{b}\xi_{ab} - \gamma_{4Z}\varphi^{i}\varphi^{c}\xi_{ic}\xi^{a}\xi^{b}\xi_{ab} - \gamma_{4Z}\varphi^{i}\xi^{c}\varphi_{ic}\xi^{a}\xi^{b}\xi_{ab} - \gamma_{4}\xi^{a}\xi^{b}\xi_{ab}\Box\varphi - 2\gamma_{4}\varphi^{i}\xi^{a}\xi^{b}{}_{i}\xi_{ab}\}$$

$$- \tfrac{1}{2}\{2\gamma_{5X}\varepsilon^{iapq}\varphi^{c}\varphi_{ci}\varphi_{p}\xi_{q}\varphi^{b}\xi_{ab} + 2\gamma_{5Y}\varepsilon^{iapq}\xi^{c}\xi_{ci}\varphi_{p}\xi_{q}\varphi^{b}\xi_{ab} + \gamma_{5Z}\varepsilon^{iapq}\xi^{c}\varphi_{ci}\varphi_{p}\xi_{q}\varphi^{b}\xi_{ab} + \gamma_{5}\varepsilon^{iapq}\varphi_{p}\xi_{q}\varphi^{b}{}_{i}\xi_{ab} +$$

$$\tfrac{1}{2}\gamma_{5}\varepsilon^{iapq}\varphi_{p}\xi_{q}\varphi^{r}\xi^{b}R_{rbia} + 2\gamma_{6X}\varepsilon^{iapq}\varphi^{c}\varphi_{ci}\varphi_{p}\xi_{q}\xi^{b}\xi_{ab} + \gamma_{6Z}\varepsilon^{iapq}\varphi^{c}\xi_{ci}\varphi_{p}\xi_{q}\xi^{b}\xi_{ab} + \gamma_{6Z}\varepsilon^{iapq}\xi^{c}\varphi_{ci}\varphi_{p}\xi_{q}\xi^{b}\xi_{ab}\} + g^{\frac{1}{2}}\{\delta_{1\varphi}X +$$

$$\delta_{1\xi}Z + 2\delta_{1X}\varphi^{i}\varphi^{j}\varphi_{ij} + 2\delta_{1Y}\varphi^{i}\xi^{j}\xi_{ij} + \delta_{1Z}\varphi^{i}\varphi^{j}\xi_{ij} + \delta_{1Z}\varphi^{i}\xi^{j}\varphi_{ij} + \delta_{2\varphi}Z + \delta_{2\xi}Y + 2\delta_{2X}\xi^{i}\varphi^{j}\varphi_{ij} + 2\delta_{2Y}\xi^{i}\xi^{j}\xi_{ij} + \delta_{2Z}\xi^{i}\varphi^{j}\xi_{ij}$$

$$+ \delta_{2Z}\xi^{i}\xi^{j}\varphi_{ij} + \delta_{1}\Box\varphi + \delta_{2}\Box\xi\}, \tag{150}$$

where I have assumed that the metric is Lorentzian and so $*\omega^{ab} := g^{ap}g^{bq} *\omega_{pq} = -\tfrac{1}{2}g^{-\frac{1}{2}}\varepsilon^{abrs}\,\omega_{rs}$.

Before we begin our analysis of $\text{Đ}_{S}{}^{i}{}_{|i} = 0$, you should note is that the $\text{Đ}_{S}{}^{i}$ presented in (131) is invariant under the transformation

$$\varphi\leftrightarrow\xi,\ \alpha_1\leftrightarrow\alpha_2,\ \alpha_3\leftrightarrow\alpha_4,\ \alpha_5\leftrightarrow-\alpha_6,\ \beta_1\leftrightarrow\gamma_2,\ \beta_2\leftrightarrow\gamma_1,\ \beta_3\leftrightarrow\gamma_4,\ \beta_4\leftrightarrow\gamma_3,\ \beta_5\leftrightarrow-\gamma_6,\ \beta_6\leftrightarrow-\gamma_5,\ \delta_1\leftrightarrow\delta_2. \tag{151}$$

This will prove to be very helpful in our analysis of (150). For example once we determine the conditions implied by setting the terms involving $\varphi_{ab}R_{cdef}$ equal to zero we can obtain the conditions necessary for the terms involving $\xi_{ab}R_{cdef}$ to vanish using (151) along with the fact that this transformation requires $\partial_X\leftrightarrow\partial_Y$. Let's now get to work.

(A.1) is a polynomial in the set

$$\{\varphi_{ab}R_{cdef},\ \xi_{ab}R_{cdef},\ \varphi_{ab}\varphi_{cd},\ \varphi_{ab}\xi_{cd},\ \xi_{ab}\xi_{cd},\ \varphi_{ab},\ \xi_{ab},\ R_{abcd}\}$$

with coefficients involving $\varphi_a$, $\xi_a$ and $g_{ab}$. So we can differentiate (150) with respect to $\varphi_{ab}$, $\xi_{cd}$, and $g_{rs,tu}$ to obtain eight multi-index tensor equations relating the twenty coefficient functions. Let us now see what these equations tell us. First if we differentiate the terms involving $\varphi_{ab}R_{cdef}$ in (150) with respect $\varphi_{ij}$ we obtain

$$0 = \alpha_{1X}\,\delta^{aeh}_{pqr}\,\varphi_a\varphi^{i}g^{pj}R_{eh}{}^{qr} + \alpha_{1X}\,\delta^{aeh}_{pqr}\,\varphi_a\varphi^{j}g^{pi}R_{eh}{}^{qr} + \tfrac{1}{2}\alpha_{1Z}\,\delta^{aeh}_{pqr}\,\varphi_a\xi^{i}g^{pj}R_{eh}{}^{qr} + \tfrac{1}{2}\alpha_{1Z}\,\delta^{aeh}_{pqr}\,\varphi_a\xi^{j}g^{pi}R_{eh}{}^{qr} +$$

$$+ \tfrac{1}{2}\alpha_{1}\,\delta^{ieh}_{pqr}\,g^{pj}R_{eh}{}^{qr} + \tfrac{1}{2}\alpha_{1}\,\delta^{jeh}_{pqr}\,g^{pi}R_{eh}{}^{qr} + \alpha_{2X}\,\delta^{aeh}_{pqr}\,\xi_a\varphi^{i}g^{pj}R_{eh}{}^{qr} + \alpha_{2X}\,\delta^{aeh}_{pqr}\,\xi_a\varphi^{j}g^{pi}R_{eh}{}^{qr} +$$

$$+\tfrac{1}{2}\alpha_{2Z}\,\delta^{aeh}_{pqr}\,\xi_a\xi^i g^{pj}R_{eh}{}^{qr}+\tfrac{1}{2}\alpha_{2Z}\,\delta^{aeh}_{pqr}\,\xi_a\xi^j g^{pi}R_{eh}{}^{qr}+\alpha_{3X}\,\delta^{abeh}_{cpqr}\,\varphi_a\xi_b\varphi^c\varphi^i g^{pj}R_{eh}{}^{qr}+\alpha_{3X}\,\delta^{abeh}_{cpqr}\,\varphi_a\xi_b\varphi^c\varphi^j g^{pi}R_{eh}{}^{qr}+$$

$$+\tfrac{1}{2}\alpha_{3Z}\,\delta^{abeh}_{cpqr}\,\varphi_a\xi_b\varphi^c\xi^i g^{pj}R_{eh}{}^{qr}+\tfrac{1}{2}\alpha_{3Z}\,\delta^{abeh}_{cpqr}\,\varphi_a\xi_b\varphi^c\xi^j g^{pi}R_{eh}{}^{qr}+\tfrac{1}{2}\alpha_{3}\,\delta^{ibeh}_{cpqr}\,g^{pj}\xi_b\varphi^c R_{eh}{}^{qr}+\tfrac{1}{2}\alpha_{3}\,\delta^{jbeh}_{cpqr}\,g^{pi}\xi_b\varphi^c R_{eh}{}^{qr}$$

$$+\alpha_{4X}\,\delta^{abeh}_{cpqr}\,\xi_a\varphi_b\xi^c\varphi^i g^{pj}R_{eh}{}^{qr}+\alpha_{4X}\,\delta^{abeh}_{cpqr}\,\xi_a\varphi_b\xi^c\varphi^j g^{pi}R_{eh}{}^{qr}+\tfrac{1}{2}\alpha_{4Z}\,\delta^{abeh}_{cpqr}\,\xi_a\varphi_b\xi^c\xi^i g^{pj}R_{eh}{}^{qr}+$$

$$+\tfrac{1}{2}\alpha_{4Z}\,\delta^{abeh}_{cpqr}\,\xi_a\varphi_b\xi^c\xi^j g^{pi}R_{eh}{}^{qr}+\tfrac{1}{2}\alpha_{4}\,\delta^{aieh}_{cpqr}\,g^{pj}\xi_a\xi^c R_{eh}{}^{qr}+\tfrac{1}{2}\alpha_{4}\,\delta^{ajeh}_{cpqr}\,g^{pi}\xi_a\xi^c R_{eh}{}^{qr}+\alpha_{5X}\varphi^i\varphi^e\xi^h\varphi_b\varepsilon^{bjfm}R_{ehfm}+$$

$$+\alpha_{5X}\varphi^j\varphi^e\xi^h\varphi_b\varepsilon^{bifm}R_{ehfm}+\tfrac{1}{2}\alpha_{5Z}\xi^i\varphi^e\xi^h\varphi_b\varepsilon^{bjfm}R_{ehfm}+\tfrac{1}{2}\alpha_{5Z}\xi^j\varphi^e\xi^h\varphi_b\varepsilon^{bifm}R_{ehfm}+\tfrac{1}{2}\alpha_{5}g^{ei}\xi^h\varphi_b\varepsilon^{bjfm}R_{ehfm}+$$

$$+\tfrac{1}{2}\alpha_{5}g^{ej}\xi^h\varphi_b\varepsilon^{bifm}R_{ehfm}+\alpha_{6X}\varphi^i\varphi^e\xi^h\xi_b\varepsilon^{bjfm}R_{ehfm}+\alpha_{6X}\varphi^j\varphi^e\xi^h\xi_b\varepsilon^{bifm}R_{ehfm}+\tfrac{1}{2}\alpha_{6Z}\xi^i\varphi^e\xi^h\xi_b\varepsilon^{bjfm}R_{ehfm}+$$

$$+\tfrac{1}{2}\alpha_{6Z}\xi^j\varphi^e\xi^h\xi_b\varepsilon^{bifm}R_{ehfm}+\tfrac{1}{2}\alpha_{6}g^{ei}\xi^h\xi_b\varepsilon^{bjfm}R_{ehfm}+\tfrac{1}{2}\alpha_{6}g^{ej}\xi^h\xi_b\varepsilon^{bifm}R_{ehfm}\,. \qquad (152)$$

At this juncture one could differentiate (152) with respect to $g_{rs,tu}$ to obtain a 6 index tensor concomitant to analyze to determine relations between the various α's. But I would prefer to avoid doing that and rather look at (152) for various possible choices of geometries. Doing that will not affect the generality of our analysis since we shall not be restricting the values of φ or ξ, and hence the values of X, Y and Z will also remain quite arbitrary.

If we now expand the generalized Kronecker deltas appearing in (152) we find that

$$-4\alpha_{1X}\varphi^i\varphi_aG^{aj}-4\alpha_{1X}\varphi^j\varphi_aG^{ai}-2\alpha_{1z}\xi^i\varphi_aG^{aj}-2\alpha_{1z}\xi^j\varphi_aG^{ai}-4\alpha_1G^{ij}-4\alpha_{2X}\varphi^i\xi_aG^{aj}-4\alpha_{2X}\varphi^j\xi_aG^{ai}-2\alpha_{2Z}\xi^i\xi_aG^{aj}+$$

$$-2\alpha_{2Z}\xi^j\xi_aG^{ai}-4X\alpha_{3X}\varphi^i\xi_aG^{aj}-4X\alpha_{3X}\varphi^j\xi_aG^{ai}+4Z\alpha_{3X}\varphi^i\varphi_aG^{aj}+4Z\alpha_{3X}\varphi^j\varphi_aG^{ai}+8\alpha_{3X}\varphi^i\varphi^j\varphi_a\xi_bR^{ab}+$$

$$-4\alpha_{3X}(\varphi^i\xi^j+\varphi^j\xi^i)\varphi_a\varphi_bR^{ab}+4\alpha_{3X}\varphi_a\xi_b\varphi_c\varphi^iR^{abcj}+4\alpha_{3X}\varphi_a\xi_b\varphi_c\varphi^jR^{abci}-2X\alpha_{3Z}\xi^i\xi_aG^{aj}-2X\alpha_{3Z}\xi^j\xi_aG^{ai}+$$

$$+2Z\alpha_{3Z}\xi^i\varphi_aG^{aj}+2Z\alpha_{3Z}\xi^j\varphi_aG^{ai}-4Z\alpha_{3Z}\xi^i\xi^j\varphi_a\varphi_bR^{ab}+2\alpha_{3Z}(\varphi^i\xi^j+\varphi^j\xi^i)\varphi_a\xi_bR^{ab}+2\alpha_{3Z}\varphi_a\xi_b\varphi_c\xi^iR^{abcj}\;+$$

$$+2\alpha_{3Z}\varphi_a\xi_b\varphi_c\xi^jR^{abci}-2\alpha_3\varphi^i\xi_aG^{aj}-2\alpha_3\varphi^j\xi_aG^{ai}+4Z\alpha_3G^{ij}+4\alpha_3g^{ij}\varphi_a\xi_bR^{ab}-2\alpha_3\xi^j\varphi_aR^{ai}-2\alpha_3\xi^i\varphi_aR^{aj}+$$

$$+2\alpha_3\xi_a\varphi_bR^{aijb}+2\alpha_3\xi_a\varphi_bR^{ajib}-4Y\alpha_{4X}(\varphi^i\varphi_aG^{aj}+\varphi^j\varphi_aG^{ai})+4Z\alpha_{4X}(\varphi^i\xi_aG^{aj}+\varphi^j\xi_aG^{ai})+4\alpha_{4X}(\varphi^i\xi^j+\varphi^j\xi^i)\varphi_a\xi_bR^{ab}$$

$$-8\alpha_{4X}\varphi^i\varphi^j\xi_a\xi_bR^{ab}+4\alpha_{4X}\xi_a\varphi_b\xi_c\varphi^iR^{abcj}+4\alpha_{4X}\xi_a\varphi_b\xi_c\varphi^jR^{abci}-2Y\alpha_{4Z}(\xi^i\varphi_aG^{aj}+\xi^j\varphi_aG^{ai})+2Z\alpha_{4Z}(\xi^i\xi_aG^{aj}+\xi^j\xi_aG^{ai})$$

$$+4\alpha_{4Z}\xi^i\xi^j\varphi_a\xi_bR^{ab}-2\alpha_{4Z}(\varphi^i\xi^j+\varphi^j\xi^i)\xi_a\xi_bR^{ab}+2\alpha_{4Z}\xi_a\varphi_b\xi_c\xi^iR^{abcj}+2\alpha_{4Z}\xi_a\varphi_b\xi_c\,\xi^jR^{abci}-4Y\alpha_4G^{ij}+$$

$$+2\alpha_4(\xi^i\xi_a G^{aj} + \xi^j\xi_a G^{ai}) + 2\alpha_4(\xi^i\xi_a R^{aj} + \xi^j\xi_a R^{ai}) - 4\alpha_4 g^{ij}\xi_a\xi_b R^{ab} - 4\alpha_4\xi_a\xi_b R^{aijb} + \alpha_{5X}\varphi^i\varphi^e\xi^h\varphi_b\varepsilon^{bjfm}R_{ehfm} +$$

$$+\alpha_{5X}\varphi^j\varphi^e\xi^h\varphi_b\varepsilon^{bifm}R_{ehfm} + \tfrac{1}{2}\alpha_{5Z}\xi^i\varphi^e\xi^h\varphi_b\varepsilon^{bjfm}R_{ehfm} + \tfrac{1}{2}\alpha_{5Z}\xi^j\varphi^e\xi^h\varphi_b\varepsilon^{bifm}R_{ehfm} + \tfrac{1}{2}\alpha_5\xi^h\varphi_b\varepsilon^{bjfm}R^i{}_{hfm} +$$

$$+ \tfrac{1}{2}\alpha_5\xi^h\varphi_b\varepsilon^{bifm}R^j{}_{hfm} + \alpha_{6X}\varphi^i\varphi^e\xi^h\xi_b\varepsilon^{bjfm}R_{ehfm} + \alpha_{6X}\varphi^j\varphi^e\xi^h\xi_b\varepsilon^{bifm}R_{ehfm} + \tfrac{1}{2}\alpha_{6Z}\xi^i\varphi^e\xi^h\xi_b\varepsilon^{bjfm}R_{ehfm} +$$

$$+ \tfrac{1}{2}\alpha_{6Z}\xi^j\varphi^e\xi^h\xi_b\varepsilon^{bifm}R_{ehfm} + \tfrac{1}{2}\alpha_6\xi^h\xi_b\varepsilon^{bjfm}R^i{}_{hfm} + \tfrac{1}{2}\alpha_6\xi^h\xi_b\varepsilon^{bifm}R^j{}_{hfm} = 0\,. \qquad (153)$$

Let us evaluate (153) at an arbitrary point P of a 4-dimensional manifold. Since (152) must hold for all possible geometries let us consider geometries for which $R_{ij}=0$ at P. For this choice of geometry we see that at P, (153) becomes

$$0 = 4\alpha_{3X}\varphi^a\xi^b\varphi^c\varphi^i R_{abc}{}^j + 4\alpha_{3X}\varphi^a\xi^b\varphi^c\varphi^j R_{abc}{}^i + 2\alpha_{3Z}\varphi^a\xi^b\varphi^c\xi^i R_{abc}{}^j + 2\alpha_{3Z}\varphi^a\xi^b\varphi^c\xi^j R_{abc}{}^i + 2\alpha_3\xi^a\varphi^b R_a{}^{ij}{}_b +$$

$$+ 2\alpha_3\xi^a\varphi^b R_a{}^{ji}{}_b + 4\alpha_{4X}\xi^a\varphi^b\xi^c\varphi^i R_{abc}{}^j + 4\alpha_{4X}\xi^a\varphi^b\xi^c\varphi^j R_{abc}{}^i + 2\alpha_{4Z}\xi^a\varphi^b\xi^c\xi^i R_{abcj} + 2\alpha_{4Z}\xi^a\varphi^b\xi^c\xi^j R_{abc}{}^i - 4\alpha_4\xi^a\xi^b R_a{}^{ij}{}_b$$

$$+ \alpha_{5X}\varphi^i\varphi^e\xi^h\varphi_b\varepsilon^{bjfm}R_{ehfm} + \alpha_{5X}\varphi^j\varphi^e\xi^h\varphi_b\varepsilon^{bifm}R_{ehfm} + \tfrac{1}{2}\alpha_{5Z}\xi^i\varphi^e\xi^h\varphi_b\varepsilon^{bjfm}R_{ehfm} + \tfrac{1}{2}\alpha_{5Z}\xi^j\varphi^e\xi^h\varphi_b\varepsilon^{bifm}R_{ehfm} +$$

$$+ \tfrac{1}{2}\alpha_5\xi^h\varphi_b\varepsilon^{bjfm}R^i{}_{hfm} + \tfrac{1}{2}\alpha_5\xi^h\varphi_b\varepsilon^{bifm}R^j{}_{hfm} + \alpha_{6X}\varphi^i\varphi^e\xi^h\xi_b\varepsilon^{bjfm}R_{ehfm} + \alpha_{6X}\varphi^j\varphi^e\xi^h\xi_b\varepsilon^{bifm}R_{ehfm} +$$

$$+ \tfrac{1}{2}\alpha_{6Z}\xi^i\varphi^e\xi^h\xi_b\varepsilon^{bjfm}R_{ehfm} + \tfrac{1}{2}\alpha_{6Z}\xi^j\varphi^e\xi^h\xi_b\varepsilon^{bifm}R_{ehfm} + \tfrac{1}{2}\alpha_6\xi^h\xi_b\varepsilon^{bjfm}R^i{}_{hfm} + \tfrac{1}{2}\alpha_6\xi^h\xi_b\varepsilon^{bifm}R^j{}_{hfm}\,. \qquad (154)$$

To assist in our analysis of (154) let us assume, without loss of generality, that the set of two vectors $\{\varphi^i, \xi^i\}$ is linearly independent. For $\alpha=1,\ldots,4$ I introduce a frame of four vectors $\{X_\alpha{}^i\} := \{\varphi^i, \xi^i, V^i, W^i\}$ where $V^i$ and $W^i$ are unit vectors perpendicular to each other and perpendicular to the plane spanned by $\{\varphi^i, \xi^i\}$. Set $\varepsilon_V := g_{ab}V^aV^b$ and $\varepsilon_W := g_{ab}W^aW^b$ where $\varepsilon_V$ and $\varepsilon_W$ are either $=1$ or $-1$. If we now contract (154) with $V_iV_j$ we discover that

$$0 = 4\alpha_3\xi^a\varphi^b V^iV^j R_{aijb} - 4\alpha_4\xi^a\xi^b V^iV^j R_{aijb} + \alpha_5\xi^h V^i\varphi_b V_j\varepsilon^{bjfm}R_{ihfm} + \alpha_6\xi^h V^i\xi_b V_j\varepsilon^{bjfm}R_{ihfm}\,. \qquad (155)$$

In terms of frame components, where, *e.g.,* $\xi^a\varphi^b V^iV^j R_{aijb} := R_{2331}$, (155) becomes

$$0 = 4\alpha_3 R_{2331} - 4\alpha_4 R_{2332} + \alpha_5\varphi_b V_j\varepsilon^{bjfm}R_{32fm} + \alpha_6\xi_b V_j\varepsilon^{bjfm}R_{32fm}\,. \qquad (156)$$

For the frame $\{X_\alpha\}$ the non-zero frame components of the metric tensor $g_{\alpha\beta}$ are $g_{11}=X$, $g_{12}=g_{21}=Z$, $g_{22}=Y$, $g_{33}=\varepsilon_V$ and $g_{44}=\varepsilon_W$. When $R_{\alpha\beta}=0$, the curvature tensor has only 10 independent components

which we take to be

$$R_{1213}, R_{1214}, R_{1223}, R_{1224}, R_{1313}, R_{1314}, R_{1324}, R_{1414}, R_{2323}, R_{2324} \tag{157}$$

with the dependent components taken as

$$R_{1212}, R_{1234}, R_{1323}, R_{1334}, R_{1424}, R_{1434}, R_{2334}, R_{2424}, R_{2434}, R_{3434} \quad . \tag{158}$$

To analyze (156) we shall need to know that in terms of independent components (*see*, Appendix C in Horndeski [1])

$$R_{1432} = R_{1324} - YZ^{-1}R_{1314} - XZ^{-1}R_{2324} \tag{159}$$

$$R_{1323} = -\tfrac{1}{2}D(XZ)^{-1}\varepsilon_V\varepsilon_W R_{1414} + \tfrac{1}{2}XZ^{-1}R_{2323} + \tfrac{1}{2}ZX^{-1}R_{1313} \tag{160}$$

where $D := XY - Z^2$. It is a straightforward matter to show that

$$\varphi_b V_j \varepsilon^{bjfm} R_{32fm} = 2Z\varepsilon_v\varepsilon^{1423}R_{1324} - 2Y\varepsilon_v\varepsilon^{1423}R_{1314} \tag{161}$$

$$\xi_b V_j \varepsilon^{bjfm} R_{32fm} = 2DZ^{-1}\varepsilon_v\varepsilon^{1324}R_{2324} + 2Y\varepsilon_v\varepsilon^{1423}R_{1324} - 2Y^2Z^{-1}\varepsilon_v\varepsilon^{1423}R_{1314} \quad . \tag{162}$$

Upon combining (156), (160), (161) and (162), noting that $R_{1313}$, $R_{1314}$, $R_{1324}$, $R_{1414}$ and $R_{2324}$ are independent components we discover that

$$\alpha_3 = \alpha_4 = \alpha_5 = \alpha_6 = 0. \tag{163}$$

Armed with this information we now return to (153), with $R_{ij}$ arbitrary. We evaluate this equation at the arbitrary point P and then contract it with $V_iV_j$ to conclude that

$$\alpha_1 = 0, \tag{164}$$

and hence due to the transformation (152) we obtain

$$\alpha_2 = 0. \tag{165}$$

Thus we have demonstrated that all of the α coefficients in the expression for $\text{Đ}_{Si}$ presented in (131) must vanish.

Although the α coefficients in (150) are gone, we still have terms involving the curvature tensor to consider. Under the assumption that $R_{ij} = 0$ at the point P where we are working, the requirement that these curvature terms in (150) vanish leads to the following equation

$$\beta_4\varphi^a\xi^b\varphi^c\xi^d R_{abcd} + \tfrac{1}{4}\beta_6\varepsilon^{iapq}\varphi_p\xi_q\varphi^r\xi^b R_{rbia} + \gamma_3\varphi^a\xi^b\varphi^c\xi^d R_{abcd} - \tfrac{1}{4}\gamma_5\varepsilon^{iapq}\varphi_p\xi_q\varphi^r\xi^b R_{rbia} = 0 . \qquad (166)$$

We easily find that at P

$$\varepsilon^{iapq}\varphi_p\xi_q\varphi^r\xi^b R_{rbia} = 2D\varepsilon^{1234}R_{1234}$$

and so (166) becomes

$$(\beta_4 + \gamma_3)R_{1212} + \tfrac{1}{2}D(\beta_6 - \gamma_5)\varepsilon^{1234}R_{1234} = 0. \qquad (167)$$

In **Appendix C** of [1] I point out that

$$R_{1212} = -DX^{-1}(\varepsilon_V R_{1313} + \varepsilon_W R_{1414}) . \qquad (168)$$

Using (168) along with the fact that $R_{1234} = R_{1432} + R_{1324}$, we see that (167) implies that

$$\beta_4 = -\gamma_3 \text{ and } \beta_6 = \gamma_5 . \qquad (169)$$

There is no further information that we can obtain from (169) through use of the transformation (151), since (169) is invariant under that transformation.

There are still terms involving curvature in (150) that we have not addressed and those are the terms involving the Ricci tensor. These terms must vanish and yield the following equation

$$\beta_1\varphi^a\varphi^b R_{ab} + (\beta_2 + \gamma_1)\varphi^a\xi^b R_{ab} + \gamma_2\xi^a\xi^b R_{ab} = 0. \qquad (170)$$

Since, in general the set of frame components $\{R_{11}, R_{12}. R_{22}\}$ is linearly independent, (170) tells us that

$$\beta_1 = 0,\ \beta_2 = -\gamma_1 \text{ and } \gamma_2 = 0 . \qquad (171)$$

(171) is invariant under the transformation (151) and so we can not get more information from it.

(163), (164), (165), (169) and (171) provide us with all the information that we can obtain through an examination of the coefficients of the terms involving, $\varphi_{ab}R_{cdef}$, $\xi_{ab}R_{cdef}$ and $R_{abcd}$. We shall now look at those terms involving $\varphi_{ab}\varphi_{cd}$. These terms must vanish and provide, after some simplification, the following equation

$$(2\beta_{2X}-\beta_3)\xi^a\varphi^b\varphi_{bc}\varphi^c{}_a + (\beta_{2Z}+\beta_4)\xi^a\xi^b\varphi_{bc}\varphi^c{}_a - (2\beta_{2X}-\beta_3)\xi^b\varphi^c\varphi_{bc}\Box\varphi - (\beta_{2Z}+\beta_4)\,\xi^b\xi^c\varphi_{bc}\Box\varphi +$$

$$+(\beta_{3Z}+2\beta_{4X})\varphi^c\xi^d\varphi_{cd}\varphi^a\xi^b\varphi_{ab} - (\beta_{3Z}+2\beta_{4X})\,\xi^c\xi^d\varphi_{cd}\varphi^a\varphi^b\varphi_{ab} + g^{-\frac{1}{2}}(\beta_{6X}-\tfrac{1}{2}\beta_{5Z})\varepsilon^{capq}\xi^d\varphi_{cd}\varphi_p\xi_q\varphi^b\varphi_{ab}$$

$$= 0. \tag{172}$$

We shall evaluate this equation at an arbitrary point P of our four-dimensional manifold and differentiate it with respect to $\varphi_{,hi}$ and $\varphi_{,jk}$. Doing so will give us a four index tensorial concomitant from which we shall be able to deduce that all of the coefficient functions in (172) must vanish. Now for the details.

Differentiating (172) with respect to $\varphi_{,hi}$ give us

$$(2\beta_{2X}-\beta_3)[\xi^a\varphi^h\varphi^i{}_a+\xi^a\varphi^i\varphi^h{}_a] + (2\beta_{2X}-\beta_3)[\xi^h\varphi^b\varphi_b{}^i + \xi^i\varphi^b\varphi_b{}^h] + 2(\beta_{2Z}+\beta_4)[\xi^a\xi^h\varphi^i{}_a + \xi^a\xi^i\varphi^h{}_a] +$$

$$-(2\beta_{2X}-\beta_3)\,[\xi^h\varphi^i+\xi^i\varphi^h]\Box\varphi - 2(2\beta_{2X}-\beta_3)\xi^b\varphi^c\varphi_{bc}g^{hi} - 2(\beta_{2Z}+\beta_4)\,\xi^h\xi^i\Box\varphi +$$

$$-2(\beta_{2Z}+\beta_4)\xi^b\xi^c\varphi_{bc}g^{hi} + 2(\beta_{3Z}+2\beta_{4X})[\varphi^h\xi^i+\varphi^i\xi^h]\varphi^a\xi^b\varphi_{ab} - 2(\beta_{3Z}+2\beta_{4X})\,\xi^h\xi^i\varphi^a\varphi^b\varphi_{ab} +$$

$$-2(\beta_{3Z}+2\beta_{4X})\,\xi^c\xi^d\varphi_{cd}\varphi^h\varphi^i + g^{-\frac{1}{2}}(\beta_{6X}-\tfrac{1}{2}\beta_{5Z})[\varepsilon^{hapq}\xi^i+\varepsilon^{iapq}\xi^h]\varphi_p\xi_q\varphi^b\varphi_{ab} +$$

$$+g^{-\frac{1}{2}}(\beta_{6X}-\tfrac{1}{2}\beta_{5Z})\,[\varepsilon^{chpq}\varphi^i+\varepsilon^{cipq}\varphi^h]\xi^d\varphi_{cd}\varphi_p\xi_q\,. \tag{173}$$

We now need to differentiate (173) with respect to $\varphi_{,jk}$. Doing so yields the following equation

$$(2\beta_{2X}-\beta_3)[\xi^j\varphi^h g^{ik} + \xi^k\varphi^h g^{ij} + \xi^j\varphi^i g^{hk} + \xi^k\varphi^i g^{hj} + \xi^h\varphi^j g^{ik} + \xi^h\varphi^k g^{ij} + \xi^i\varphi^j g^{hk} + \xi^i\varphi^k g^{hj} - 2\,\xi^h\varphi^i g^{jk} - 2\,\xi^i\varphi^h g^{jk} +$$

$$-2\,\xi^j\varphi^k g^{hi} - 2\,\xi^k\varphi^j g^{hi}] + 2(\beta_{2Z}+\beta_4)[\xi^j\xi^h g^{ik} + \xi^k\xi^h g^{ij} + \xi^j\xi^i g^{hk} + \xi^k\xi^i g^{hj} - 2\xi^h\xi^i g^{jk} - 2\xi^j\xi^k g^{hi}] +$$

$$+2(\beta_{3Z}+2\beta_{4X})[\varphi^h\xi^i\varphi^j\xi^k + \varphi^i\xi^h\varphi^j\xi^k + \varphi^h\xi^i\varphi^k\xi^j + \varphi^i\xi^h\varphi^k\xi^j - 2\varphi^h\varphi^i\xi^j\xi^k - 2\varphi^j\varphi^k\xi^h\xi^i] +$$

$$+g^{-\frac{1}{2}}(\beta_{6X}-\tfrac{1}{2}\beta_{5Z})[\varepsilon^{hjpq}\xi^i\varphi^k\varphi_p\xi_q + \varepsilon^{hkpq}\xi^i\varphi^j\varphi_p\xi_q + \varepsilon^{ijpq}\xi^h\varphi^k\varphi_p\xi_q + \varepsilon^{ikpq}\xi^h\varphi^j\varphi_p\xi_q +$$

$$-\varepsilon^{hjpq}\varphi^i\xi^k\varphi_p\xi_q - \varepsilon^{hkpq}\varphi^i\xi^j\varphi_p\xi_q - \varepsilon^{ijpq}\varphi^h\xi^k\varphi_p\xi_q - \varepsilon^{ikpq}\varphi^h\xi^j\varphi_p\xi_q] = 0\,. \tag{174}$$

As usual, we shall examine (174) at an arbitrary point P of our manifold in terms of the frame $\{X_\alpha{}^i\} = \{\varphi^i, \xi^i, V^i, W^i\}$. If we contract (174) with $V_i W_j \varphi_h \xi_k$ we discover that

$$-g^{-\frac{1}{2}}(\beta_{6X} - \tfrac{1}{2}\beta_{5Z})D\, V_i W_j \varphi_p \xi_q \varepsilon^{ijpq} = 0,$$

and hence

$$\beta_{6X} = \tfrac{1}{2}\beta_{5Z}\,. \tag{175}$$

Next contract (174) with $V_h V_i$ to get

$$-2\varepsilon_V(2\beta_{2X} - \beta_3)[\xi^j\varphi^k + \xi^k\varphi^j] - 4\varepsilon_V(\beta_{2Z} + \beta_4)\xi^j\xi^k = 0,$$

which tells us that

$$2\beta_{2X} = \beta_3 \text{ and } \beta_{2Z} = -\beta_4\,. \tag{176}$$

Lastly, combining (174)−(176) tells us that

$$\beta_{3Z} = -2\beta_{4X}\,. \tag{177}$$

If we apply the transformation (151) to (175)−(177) we obtain the following additional constraints on the coefficient functions in (150)

$$\gamma_{5Y} = \tfrac{1}{2}\gamma_{6Z}\,,\ \gamma_4 = 2\gamma_{1Y}\,,\ \gamma_3 = -\gamma_{1Z} \text{ and } \gamma_{4Z} = -2\gamma_{3Y}\,. \tag{178}$$

In a similar way we can analyze the terms in (150) involving $\varphi_{ab}\xi_{cd}$ to find repeats of the above constraints along with

$$\gamma_4 = -2\beta_{2Y},\ \beta_3 = -2\gamma_{1X},\ 2\beta_{6Y} = \gamma_{6Z} \text{ and } \beta_{5Y} = \gamma_{6X}\,. \tag{179}$$

If we now use the conditions presented in (163)-(165), (169), (171), (175)–(179) in the expression for $\text{Đ}_S{}^i$ given in (131) we discover that $\text{Đ}_S{}^i$ has reduced to

$$\begin{aligned}\text{Đ}_S{}^i = {} & g^{\frac{1}{2}}\{\beta_2[\varphi^i{}_a\xi^a - \xi^i\Box\varphi + \varphi^i\Box\xi - \xi^i{}_a\varphi^a] + 2\beta_{2X}[\varphi^i\varphi^a\xi^b\varphi_{ab} - \xi^i\varphi^a\varphi^b\varphi_{ab}] + 2\beta_{2Y}[\varphi^i\xi^a\xi^b\xi_{ab} - \xi^i\xi^a\varphi^b\xi_{ab}] \\ & + \beta_{2Z}[\varphi^i\xi^a\xi^b\varphi_{ab} - \xi^i\xi^a\varphi^b\varphi_{ab} + \varphi^i\varphi^a\xi^b\xi_{ab} - \xi^i\varphi^a\varphi^b\xi_{ab}]\} + \beta_5{*\omega^{ia}}\varphi^b\varphi_{ab} + \gamma_6{*\omega^{ia}}\xi^b\xi_{ab} + \beta_6[{*\omega^{ia}}\xi^b\varphi_{ab} + {*\omega^{ia}}\varphi^b\xi_{ab}] + \\ & + g^{\frac{1}{2}}(\delta_1\varphi^i + \delta_2\xi^i)\,,\end{aligned} \tag{180}$$

where $\beta_2$ is arbitrary while $\beta_5$, $\beta_6$ and $\gamma_6$ must satisfy

$$\beta_{5Y} = \gamma_{6X}\,,\ 2\beta_{6Y} = \gamma_{6Z}\,,\ 2\beta_{6X} = \beta_{5Z}\,. \tag{181}$$

Constraints on $\delta_1$ and $\delta_2$ are yet to be determined, and come from an analysis of the $\varphi_{ab}$ and $\xi_{ab}$ terms in (150). In order to understand the implications of (180) let us examine the two Lagrangians

$$L_1 := g^{1/2}\sigma_1\omega^{ab}F_{ab} \text{ and } L_2 := g^{1/2}\sigma_2\omega_{ab}{}^*F^{ab} \tag{182}$$

where $\sigma_1$ and $\sigma_2$ are functions of $\varphi$, $\xi$, X, Y and Z. Using (27) we easily find that

$$E^i(L_1) = g^{1/2}\{\sigma_1[\varphi^i{}_a\xi^a - \xi^i\Box\varphi + \varphi^i\Box\xi - \xi^i{}_a\varphi^a] + 2\sigma_{1X}[\varphi^i\varphi^a\xi^b\varphi_{ab} - \xi^i\varphi^a\varphi^b\varphi_{ab}] + 2\sigma_{1Y}[\varphi^i\xi^a\xi^b\xi_{ab} - \xi^i\xi^a\varphi^b\xi_{ab}] +$$

$$+ \sigma_{1Z}[\varphi^i\xi^a\xi^b\varphi_{ab} - \xi^i\xi^a\varphi^b\varphi_{ab} + \varphi^i\varphi^a\xi^b\xi_{ab} - \xi^i\varphi^a\varphi^b\xi_{ab}] + \varphi^i[\sigma_{1\varphi}Z + \sigma_{1\xi}Y] - \xi^i[\sigma_{1\varphi}X + \sigma_{1\xi}Z]\}, \tag{183}$$

$$E^i(L_2) = 4\sigma_{2X}{}^*\omega^{ij}\varphi^c\varphi_{cj} + 4\sigma_{2Y}{}^*\omega^{ij}\xi^c\xi_{cj} + 2\sigma_{2Z}[{}^*\omega^{ij}\varphi^c\xi_{cj} + {}^*\omega^{ij}\xi^c\varphi_{cj}]. \tag{184}$$

Upon comparing (180) with (183), (184) we see that if choose $\sigma_1=\beta_2$ and can find a function $\sigma_2$ which is such that

$$4\sigma_{2X} = \beta_5,\ \ 4\sigma_{2Y} = \gamma_6 \text{ and } 2\sigma_{2Z} = \beta_6 \tag{185}$$

then

$$\text{Đ}_S{}^i = E^i(L_1) + E^i(L_2) + g^{1/2}(\delta'_1\varphi^i + \delta'_2\xi^i) \tag{186}$$

where $\delta'_1$ and $\delta'_2$ are yet to be determined. However $\delta'_1$ and $\delta'_2$ must vanish. This is so since (186) tells us that $g^{1/2}(\delta'_1\varphi^i + \delta'_2\xi^i)$ must be a divergence free concomitant which is of zeroth degree in the second derivatives of the scalar fields. We previously demonstrated that all such concomitants vanish in a space of four dimensions. To complete our proof that

$$\text{Đ}_S{}^i = E^i(L_1) + E^i(L_2) \tag{187}$$

I need to prove that we can always solve (185). This can be done as follows.

Suppose we are given three functions $\beta_5$, $\beta_6$ and $\gamma_6$ of $\varphi$, $\xi$, X, Y and Z which satisfy (185). Let us define $\sigma_2$ by an indefinite integral with respect to X as

$$\sigma_2 := \tfrac{1}{4}\textstyle\int\beta_5 dX + F(\varphi,\xi,Y,Z) \tag{188}$$

where F is an “integration constant” to be determined. So right now $4\sigma_{2X} = \beta_5$ as required. Upon differentiating (188) with respect to Y, noting (182), we get

$$4\sigma_{2Y} = \int\gamma_{6X}dX + F_Y\,. \tag{189}$$

Since we know what $\gamma_6$ is we take $\int\gamma_{6X}dX = \gamma_6$. Thus (189) and the requirement that $4\sigma_{2Y} = \gamma_6$ implies that $F_Y$ vanishes. Continuing we differentiate (188) with respect to Z, noting (182), to deduce that the requirement that $2\sigma_{2Z} = \beta_6$ requires $F_Z = 0$. Thus we have shown that the function $\sigma_2$ defined by (188) satisfies (185) when F is an arbitrary function of $\varphi$ and $\xi$.

Consequently we have shown that $\text{Đ}_S{}^i$ is indeed given by (187), where $\sigma_1$ and $\sigma_2$ are essentially arbitrary functions of $\varphi$, $\xi$, X, Y and Z. In passing one should note that when $\sigma_2$ is only a function of $\varphi$ and $\xi$, the Lagrangian $L_2$ is a divergence, and hence its Euler-Lagrange tensors vanish. As a result if you were to add an arbitrary function of $\varphi$ and $\xi$ to $\sigma_2$ it would not change the Euler-Lagrange equations obtained from $L_2$.